\documentclass[11pt]{article}
\title{Universal BPS structure of stationary supergravity solutions}

\global\arraycolsep=1pt
\usepackage[colorlinks=true,backref=true,linkcolor=black,anchorcolor=black,citecolor=black,filecolor=black,menucolor=black,pagecolor=black,urlcolor=black]{hyperref}

\usepackage{amsmath}
\usepackage{amsthm}
\usepackage{amssymb}
\usepackage{mathrsfs}
\usepackage[Symbol]{upgreek}
\usepackage{dsfont}
\usepackage[vcentermath]{youngtab}
\usepackage{textcomp}

\usepackage{latexsym}
\usepackage{graphicx}

\newcommand{\eprint}[1]{{\href{http://arxiv.org/abs/#1}{\texttt{[#1}]}}}
\newcommand{\eprintN}[1]{{\href{http://arxiv.org/abs/#1}{\texttt{#1 [hep-th]}}}}

\def\half{\frac{1}{2}}
\def\bsh{\backslash}

\newfont{\bbbold}{msbm10 scaled \magstep1}

\def\bbC{\mbox{\bbbold C}}

\def\bbR{\mbox{\bbbold R}}

\def\cA{{\cal A}}

\def\cD{{\cal D}}
\def\cE{{\cal E}}
\def\cF{{\cal F}}

\def\cL{{\cal L}}

\def\cN{{\cal N}}
\def\cO{{\cal O}}

\def\cS{{\cal S}}

\def\cU{{\cal U}}
\def\cV{{\cal V}}
\def\cW{{\cal W}}

\newfont{\goth}{eufm10 scaled \magstep1}

\def\go{\mbox{\goth o}}
\def\gp{\mbox{\goth p}}

\def\gs{\mbox{\goth s}}

\def\gu{\mbox{\goth u}}

\def\a{\alpha}\def\adt{\dot \alpha}
\def\b{\beta}\def\bdt{\dot \beta}
\def\c{\gamma}\def\cdt{\dot\gamma}
\def\d{\delta}\def\ddt{\dot\delta}
\def\ve{\varepsilon}
\def\F{\Phi}
\def\h{\eta}
\def\k{\kappa}
\def\l{\lambda}

\def\th{\theta}

\def\be{\begin{equation}}\def\ee{\end{equation}}
\def\bea{\begin{eqnarray}}\def\eea{\end{eqnarray}}
\def\barr{\begin{array}}\def\earr{\end{array}}

\def\O{\Omega}
\def\del{\partial}

\def\xz{\times}

\let\la=\label

\def\nn{\nonumber}
\def\bd{\begin{document}}
\def\ed{\end{document}}
\def\ba{\begin{array}}
\def\ea{\end{array}}
\def\bea{\begin{eqnarray}}
\def\eea{\end{eqnarray}}
\def\ft#1#2{{\frac{\scriptstyle #1}{\scriptstyle #2}}}
\def\fft#1#2{\frac{#1}{#2}}
\def\sst#1{{\scriptscriptstyle #1}}
\def\oneone{\rlap 1\mkern4mu{\rm l}}

\newcommand{\eq}[1]{(\ref{#1})}
\newcommand{\w}[1]{\\[0.#1cm]}
\def\eqs#1#2{(\ref{#1}-\ref{#2})}
\def\det{{\rm det\,}}
\def\tr{{\rm tr}}

\newcommand{\hoch}[1]{$\, ^{#1}$}
\newcommand{\imperial}{\it\small Theoretical Physics Group, Imperial College London\\ Prince Consort Road, London SW7 2AZ, UK}
\newcommand{\kings}
{\it\small Department of Mathematics, King's College, University of London\\ Strand, London WC2R 2LS, UK}
\newcommand{\uu}
{\it\small Department of Theoretical Physics, Uppsala, Sweden}
\newcommand{\hip}
{\it\small HIP-Helsinki Institute of Physics, P.O. Box 64 FIN-00014
University of Helsinki, Suomi-Finland}
\newcommand{\stock}
{\it\small Department of Theoretical Physics, Stockholm, Sweden}
\newcommand{\cpht}
{\it\small Centre de Physique Th{\'e}orique, Ecole Polytechnique, CNRS\\ 91128 Palaiseau Cedex, France}
\makeatletter
\renewcommand\theequation{\thesection.\arabic{equation}}
\@addtoreset{equation}{section} \makeatother

\newcommand{\sa}{/ \hspace{-1.2ex}}
\newcommand{\saa}{/ \hspace{-1.4ex}}
\newcommand{\saaa}{\, / \hspace{-1.6ex}}
\newcommand{\Scal}[1]{\Bigl ({#1} \Bigr )}
\newcommand{\scal}[1]{\bigl ({#1} \bigr )}

\newcommand{\CR}{\nonumber \\*}

\newcommand{\trace}{\hbox {tr}~}
\newcommand{\traceS}{\hbox {tr}_{\scriptscriptstyle \mathfrak{S}}~}

\DeclareMathAlphabet{\mathpzc}{OT1}{pzc}{m}{it}
\def\BRST{\,\mathpzc{s}\,}
\def\aBRST{{\scriptstyle (\mathpzc{s})}}
\def\q{{{\scriptscriptstyle (Q)}}}
\def\qs{{\scriptscriptstyle (Q\mathpzc{s})}}
\def\Qsla{{\mathcal{S}_{\q}}}
\def\Slav{{\mathcal{S}_\aBRST}}
\def\epsilonb{{\overline{\epsilon}}}
\def\bulletup{{\scriptstyle \bullet}}

\newcommand{\gra}[2]{{\scriptscriptstyle (#1 , #2 )}}
\newcommand{\ord}[1]{{\scriptscriptstyle (#1)}}

\def\cL{{\cal L}}
\def\cN{\mathcal{N}}
\def\cO{\mathcal{O}}

\def\ie{{\it i.e.}\ }
\def\eg{{\it e.g.}\ }

\newcommand{\sfrac}[2]{{\scriptstyle \frac{#1}{#2}}}
\newcommand{\stfrac}[2]{{\scriptscriptstyle \frac{#1}{#2}}}

 \def\balpha{{\overline{\alpha}}}
 \def\bbeta{{\overline{\beta}}}
 \def\bgamma{{\overline{\gamma}}}
 \def\bdelta{{\overline{\delta}}}
 \def\bepsilon{{\overline{\epsilon}}}
 \def\bvarepsilon{{\overline{\varepsilon}}}
 \def\bzeta{{\overline{\zeta}}}
 \def\bareta{{\overline{\eta}}}
 \def\btheta{{\overline{\theta}}}
 \def\bvartheta{{\overline{\vartheta}}}
 \def\biota{{\overline{\iota}}}
 \def\bkappa{{\overline{\kappa}}}
 \def\blambda{{\overline{\lambda}}}
 \def\bmu{{\overline{\mu}}}
 \def\bnu{{\overline{\nu}}}
 \def\bxi{{\overline{\xi}}}
 \def\bpi{{\overline{\pi}}}
 \def\brho{{\overline{\rho}}}
 \def\bvarrho{{\overline{\varrho}}}
 \def\bsigma{{\overline{\sigma}}}
 \def\bvarsigma{{\overline{\varsigma}}}
 \def\btau{{\overline{\tau}}}
 \def\bphi{{\overline{\phi}}}
 \def\bvarphi{{\overline{\varphi}}}
 \def\bchi{{\overline{\chi}}}
 \def\bpsi{{\overline{\psi}}}
 \def\bomega{{\overline{\omega}}}

\def\thalf{{\textrm{\tiny\textonehalf}}}
\def\tquarter{{\textrm{\tiny\textonequarter}}}
\def\Ko{{\scriptscriptstyle K}}
\def\tKo{\scriptscriptstyle k }
\def\N{{\mathcal{N}}}
\def\csN{{\fontsize{9.35pt}{9pt}\selectfont \mbox{$\cN$} \fontsize{12.35pt}{12pt}\selectfont }}
\def\cssN{{\fontsize{6.35pt}{6pt}\selectfont \mbox{$\cN$} \fontsize{12.35pt}{12pt}\selectfont }}
\def\csssN{{\fontsize{4.35pt}{4pt}\selectfont \mbox{$\cN$} \fontsize{12.35pt}{12pt}\selectfont }}
\def\UT{{\fontsize{9.35pt}{12pt}\selectfont   \mbox{$\frac{1}{\sqrt{1-T \bar T}}$}\fontsize{12.35pt}{12pt}\selectfont }}
\def\Utau{{\fontsize{9.35pt}{12pt}\selectfont   \mbox{$\frac{ i + \uptau}{2 \sqrt{\rm{Im}[\uptau]}}$}\fontsize{12.35pt}{12pt}\selectfont }}
\def\ai{{\hat{\imath}}}
\def\aj{{\hat{\jmath}}}
\def\ak{{\hat{k}}}

\def\un{{\mathpzc{1}}}
\def\deux{{\mathpzc{2}}}
\def\trois{{\mathpzc{3}}}
\def\quatre{{\mathpzc{4}}}
\def\cinq{{\mathpzc{5}}}

\newcommand{\red}[1]{ {\color{red} #1 }} 
\newcommand{\blue}[1]{{\color{blue} #1 }}
\newcommand{\green}[1]{{\color{green} #1 }}
\newcommand{\bleu}[1]{ {\color{cyan} #1 }} 

\renewcommand{\thefootnote}{\arabic{footnote}}

\begin{document}

\newcommand{\auth}{\large G.\ Bossard\footnote{email: bossard@cpht.polytechnique.fr}, P.S.\ Howe\footnote{email: paul.howe@kcl.ac.uk} and K.S.\ Stelle\footnote{email: k.stelle@imperial.ac.uk}}

\thispagestyle{empty}

\renewcommand{\thefootnote}{\arabic{footnote}}

\null
\begin{flushright}
{\small CPHT-RR027.0413}\\
{\small KCL-MTH-13-05}\\
{\small Imperial/TP/13/KSS/01}\\
\vskip 1 cm
\end{flushright}

\begin{center}
{\Large{\bf  Invariants and divergences in half-maximal supergravity theories}}
\vspace{.75cm}

\auth

\vspace{.5cm}

\begin{itemize}
\item [$^1$]\cpht\item[$^2$] \kings \item [$^3$] \imperial
\end{itemize}
\vspace{.4cm}

%%%%%%%%%%%%%%%%%%%%%%%%%
{\bf Abstract}
\end{center}
%%%%%%%%%%%%%%%%%%%%%%
The invariants in half-maximal supergravity theories in $D=4,5$ are discussed in detail up to dimension eight (\eg $R^4$). In $D=4$, owing to the anomaly in the rigid $SL(2,\bbR)$ duality symmetry, the restrictions on divergences need careful treatment. In pure $\cN=4$ supergravity, this anomalous symmetry still implies duality invariance of candidate counterterms at three loops. Provided one makes the additional assumption that there exists a full 16-supercharge off-shell formulation of the theory, counterterms at $L\ge2$ loops would also have to be writable as full-superspace integrals. At the three-loop order such a duality-invariant full-superspace integral candidate counterterm exists, but its duality invariance is marginal in the sense that the full-superspace counter-Lagrangian is not itself duality-invariant. We show that such marginal invariants are not allowable as counterterms in a 16-supercharge off-shell formalism. It is not possible to draw the same conclusion when vector multiplets are present because of the appearance of $F^4$ terms in the $SL(2,\bbR)$ anomaly. In $D=5$ there is no one-loop anomaly in the shift invariance of the dilaton, and we argue that this implies finiteness at two loops, again subject to the assumption that 16 supercharges can be preserved off-shell.
%%%%%%%%%%%%%%%%%%%%%%%%%%

\vspace{.5cm}

\renewcommand{\thefootnote}{\arabic{footnote}}
\setcounter{footnote}{0}

\pagebreak
\tableofcontents
\setcounter{page}{1}

%%%%%%%%%%%%%%%%%%%%%%%%%%%%%%%%%%%%%%%%%%%%%%%%%%%%%%%%%%%%%%%%%%%%%%%%%%%%%%%%%%%%%%
\section{Introduction}
%%%%%%%%%%%%%%%%%%%%%%%%%%%%%%%%%%%%%%%%%%%%%%%%%%%%%%%%%%%%%%%%%%%%%
Developments in the evaluation of scattering amplitudes using unitarity methods over the past decade or so have made it possible to push the investigation of the onset of ultra-violet divergences in maximal supergravity theories to higher loop orders than would have been possible using conventional Feynman diagram techniques. In particular, it has been shown that $D=4,\  \cN=8$ supergravity is finite at three loops ($R^4$) \cite{Bern:2007hh}, and that $D=5$ maximal supergravity is finite at four loops ($\partial^6 R^4$) \cite{Bern:2009kd}, despite the existence of corresponding counterterms, at least at the linearised level \cite{Kallosh:1980fi,Howe:1981xy,Bossard:2009sy}. Since these invariants are of F-type, \ie correspond to integrals over fewer than the maximal number of odd superspace coordinates, it might have been thought that they should be protected by superspace non-renormalisation theorems \cite{Bossard:2009sy}, but it is difficult to justify this argument because there are no known off-shell versions of maximal supergravity that realise all of the supersymmetries linearly. Indeed, such off-shell versions cannot exist in every dimension because it is known that divergences do occur for F-type counterterms in $D=6$ and $D=7$ above one loop \cite{Bern:1998ug}. However, these finiteness results can be explained instead by duality-based arguments. $E_{7(7)}$ Ward identities can be defined at the cost of manifest Lorentz covariance \cite{Hillmann:2009zf,Bossard:2010dq}, and can be shown to be non-anomalous.\footnote{The absence of a supersymmetric anomaly for the $E_{7(7)}$ Ward identities that cannot be removed by supersymmetric non-invariant counterterms has not been rigourously established at all orders in perturbations theory. Nonetheless, the complete characterisation of the supersymmetry invariants of type $R^4$, $\partial^4 R^4$, $\partial^6 R^4$, $\partial^8 R^4$   
\cite{Elvang:2010kc,Drummond:2003ex,Bossard:2011tq} allows one to prove that such an anomaly cannot appear before eight loops.} These Ward identities imply that the counterterms associated to logarithmic divergences must be $E_{7(7)}$ invariant. The unique $SU(8)$ invariant $R^4$ candidate counterterm can be proved to violate $E_{7(7)}$ symmetry from a perturbative scattering amplitude approach \cite{Elvang:2010kc} and from a direct field-theoretic argument \cite{Bossard:2010bd} that makes use of dimensional reduction and of the uniqueness of the $D=4$ counterterms at the linearised level \cite{Drummond:2003ex}.  In addition, there is no superspace measure for the $R^4$ invariant at the full non-linear level, while an analysis of the closed super-four-form that does define this supersymmetric invariant leads to the same conclusion: there is no three-loop acceptable counterterm that is both $\cN=8$ supersymmetric and $E_{7(7)}$ duality invariant \cite{Bossard:2010bd}. Furthermore, these arguments can be extended to the other two F-term invariants in $D=4$ arising at the five and six-loop orders \cite{Bossard:2010bd,Beisert:2010jx}, there being no four-loop invariant \cite{Drummond:2003ex}.  One can then use dimensional reduction and the known divergences at one, two and three loops in $D=8,7$ and 6, respectively,  to show that these are the only F-term divergences that can arise in maximal supergravity in any dimension. This result can also be seen  from an analysis of the conjectured duality properties of superstring theory \cite{Green:2010sp,Green:2010kv}.

It therefore seems that maximal supergravity must be ultra-violet finite up through at least six loops in $D=4$, and that there are no divergences that correspond to the known linearised BPS counterterms (F-terms) \cite{Kallosh:1980fi,Howe:1981xy,Drummond:2003ex,Elvang:2010jv,Drummond:2010fp}. At the seven-loop order, we reach the borderline between F-term and D-term invariants. At this order there would seem to be a candidate D-term invariant, the volume of superspace, which is manifestly symmetric with respect to all symmetries and which would be difficult to protect by conventional field-theoretic arguments. However, it is now known that the volume of superspace vanishes on-shell for any $\cN$ in $D=4$, although there is still an $\cN=8$ seven-loop invariant that can be written as a manifestly duality-invariant harmonic-superspacee integral over $28$ odd coordinates \cite{Bossard:2011tq}. The situation at this order is therefore somewhat ambiguous, although it is unlikely that there is an off-shell formulation of the maximal supergravity theory preserving all the supersymmetries linearly which could be used to try to justify the absence of a seven-loop divergence. A direct computational resolution of this ambiguity would seem to be a tall order, at least in the near future, but a similar situation arises in the half-maximal case which is more tractable from both the computational and formal points of view.

In $D=4,\ \cN=4$ supergravity the F/D borderline occurs at the three-loop level, \ie for $R^4$ type counterterms. It has recently been shown that half-maximal supergravity is finite at this order \cite{Bern:2012cd,Tourkine:2012ip,Tourkine:2012vx} and that this state of affairs persists in $D=5$ \cite{Bern:2012gh}  (where the relevant loop order is two) and in the presence of vector multiplets \cite{Tourkine:2012ip,Tourkine:2012vx}. These finiteness results have been obtained from scattering amplitude computations \cite{Bern:2012cd,Bern:2012gh} in pure supergravity and from string theory \cite{Tourkine:2012ip,Tourkine:2012vx} in supergravity coupled to vector multiplets. Field-theoretic arguments in support of these results have been given using duality arguments\footnote{A similar controversial argument, that $E_{7(7)}$ symmetry could be much more restrictive and that $\cN=8$ supergravity might in consequence be finite at all orders, was made in \cite{Kallosh:2011dp,Kallosh:2011qt}, despite the existence of an infinite number  of duality-invariant full-superspace counterterms starting at eight loops \cite{Kallosh:1980fi,Howe:1980th}.} \cite{Kallosh:2012ei} and conformal symmetry \cite{Ferrara:2012ui}. From the counterterm point of view, the situation resembles seven loops in $\cN=8$ because the natural candidate for the $R^4$ invariant would be the volume of superspace. As in $\cN=8$, this turns out to vanish in both $D=4$ and $5$, but in both cases one can also construct $R^4$ invariants as harmonic-superspacee integrals over twelve odd coordinates instead of the full sixteen. As we shall show, duality-invariant  counterterms of this type can be re-expressed as full-superspace integrals with integrands that are not themselves duality invariant. The issue is therefore to understand if this property is enough to rule out these counterterms as possible divergences. In order to do so, one has to assume that there are off-shell formulations of the theories under consideration that preserve all the supersymmetries as well as the duality symmetries, or at least a large enough subgroup thereof.

Perhaps the simplest case to consider is half-maximal $D=5$ supergravity which has only one scalar, the dilaton $\F$, and for which the duality symmetry is simply a shift of this field. As mentioned above, the superspace volume vanishes, but the full-superspace integral of any function of $\F$ will give rise to a supersymmetric $R^4$ invariant. The only choice of function for which this integral is shift-invariant is a linear one, because under a shift this would give rise to the volume, which vanishes. Moreover, one can rewrite this invariant as a twelve-theta harmonic-superspacee integral of a Grassmann-analytic function that  is manifestly shift-invariant. In a sense, one can regard $\F$ as a zero-form potential for a gauge transformation with a closed zero-form parameter, \ie a constant, and so the integral could be thought of as the simplest type of Chern--Simons invariant. One could therefore rule this out as a divergence if one could show that divergences have to correspond to full-superspace integrals of integrands that are manifestly invariant under duality transformations.

For $D=4,\ \cN=4$ supergravity, the duality group is $SL(2,\bbR)$. This symmetry is anomalous \cite{Marcus:1985yy}, but we shall show that the anomalous Ward identities still require the three-loop counterterm to be duality-invariant. It turns out that this counterterm can be expressed either as a twelve-theta harmonic-superspacee integral with a manifestly duality-invariant integrand or as a full-superspace integral of the K\"ahler potential on the scalar manifold. This potential can be regarded as a local function depending on the complex scalar field $T$, the unique independent chiral superfield in $\cN=4$ supergravity, and on its complex conjugate. One would therefore need to show that the only allowable counterterms should be full-superspace integrals of integrands that are manifestly symmetric under the isometries of this space in order to rule out a possible divergence at three loops. This case has some features in common with (2,2) non-linear sigma models in $D=2$, which also have K\"ahler target spaces. In \cite{Howe:1986ys} it was shown, using the background field method in superspace, that all possible counterterms beyond one loop have to be full-superspace integrals of tensorial functions of the background field, \ie constructed from the Riemann tensor and its target-space derivatives. So in the case of a symmetric target space these functions would be constant, \ie manifestly invariant under the isometries. If we could prove a similar result for $\cN=4$ supergravity then three-loop finiteness would be a consequence.

As well as studying half-maximal supergravity theories we shall also consider the inclusion of $n$ extra vector multiplets. In $D=4$ this leads to an additional $SO(6,n)$ duality symmetry that acts on the manifold of the extra scalars, while in $D=5$ we have $SO(5,n)$. In five dimensions there are no one-loop divergences to complicate matters and we shall argue that the extra multiplets do not affect the counterterm analysis at two loops. In four dimensions, on the other hand, there is a one-loop divergence corresponding to an $F^4$ counterterm \cite{Fischler:1979yk}. We shall show that the full super-invariant corresponding to this is also invariant under $SL(2,\bbR) \xz SO(6,n)$, a fact that is non-trivial to prove. This invariant turns out to affect the analysis of divergences at three loops because the one-loop anomaly functional, in the presence of vector multiplets, necessarily includes an $F^4$ term. The presence of this term implies that the theory is no longer invariant at one loop under shifts of the dilaton, which it is in the absence of vector multiplets.

In order to argue our case, we shall have to assume the existence of a suitable off-shell formalism that preserves all of the supersymmetries linearly. It was shown some time ago by a counting argument \cite{Siegel:1981dx,Rivelles:1982gn} that, in ordinary superspace with a finite number of component fields, this is only possible for $\cN=4$ supergravity coupled to $6+8k$ vector multiplets. The case $k=0$ corresponds to dimensional reduction of the off-shell $D=10,\ \cN=1$ construction given in \cite{Howe:1982mt} which makes use of a six-form potential rather than the usual two-form. However, even if one were able to construct these multiplets for arbitrary $k$, there would still be a difficulty with duality symmetry because some of the scalars appear as antisymmetric two-form potentials off-shell. For example, for $k=0$, \ie with six vector multiplets, one finds that the $36$ scalars appear as $20+1$ scalars and $15$ two-forms. For this reason, we shall instead assume that a suitable off-shell theory can be constructed in harmonic superspace even though no one has succeeded in doing this for a single $\cN=4$ vector multiplet, owing in part to the vector multiplet's self-conjugate nature.\footnote{Notwithstanding the fact that there is an off-shell version of $\cN=3$ Yang--Mills theory, which has the same physical spectrum as $\cN=4$, in harmonic superspace \cite{Galperin:1985uw}.} This difficulty should be contrasted with the successful construction of off-shell harmonic-superspacee formulations of the 8-supercharge hypermultiplet in $D=4,5\ \&\ 6$. Similarly to the `finite-field' $\cN=4$ cases, there is a  ``no-go'' theorem for `finite-field' off-shell formulations of hypermultiplets where the physical modes involve only scalar fields \cite{Howe:1985ar}, but harmonic-superspacee constructions, with infinite numbers of component fields, nonetheless exist \cite{Galperin:1984av,Karlhede:1984vr}. So the question of whether an off-shell formulation exists for half-maximal supergravity and its vector-multiplet couplings still remains open.

To summarise, the dimension-eight ($R^4$) invariants in pure half-maximal supergravity theories in $D=4,5$ can be considered to be on the F/D borderline because they can be expressed either as integrals of duality-invariant integrands over twelve odd coordinates or as full-superspace integrals whose integrands are not themselves invariant even though the integrals are. If we make the assumption that there exist off-shell versions of these theories that preserve all of the supersymmetries linearly as well as the duality symmetries, then the divergences would have to correspond to full-superspace integrals with integrands that are duality-invariant and hence would be absent at the corresponding (3,2) loop orders. When vector multiplets are present, this result still holds in $D=5$ but is vitiated in $D=4$ owing to the fact that the anomaly is altered in the presence of the vector multiplets. In $D=4$, the requirement of duality symmetry is not compatible with manifest Lorentz invariance, so this would presumably have to be given up in superspace, as it must in the component formalism \cite{Bossard:2010dq,Henneaux:1988gg}. In superspace, it might be possible to make use of light-cone harmonic-superspacee techniques to help out in this context \cite{Sokatchev:1985tc,Delduc:1989ah,Galperin:1991gk}. In $D=5$, on the other hand, duality is fully compatible with manifest Lorentz symmetry.

Although our main motivation is the understanding of the ultra-violet behaviour of half-maximal supergravity theories, our classification of supersymmetry invariants is also of interest for understanding string-theory effective actions with half-maximal supersymmetry. In particular, the non-linear structure of the $R^2$ type supersymmetry invariant is shown to be such that the threshold function multiplying the $F^4$ type structure is not independent of the $R^2$ threshold. Our results are less conclusive for higher-derivative invariants like $\partial^2 F^4$, since we do not discuss in this paper the higher-order corrections required to promote  a supersymmetry invariant considered modulo the equations of motion to an actual supersymmetric effective action. Nonetheless, the structure of $R^4$ type invariants is clarified, shedding light on the property that the tree-level $R^4$ invariant does not get corrected at one- and two-loops in heterotic string theory.

 In the next section we give a brief summary of $D=4$ supergeometry, including vector multiplets, and then continue in Section \ref{sec3} with a comprehensive study of invariants up to dimension eight. In order to do this, we make extensive use of harmonic-superspacee methods and the ectoplasm formalism. At dimension four there are $R^2$ and $F^4$ invariants, both of which appear in the cocycle (\ie closed super-$D$-form) that determines the $SL(2,\bbR)$ anomaly in the presence of vector multiplets. The $F^4$ invariant can be written as a harmonic-superspacee integral that is not manifestly $SO(6,n)$ invariant, but that can also be expressed as a cocycle that is so invariant. At dimension six, the $\del^2 F^4$ one-quarter BPS invariant is also writable as a harmonic-superspacee integral, but in this case we have not been able  to establish $SO(6,n)$ invariance. In addition, there are invariants involving $R^2 F^2$ terms. At dimension eight, we show that the volume of superspace vanishes even in the presence of vector multiplets. As a consequence, the $R^4$ type duality invariant can be written either as a sixteen-theta full-superspace integral of a non-duality-invariant function of the $\cN=4$ supergravity complex scalar field $T$ or as a manifestly invariant twelve-theta harmonic-superspacee integral. There is also a $\partial^4 F^4$ type duality invariant  that can only be written as a twelve-theta harmonic-superspacee integral. In Section \ref{sec4}, we discuss the $SL(2,\bbR)$ anomaly. It is given by a dimension-four $(R\wedge R)$-type invariant which involves $F^4$ in the presence of vector multiplets, and in the following section we discuss the implications of the anomaly in perturbation theory. This section includes a brief recap of superspace non-renormalisation theorems as well as a discussion of the notion of superspace co-forms   that is needed in order to justify non-renormalisation of marginal counterterms. The non-renormalisation theorem is illustrated by the example of $(2,2)$ non-linear sigma models in $D=2$.  A brief discussion of pure $\cN=4, D=4$ supergravity was given in 
reference \cite{Bossard:2012xs}.

In Sections 6 and 7, we discuss the situation in $D=5$. We give some details of $D=5$ supergeometry, starting with the maximal $\cN=4$ case, and then go on to discuss the invariants for the half-maximal $\cN=2$ case, up to dimension eight. We show that the volume of $D=5$ half-maximal superspace vanishes (although it does not in the $D=5$ maximal case), and analyse the effect that this has on dimension-eight invariants. Since there is no one-loop anomaly for the $D=5$ supergravity duality symmetry, which is simply invariance under shifts of the dilaton, it is more straightforward to argue that half-maximal theories should be two-loop finite in $D=5$, given the assumption of a suitable off-shell formulation. We also show that certain $R^4$-type invariants cannot be written as full-superspace integrals, depending on the function of the dilaton that multiplies $R^4$ in the spacetime invariant. In the context of the heterotic string, we argue that this suggests that $R^4$ is only protected, \ie of BPS type, for loop-orders one to four, but not beyond. We state our conclusions in Section 8. 

The borderline F/D problem is difficult to analyse from our field-theoretic point of view, owing to the need to assume the existence of a full off-shell formalism, but for higher loops there will certainly be candidate counterterms that are purely D-type and whose integrands are invariant with respect to all known symmetries. In this sense, an unambiguous test of ``miraculous'' ultra-violet cancellations in half-maximal supergravity will require calculations at one loop higher than those that have been carried out to date.

%%%%%%%%%%%%%%%%%%%%%%%%%%%%%%%%%%%%%%%%%%%%%%%%%%%%%%%%%%%%%%%%%%%
\section{Supergeometry of $\N=4$ supergravity theories in four dimensions}
%%%%%%%%%%%%%%%%%%%%%%%%%%%%%%%%%%%%%%%%%%%%%%%%%%%%%%%%%%%%%%

In the superspace formulation of  $D=4, \cN$-extended supergravity there are preferred basis one-forms $E^A=(E^a, E^\a_i, \bar E^{\adt i})$ related to the coordinate basis forms by the supervielbein matrix $E^A= dz^M E_M{}^A$, where $a,\a,\adt$ are respectively Lorentz vector and two-component spinor indices and $i=1,\ldots \csN$ is an internal index for the fundamental representation of $U(\csN)$ ($SU(8)$ for $\cN=8$). The structure group is $SL(2,\bbC)\xz (S)U(\csN)$. The connection $\O_A{}^B$ and the curvature $R_A{}^B$ are respectively one- and two-forms that take their values in the Lie algebra of this structure group, so that there are no mixed even-odd components. 

For general $\cN$, one can impose a conformal constraint on the dimension-zero torsion requiring that it be flat, \ie that the only non-vanishing part is \cite{Howe:1981gz}

\be 
T_\alpha^i{}_{\dot{\beta} j}{}^c = - i \delta^i_j \sigma^c_{\alpha\dot{\beta}}\ . \ee
This constraint, together with the imposition of conventional ones, implies
 that the only non-vanishing component of the dimension-one-half component of the torsion is 
\be
T_\a^i{}_\b^j{}^{\cdt k}=\ve_{\a\b}\bar\chi^{ijk\cdt}
\ee
together with its complex conjugate. For $\cN\leq 4$ this leads to an off-shell conformal supergravity multiplet \cite{Howe:1980sy}, while for $\cN>4$ one finds a partially off-shell system  \cite{Howe:1981gz,Gates:1982ae}. To go on-shell in Poincar\'e supergravity, one needs to specify in addition a number of dimension-one superfields in terms of the physical fields. One can show that there are additional field strengths corresponding to the vector and scalar fields in the theory, as discussed in detail in \cite{Brink:1979nt,Howe:1981gz}. The geometry relevant for $\cN=4$ supergravity coupled to $n$ vector multiplets can be obtained by truncating  $\cN=8$ supergravity to $\cN=4$ supergravity coupled to six vector multiplets and then extending to $n$ using $SO(n)$ symmetry. 

%%%%%%%%%%%%%%%%%%
\subsection{Field-strength tensors}
%%%%%%%%%%%%%%%%%
In this subsection we give the non-vanishing components of the superspace field-strength tensors up to dimension one. As well as the torsion and the curvature we also have the field-strengths for the vectors and the scalars.
In $\cN=4$ we can replace the three-index spinor field by a one-index one, $\chi_{\a ijk}=-\ve_{ijkl} \chi^l_{\a}$, so that the dimension-one-half torsion can be rewritten as
\be T_{\alpha}^i{}_\beta^j{}^{\dot{\gamma} k} = - \varepsilon_{\alpha\beta} \varepsilon^{ijkl} \bar\chi^{\dot{\gamma}}_l \ .
\ee
The non-vanishing components of the torsion at dimension one are
\bea
T_{\a\adt,}{}_{\b ,}^j{}_{\cdt}^k&=&i\ve_{\a\b} \bar M_{\adt\cdt}^{jk} -\frac{i}{4}\ve_{\adt\cdt}\ve^{jklm} \l_{\a l}^A\l_{\b m A}\ , \nn\w1
T_{\a\adt,}{}_{\b ,}^j{}_{\c k}&=&-i\ve_{\a\b} \chi_\c^j\bar\chi_{\adt k}+ \frac{i}{8} \delta_k^j \scal{  2 \ve_{\c(\a} \chi_{\b)}^l \bar\chi_{\adt l} + 5 \ve_{\a\b} \chi_\c^l \bar \chi_{\adt l}} \nn\w1
&\phantom{=}&-i\ve_{\a\b}\bar\l^{j A}_{\adt} \l_{\c k A}-\frac{i}{24}\d^j_k(\ve_{\c(\a}\bar\l^{lA}_{\adt} \l_{\b)l A}-\frac{3}{2}\ve_{\a\b} \bar\l^{l A}_{\adt}\l_{\c l A})\ ,
\eea
where $M_{\a\b ij}$ is the field-strength for the six vectors in the supergravity multiplet and $\lambda_{\a i A}$, $A=1,\ldots n$, denote the $n$ vector-multiplet spinor fields.\footnote{The index $A$ is also used as a tangent space super-index but it should be clear from the context which is meant.} The non-vanishing dimension-one components of the Lorentz curvature are 
\bea
R_\alpha^i{}_\beta^j{}_{\gamma\delta} &=& - \frac{1}{4} \varepsilon^{ijkl} \varepsilon_{\alpha\beta} \lambda_{\gamma k A} \lambda_{\delta l}{}^A\ ,\nn\w1
R_\alpha^i{}_\beta^j{}_{\dot{\gamma}\dot{\delta}}& =& - \varepsilon_{\alpha\beta} \bar M_{\dot{\gamma}\dot{\delta}}^{ij}\ ,\nn\w1
R_\alpha^i{}_{\dot{\beta}j}{}_{\gamma\delta} &= &\varepsilon_{\alpha(\gamma} \scal{  \chi^i_{\delta)} \bar \chi_{\dot{\beta}j} + \sfrac{1}{2} \bar \lambda_{\dot{\beta}}^{i A} \lambda_{\delta)jA}} - \frac{1}{2} \delta^i_j \varepsilon_{\alpha(\gamma} \scal{   \chi^k_{\delta)} \bar \chi_{\dot{\beta}k} + \sfrac{1}{2} \bar \lambda_{\dot{\beta}}^{k A} \lambda_{\delta)kA}}
\eea
together with their complex conjugates. The dimension-one components of the $U(4)$ curvature, $R^k{}_l$, are
\be
R_\alpha^i{}_\beta^j{}^k{}_l =\frac{1}{4} \varepsilon_{\alpha\beta} \varepsilon^{ijkp} \varepsilon^{\gamma\delta} \lambda_{\gamma p A} \lambda_{\delta l}{}^A - \frac{1}{2} \delta_l^{(i} \varepsilon^{j)kpq} \lambda_{\alpha p A} \lambda_{\beta q}{}^A\ , \label{Rsu4ab} 
\ee
its complex conjugate, and
\bea R_\alpha^i{}_{\dot{\beta}j}{}^k{}_l &=& - \frac{1}{2} \delta^k_l  \chi_\alpha^i \bar \chi_{\dot{\beta}j}   - \frac{1}{2} \delta^i_l \bar \lambda_{\dot{\beta}}^{k A} \lambda_{\alpha j A } -  \frac{1}{2} \delta^k_j \bar \lambda_{\dot{\beta}}^{i A} \lambda_{\alpha l A } +\nn\w1&\phantom{=}& + \frac{1}{4} \delta^k_l\bar  \lambda_{\dot{\beta}}^{i A} \lambda_{\alpha j A} + \frac{1}{2} \delta^i_j \bar  \lambda_{\dot{\beta}}^{k A} \lambda_{\alpha l A} + \frac{1}{4} ( 2  \delta^i_l \delta^k_j - \delta^i_j \delta^k_l ) \bar \lambda_{\dot{\beta}}^{pA} \lambda_{\alpha p A} \ . \label{Rsu4abd}
\eea
In the absence of the vector multiplets, the $U(4)$ curvature collapses to $U(1)$. The dimension-one $SO(n)$ curvatures are
\be R_\alpha^i{}_\beta^j{}_{AB} = - \frac{1}{2} \varepsilon_{\alpha\beta} \varepsilon^{\gamma\delta} \varepsilon^{ijkl} \lambda_{\gamma k A} \lambda_{\delta l B }  \ , \quad R_\alpha^i{}_{\dot{\beta}j}{}_{AB}  = - \bar  \lambda_{\dot{\beta}}^i{}_{[A} \lambda_{\alpha j B]} + \delta^i_j \bar \lambda_{\dot{\beta}}^k{}_{[A}\lambda_{\alpha k B]} \ . \ee

As well as the geometrical tensors there are (complex) spin-one field strengths for the supergravity vectors, $F_{ij}$, and for the $n$ vector multiplets $F_A$. For the scalars there are one-form field strengths $P$ and $P_{ij A}$. These fields obey the Bianchi identities
\bea
D F_{ij}&=& P\wedge \bar F_{ij} + P_{ij A} \wedge \bar F^A\nn\w1
DF_A&=& P_{A ij}\wedge \bar F^{ij} + \bar P\wedge \bar F_A\ ,
\eea
where antisymmetric pairs of $SU(4)$ indices are raised or lowered with $\half\ve^{ijkl}$. The  one-forms are covariantly constant, $DP=DP_{ij A}=0$, where $D$  is covariant with respect to the local $U(4)\xz SO(n)$. These forms are given by
\bea
P&=&E^\a_i \chi_\a^i + E^a P_a \nn\w1
P_{ij A}&=& E^\a_{[i} \l_{\a j]A}-\half \bar E^{\adt k} \ve_{ijkl} \bar\l^l_{\adt A} + E^a P_{a ij A}\ ,
\la{2.10}
\eea
where the leading components of the dimension-one fields $P_a$ can be thought of as the supercovariant derivatives of the scalar fields.

The non-vanishing components of the two-forms $F_{ij}, F_A$ are given by 
\bea
F_\a^i{}_\b^j{}_{,kl}&=& -2i\d^{ij}_{[kl]}  \nn\w1
F_{\a\adt,\bdt j,kl}&=&-\ve_{\adt\bdt}\ve_{jklm}\chi_\a^m\qquad ; \qquad \qquad \qquad\  F_{\a\adt,\bdt j,A}=\ve_{\adt\bdt} \l_{\a j A}\w1
F_{\a\adt,\b\bdt,kl}&=&-i\ve_{\adt\bdt} M_{\a\b kl}+\frac{i}{4}\ve_{\a\b}\ve_{ijkl}\bar\l^k_{\adt A}\bar\l^{l A}_{\bdt} \ ;\ F_{\a\adt,\b\bdt,A}=-i\ve_{\adt\bdt} M_{\a\b A}-i\ve_{\a\b}\bar\l^k_{(\adt A}\bar\chi_{\bdt) k}\ .\nn
\eea
The $U(4)$ and $SO(n)$ curvatures can be expressed in terms of the one-forms by
\bea
R^i{}_j&=& -\half\d^i_j \bar P\wedge  P -\frac{1}{2} \varepsilon^{ikmn} P_{mn}^A\wedge P_{jk A} \ ,\nn\w1
R_{AB}&=&- \frac{1}{2} \varepsilon^{ijkl}  P_{ij[A} \wedge P_{kl|B]}\ . \label{MaurerCartan} 
\eea
%%%%%%%%%%%%%%%%%%%
\subsection{Scalar fields and supersymmetry relations}
%%%%%%%%%%%%%%%%%%%
The scalar fields in the supergravity sector take their values in the coset $SO(2)\bsh SL(2,\bbR)$ and can be described by an $SU(1,1)\ (\cong SL(2,\bbR))$ matrix $\cV$. The Maurer--Cartan form $d\cV \cV^{-1}$ can be written
\be
d\cV \cV^{-1}=\left(\barr{cc} -2Q & P\cr \bar P & 2Q\earr\right)\ ,
\la{2.4}
\ee
where $Q$ is the composite $U(1)$ connection. The scalar matrix $\cV$ can be parametrised in the form
\be
\cV=\left(\barr{cc} U & UT \cr \bar U\bar T & \bar U\earr\right)
\ee
where 
\be U \bar U ( 1 - T \bar T ) = 1 \ .\ee
The $U(1)$ gauge-invariant complex scalar superfield $T$ can be considered to be a coordinate of the $SO(2)\bsh SL(2,\mathds{R})$ coset on the unit disc. $T$ is chiral while 
\be D_\alpha^i U = \frac{1}{U} \frac{ \bar T }{1 - T \bar T } \chi_\alpha^i \qquad D_\alpha^i T = \frac{1}{U^2} \chi_\alpha^i \ee
and 
\be
P_{\alpha\dot{\beta}} = \sigma^a_{\alpha\dot{\beta}} U^2 D_a T\ .
\ee
The scalar fields in the vector multiplets parametrise the coset $( SO(6) \times SO(n))\bsh SO(6,n)$. They can be represented by an $SO(6,n)$ matrix $\cU$ obeying the constraint $\cU \h \, \cU^T=\h$, where $\h$ is the $SO(6,n)$-invariant metric in $\bbR^{6,n}$. If we set
\be
\cU=\left(\barr{cc} U_{ij}{}^I & \ V_{ij}{}^{\hat{I}}\cr
V_A{}^I &\  U_A{}^{\hat{I}} \earr\right)\ ,
\ee
where the pair $ij$ represents an $SO(6)$ vector index, $A$ is acted on by $SO(n)$, and $I$ run from $1$ to $6$ and $\hat I$ from $7$ to $n+6$ and together make up an $SO(6,n)$ vector index, and where  
\be U_{ij}{}^I = \frac{1}{2} \varepsilon_{ijkl} \bar U^{kl I}   \ , \qquad V_{ij}{}^{\hat{I}} = \frac{1}{2} \varepsilon_{ijkl} \bar V^{kl \hat{I}}\ ,  \ee
then the $SO(6,n)$ constraint can be written 
\be \begin{split} \delta_{IJ} U_{ij}{}^{I} U_{kl}^J - \delta_{\hat{I}\hat{J}} V_{ij}{}^{\hat{I}} V_{kl}{}^{\hat{J}} &= \tfrac{1}{2} \varepsilon_{ijkl}\ ,  \\
\delta_{IJ} U_{ij}{}^I V_{A}{}^J -\delta_{\hat{I}\hat{J}}  V_{ij}{}^{\hat{I}} U_B{}^{\hat{J}} &= 0 \ , \\
\delta_{\hat{I}\hat{J}} U_A{}^{\hat{I}} U_{B}{}^{\hat{J}} - \delta_{IJ} V_A{}^{I} V_B{}^J &= \delta_{AB} \ , \end{split}\hspace{5mm}\begin{split}
\tfrac{1}{2} \varepsilon^{ijkl} U_{ij}{}^I U_{kl}{}^J - \delta^{AB} V_A{}^I V_B{}^J &= \delta^{IJ} \ , \\
\tfrac{1}{2} \varepsilon^{ijkl} U_{ij}{}^I V_{kl}{}^{\hat{J}} - \delta^{AB} U_A{}^{\hat{J}} V_B{}^I &= 0 \  , \\
\delta^{AB} U_A{}^{\hat{I}} U_B{}^{\hat{J}} - \tfrac{1}{2} \varepsilon^{ijkl} V_{ij}{}^{\hat{I}} V_{kl}{}^{\hat{J}} &= \delta^{\hat{I}\hat{J}}\ .
\end{split}\label{SO5nConst} \ee
Here, and elsewhere in the text, the indices $A,I,\hat I$ and $ij$ are raised and lowered with the appropriate Euclidean metrics (and $\half\ve^{ijkl}$ for the $ij$). 
We have
\be
D\cU\,\cU^{-1}=\left(\barr{cc} 0 & \  P\cr\  P^T & 0\earr\right)
\ee
where $P=P_{ij A}$ is the one-form given in (2.10). In components, these Maurer-Cartan equations read
\bea D_\alpha^k U_{ij}{}^I =  \delta^{k}_{[i} \lambda_{\alpha j ]A}  V^{A I} \ , \qquad D_\alpha^k V_{ij}{}^{\hat{I}} = \delta^{k}_{[i} \lambda_{\alpha j ]A} U^{A {\hat{I}}} \ , \CR
\bar D_{\dot{\alpha} k } U_{ij}{}^I = \frac{1}{2} \varepsilon_{ijkl} \bar\lambda_{\dot{\alpha} A}^l V^{AI} , \quad \bar D_{\dot{\alpha} k } V_{ij}{}^{\hat{I}} = \frac{1}{2} \varepsilon_{ijkl} \bar\lambda_{\dot{\alpha} A}^l U^{A\hat{I}} \ , 
\eea
and 
\bea D_\alpha^k U_A{}^{\hat{I}} = \frac{1}{2} \varepsilon^{ijkl} \lambda_{\alpha l A} V_{ij}{}^{\hat{I}} \ , \qquad  D_\alpha^k V_A{}^{{I}} = \frac{1}{2} \varepsilon^{ijkl} \lambda_{\alpha l A} U_{ij}{}^{{I}}\ ,  \CR
\bar D_{\dot{\alpha} i} U_A{}^{\hat{I}} = \bar\lambda_{\dot{\alpha} A}^j V_{ij}{}^{\hat{I}} \ , \qquad  \bar D_{\dot{\alpha} i} V_A{}^{{I}} = \bar\lambda_{\dot{\alpha} A}^j U_{ij}{}^{{I}}\ .
\eea 
The supersymmetry variations of the dimension-one-half fermions are given by
\be \bar D_{\dot{\alpha} i} \chi_\beta^j = - i \delta_i^j P_{\beta\dot{\alpha}} \ , \qquad D_\alpha^i \chi_\beta^j  = - \frac{1}{2} \varepsilon^{ijkl} M_{\alpha\beta kl} - \frac{1}{4} \varepsilon_{\alpha\beta} \varepsilon^{\dot{\alpha} \dot{\beta}} \bar \lambda_{\dot{\alpha}}^{i A} \bar \lambda_{\dot{\beta}}^j{}_A \ , \ee
and 
\be \bar D_{\dot{\alpha} i} \lambda_{\beta j A} = 2  i  P_{\beta\dot{\alpha} ij A } \ , \qquad D_\alpha^i \lambda_{\beta j A}   =\delta^i_j  M_{\alpha\beta A}  + \varepsilon_{\alpha\beta} \varepsilon^{\dot{\alpha} \dot{\beta}} \scal{\bar  \lambda_{\dot{\alpha}}^{i}{}_{A} \bar  \chi_{\dot{\beta} j}- \sfrac{1}{2} \delta^i_j  \bar  \lambda_{\dot{\alpha}}^{k}{}_{A} \bar \chi_{\dot{\beta} k}} \; . \ee
In this paper we shall also need the spinorial derivatives of the some of the dimension-one fields. These are
\bea \bar D_{\dot{\gamma}k} \bar M_{\dot{\alpha}\dot{\beta}}^{ij} &=&    \bar \chi_{(\adt k} \bar \lambda^{[i}_{\bdt) A} \bar \lambda_{\cdt}^{j] A} - \frac{1}{2}  \bar \chi_{\cdt k} \bar \lambda^{[i}_{\adt A} \bar \lambda_{\bdt}^{j] A} + \frac{i}{2} \varepsilon^{ijpq} \varepsilon_{\cdt(\adt} P_{\bdt)\delta pq A} \lambda^{\delta A}_k  \CR
&&+ 2 \delta_k^{[i} \Scal{\bar  \rho^{j]}_{\dot{\alpha}\dot{\beta}\dot{\gamma}} + 2i  \varepsilon_{\dot{\gamma}(\dot{\alpha}} \bar P_{\dot{\beta})\delta} \chi^{j]\delta} + \dots }  \hspace{5mm}   \ , \CR
 D_\gamma^k \bar M_{\dot{\alpha}\dot{\beta}}^{ij} &=& \varepsilon^{ijkl} \Scal{ i D_{\gamma(\dot{\alpha}} \bar  \chi_{\dot{\beta})l}  + \tfrac{1}{4} \chi_\gamma^p \bar \chi_{(\dot{\alpha} l} \bar \chi_{\dot{\beta}) p} + \dots }  + \chi_\gamma^k \varepsilon^{ijpq} \bar  \chi_{\dot{\alpha}p}  \bar \chi_{\dot{\beta}q}  + \frac{1}{2} \varepsilon^{ijpq} \bar \lambda^{k A}_{(\adt} \lambda_{\gamma p A} \bar \chi_{\bdt) q}  \ , \CR
 D_\gamma^k P_{\alpha\bdt ij A} &=&\frac{i}{2} \varepsilon_{\gamma\alpha} \Scal{ \tfrac{1}{2} \varepsilon^{ijpq} \bar M_{\bdt\ddt pq} \bar \lambda^{\ddt k}_{A} + 2  \bar \chi_{\bdt [i} \lambda_{\delta j] A} \chi^{\delta k}  - \bar\lambda^{k }_{\bdt B} \lambda_{\delta [i }^B \lambda^\delta_{j] A} } -  \frac{i}{4} \lambda^B_{\gamma[i} \lambda_{\alpha j] B } \bar\lambda^k_{\bdt A} \CR
 && + \delta^k_{[i} \Scal{ D_{\bdt(\alpha} \lambda_{\beta) j] A} + \dots }  \hspace{5mm}   \ ,
  \eea
where the dots state for terms involving the matter fields that will not be relevant for our computations and $\bar \rho^{i}_{\dot{\alpha}\dot{\beta}\dot{\gamma}}$ is the gravitino field strength. To compute these derivatives, one uses the supersymmetry algebra as defined in \cite{Howe:1981gz} for $\cN=4$ supergravity coupled to six vector multiplets, and one extends the corresponding expressions to $n$ vector multiplets using the $SO(n)$ symmetry. It turns out to be helpful to use such expressions to compute first the torsion \cite{Howe:1981gz}, and then to make use of the $U(4) \times SO(n)$ covariant supersymmetry algebra in order to obtain the expressions for $n$ vector multiplets.

%%%%%%%%%%%%%%
\subsection{Harmonic superspaces}
%%%%%%%%%%%%%
Since we shall be interested in  investigating F-term invariants in the full non-linear theory, we shall need to know which sub-superspace measures are available. The simplest possibility is chiral superspace, and we know that the supergravity scalar fields $T,U$ are indeed chiral. However, these are in fact the only independent chiral superfields due to the the fact that the dimension-one-half torsion is non-zero. This is similar to  IIB supergravity, where it is known that there is no chiral measure for similar reasons. The other possible measures can be investigated using harmonic-superspacee methods. In flat superspace, we recall that a G-analytic (G for Grassmann) structure of type $(p,q)$ consists of a set of $p\ D$s and $q\ \bar D$s that mutually anti-commute, and that such sets can be parametrised by the coset spaces $(U(p)\xz U(\csN-(p+q))\xz U(q))\bsh U(\csN)$, which are compact, complex manifolds (flag manifolds) \cite{Rosly:1982,Galperin:1984av, Howe:1995md}. Harmonic superspaces of type $(\csN,p,q)$ consist of ordinary superspaces augmented by the above cosets.  Analytic superspaces, which have $2(\csN-p)+2(\csN-q)$ odd coordinates as well as the harmonic cosets, are generalisations of chiral superspace on which $(p,q)$ G-analytic fields can be defined in a natural way.\footnote{Further discussions of harmonic or projective superspace techniques in $\cN=2$ supergravity theories can be found \eg in \cite{Galperin:1987ek,Kuzenko:2008ep}.} In curved superspace one has to check that the corresponding sets of odd derivatives, suitably extended to include the harmonic directions, remain involutive in the presence of the non-trivial geometry. It turns out that this is only possible when both $p,q\leq 1 $ for $\cN>4$, but for $\cN=4$ one can also have $(p,q)=(1,1),\,(2,2)\ {\rm or}\ (1,0)$, $(0,1)$ \cite{Hartwell:1994rp}.

Let us consider first  $(4,1,1)$ harmonic superspace. The harmonic variables, $u^1{}_i, u^r{}_i, u^4{}_i$ (and their inverses $u^i{}_1, u^i{}_r, u^i{}_4$), where $r=2,3$, can be used to parametrise the coset of $U(4)$ with isotropy group $\scal{U(1) \times U(2) \times U(1)}$ in an equivariant fashion. The first obstruction to involutivity vanishes because $T_\a^1{}_\b^1{}^{\cdt k}=0$ (where quantities with indices $(1,r,4)$ are projected by means of the appropriate harmonic matrices), but we still need to check that the dimension-one curvatures do not cause problems. It is straightforward to show, using (\ref{Rsu4ab},\ref{Rsu4abd}), that no further obstructions to involutivity arise because
\be
R_\a^1{}_{\b,}^1{}^1{}_r=R_\a^1{}_{\b,}^1{}^1{}_4=R_\a^1{}_{\b,}^1{}^r{}_4=0\ ,
\ee
and similarly for $R_\a^1{}_{\bdt 4,}{}^k{}_l$ and $R_{\adt 4\bdt 4,}{}^k{}_l$ , for the same projections of the Lie algebra indices. The $(4,1,0)$ (and $(4,0,1)$) cases are also fine. For the former, one splits the index into $(1,r),\  r=2,3,4$, and the only possible curvature 
obstruction involves $R_\a^1{}_{\b,}^1{}^1{}_r$, which clearly vanishes.

For $(4,2,2)$ harmonic superspace, the relevant harmonic coset
is $\scal{ U(2) \times U(2) } \setminus U(4)$ for which we define harmonic variables $(u^r{}_i,u^{\hat{r}}{}_i)$ and their inverses $(u^i{}_r,u^i{}_{\hat{r}})$ where $r = 1 , 2 $ and $\hat{r} = 3,4$, with
\be u^r{}_i u^i{}_s{} = \delta^r_s \ , \qquad u^r{}_i  u^i{}_{\hat{s}} = 0 \ , \qquad  u^{\hat{r}}{}_i u^i{}_{\hat{s}}{} = \delta^{\hat{r}}_{\hat{s}} \ ,\ee  
where unitarity implies that, \eg, $u^i{}_r=\bar u_r{}^i$.

G-analyticity of type $(2,2)$ is allowed \cite{Hartwell:1994rp} because the appropriate projections of the dimension one-half component of the torsion are
 \be T_\alpha^r{}_\beta^s{}^{ \dot{\gamma}t} \equiv u_i{}^r u_j{}^s u_k{}^t T_\alpha^i{}_\beta^j{}^{\dot{\gamma} k} = 0 \ee
 while the dimension-one components of the Riemann tensor satisfy 
\be 
R_\alpha^r{}_\beta^s{}^t{}_{\hat{u}} = R_{\adt \hat{r}\bdt\hat{s}}{}^t{}_{\hat{u}}= R_\alpha^r{}_{\dot{\beta} \hat{s}}{}^t{}_{\hat{u}} = 0\ . 
\ee

Although the torsion and curvature tensors are compatible with involutivity for each of the above cases, harmonic-analytic fields will transform under non-trivial representations of the isotropy groups and there are constraints on the allowed representations that arise because of the non-vanishing components of the dimension-one curvature components restricted to the isotropy sub-algebra. It is straightforward to check that the representations under which putative integrands must transform are allowed. For example, in the $(4,1,1)$ case, a G-analytic integrand should be of the form $\cO^{11}_{44}$. Such fields are allowed because $R_\a^1{}_{\bdt 4}{}^1{}_1= R_\a^1{}_{\bdt 4}{}^4{}_4$. In the $(4,2,2)$ case there is again no obstruction to the existence of a G-analytic superfield of the correct $U(1)$ weight, because 
\be R_\alpha^r{}_{\beta}^s{}^t{}_t = 0 \ , \quad R_\alpha^r{}_{\beta}^s{}^{\hat{t}}{}_{\hat{t}} = 0 \ , \quad R_\alpha^r{}_{\dot{\beta}\hat{s}}{}^t{}_t = - \chi_\alpha^r{} \chi_{\dot{\beta} \hat{s}} \ , \quad R_\alpha^r{}_{\dot{\beta}\hat{s}}{}^{\hat{t}}{}_{\hat{t}} = - \chi_\alpha^r{} \chi_{\dot{\beta} \hat{s}}\ .  \ee
In other words, the $\mathfrak{su}(4)$ curvature decomposes as the sum of two $\mathfrak{su}(2)$ curvatures. 

In all of the above cases one can define measures for G-analytic fields by means of expansions of the Berezinian of the supervielbein with respect to $2p+2q$ normal coordinates. This will be done explicitly for $(4,1,1)$ case later on; it is more difficult for $(4,2,2)$ but it is clear from the construction that such a measure does exist. These integrals are the curved-space versions of integrals over analytic superspaces in the flat case.
%%%%%%%%%%%%%%%%%%%%
\section{Invariants in $\cN=4$ supergravity theories}\label{sec3}
%%%%%%%%%%%%
\subsection{Linearised invariants}
%%%%%%%%%%
We begin with a brief survey of invariants constructed from the on-shell linearised field-strength superfields that transform as primary superconformal fields. For $\cN=4$ the superconformal algebra can be taken to be $\gs\gu(2,2|4)$ but the abelian subalgebra generated by the unit matrix, \ie R-symmetry, does not act on superspace -- only the quotient algebra $\gp\gs\gu(2,2|4)$ does. The quantum numbers associated with the latter are the dilation weight, $L$, the two spin quantum numbers, $J_1,J_2$, and three Dynkin labels $(r_1,r_2,r_3)$ specifying an $\gs\gu(4)$ representation.\footnote{The superconformal dimension $L$ of a field is its canonical one, while the dimensions used in the full theory are geometrical, \eg scalars have $L=1$ but geometrical dimension zero.} The chiral supergravity field strength superfield $W$ has $L=1$ and will also be assigned $R=-1$. ($W$ is the linearised limit of $T$ and we use the $U(1)$ gauge in which $U=\UT = 1 + \mathcal{O}(T\bar T)$.) The vector-multiplet fields strengths $W_A$ have $L=1,R=0$ and transform in the $[0,1,0]$ of $\gs\gu(4)$. Note that the $U(1)$ symmetry of supergravity is not part of the superconformal symmetry for $\cN=4$; it has been referred to as $U(1)_Y$ in the literature \cite{Intriligator:1998ig}. It does act on superspace, however, since it can be considered as a subgroup of $U(2,2|4)$ \cite{Heslop:2003xu}. 

The invariants are integrals of monomials in the field strengths and their derivatives. These can be put into representations of the superconformal group, either primaries or descendants. Because we are integrating, we need only consider the primaries, and since primaries with non-zero spin quantum numbers do not have Lorentz scalar top components, we can restrict our attention to primaries with $J_1=J_2=0$ \cite{Drummond:2010fp}. An important class of invariants consists of those that are short, or BPS. These correspond to the unitary series B and C representations with $J_1=J_2=0$. The series B unitarity bounds are $L+R=\half m$, or $L-R=2 m_1 -\half m$, together with $L\geq1+m_1$, while for series $C$ we have $L=m_1,\ R=\half m-m_1$, where $m_1=\sum_k r_k$ is the number of boxes in the first row of the $\gs\gu(4)$ Young tableau and $m=\sum_k k r_k$ is the total number of boxes \cite{DobrevSC}. In fact, the series C representations do not involve the supergravity field strength and thus have $R=0$. 

The simplest pure supergravity invariants are the chiral ones, $W^p$, and of these only $p=2$ gives rise to a $U(1)$ invariant integral. This is the linearised $R^2$ invariant and is in fact a total derivative on-shell. We can also construct $(4,0,2)$ invariants with integrands of the form $M_{\a\b 12} M^{\a\b}_{12} W^{p+2}$, for $p\ge 0$, although only the one with $p=-2$ would be $U(1)$ invariant (and is also a total derivative). The corresponding chiral primaries are of the form
\be u^i{}_1 u^j{}_2 u^k{}_1 u^l{}_2  \Scal{ M_{\a\b ij} M^{\a\b}_{kl}\, W^{p+2}  - \varepsilon_{ijmn} \chi_\a^m \chi_\b^n M^{\a\b}_{kl} \, W^{p+1}   + \tfrac{1}{4} \varepsilon_{ijmn} \varepsilon_{klrs} \chi_{(\a}^m \chi_{\b)}^n \chi^{\a r} \chi^{\b s}\, W^p }  \; , \ee
where the three terms define integrands that are equivalent up to total derivatives. These are series B with $L+R=2$ and $\gs\gu(4)$ representation $[0,2,0]$. The pure vector-multiplet invariants were considered in \cite{Drummond:2003ex}. Here, we must insist on $\gs\gu(4)$ symmetry (and $\gs\go(n)$), and there is only one one-half BPS invariant $W_A W^A W_B W^B$ in the $[0,4,0]$ representation of $\gs\gu(4)$. It can be integrated over an eight-theta measure, either as a super-invariant, or in $(\cN,p,q)=(4,2,2)$ analytic superspace.  There is also only one one-quarter BPS invariant. It is $W_{[A} W_{B]} W^{A} W^B$ in the representation $[2,0,2]$. It can be integrated with respect to a twelve-theta measure as a superaction or as an integral over $(4,1,1)$ analytic superspace.

We can also have mixed supergravity-vector invariants. Since $W$ is chiral the short ones in this series will be chiral with respect to fewer than four dotted derivatives. The simplest case is $(4,0,2)$. Invariance with respect to $\gs\gu(4)$ means that we should have $W_A W^A$ in the $[0,2,0]$ representation. This superfield is the energy-momentum tensor for the vector multiplet. It can be defined on (4,2,2) superspace, but if we multiply it by some power $p$ of $W$, then we get superfields of the required $(4,0,2)$ type. This series of superfields has $L=p+2, \, R=-p, \, m=4,\, m_1=2$ and saturates the series B bound. Note that one must have $p\geq 2$ since otherwise the integral would give zero due to the second-order constraint on $W$. 

The final possibility for BPS invariants is $(4,0,1)$. In order to ensure $\gs\gu(4)$ symmetry we require two powers of $W_A$, but also two undotted spinorial derivatives. The representation has $L=p+3$, for a factor of $W^p$, $p\geq 2$, and must be in the $[0,0,2]$ representation which has $m_1=2,\,  m=6$. These fields again satisfy the series B bound. If we define the energy-momentum tensor superfield by

\be
T_{ij,kl}= -W^A_{k(i} W_{j)l A}
\ee
then the primary linear combination is

\be
\cO_{ij}:=2 T_{ij,kl}\, D^{kl} W^p+  (p-1) D_\a^k T_{ij,kl}\, D^{\a l} W^p + \frac{1}{4}  p (p-1) D^{kl} T_{ij,kl}\, W^p\ , \label{TDDW} 
\ee
where $D^{ij}:= D_\a^i D^{\a j}$. The analytic superspace integrand is $\cO_{44}=\cO_{ij} u^i{}_4 u^j{}_4$. 

Note that the $(4,0,1)$ and $(4,0,2)$ chiral primaries give rise to invariants that reproduce $(4,2,2)$ harmonic integrals discussed in 
\cite{Antoniadis:2007cw}

\subsection{Dimension-four invariants: $R^2$ and $F^4$}

We shall now analyse the various nonlinear invariants in $\cN=4$ supergravity starting at dimension four.

\subsubsection{$R^2$ invariants in pure supergravity}
\label{Rdeux} 
%%%%%%%%
The first possible on-shell invariants in pure $\cN=4$ supergravity are all of generic $R^2$ structure. Although the relevant $SL(2,\bbR)$ invariants vanish on-shell for trivial spacetime topology \cite{Deser:1977nt}, there do exist non-vanishing invariants with a non-trivial dependence on the complex scalar field. 

Because both $\bar U$ and $\bar  T$ are antichiral, one might think that one could define $R^2$ type invariants from an anti-chiral superspace integral of the kind 
\be \int d^4 x d^{8} \bar \theta \cE\,  \bar U^{4} \cF^\ord{-2}(\bar T) \ee 
for some antichiral measure $d^{8} \bar \theta\, \cE$. However, owing to the presence of the dimension one-half torsion component, $\bar U$ and $\bar  T$ are the only fields satisfying the antichirality constraint, and such a na\"{\i}ve measure does not exist. This should be compared to the attempt to construct a superspace-integral $R^4$ invariant in type IIB supergravity \cite{Skenderis}, a construction that similarly failed owing to the lack of a chiral measure, again due to the presence of a non-trivial dimension one-half superspace torsion component \cite{Howe:1983sra,Berkovits:2001ue}. Nevertheless, in $D=4,\ \cN=4$ one can define invariants involving $\bar U$ and $\bar T$ from closed super-four-forms using the ectoplasm formalism \cite{Voronov,Gates:1997kr,Gates:1997ag}. These depend on an arbitrary holomorphic function $\cF^\ord{0}$ which one can intuitively think of as being the second derivative of the function $\cF^\ord{-2}$. In pure supergravity, super-four-forms can be expanded in terms of the supervielbeins as follows 
\begin{multline} \cL[\cF] =  \frac{1}{24} \varepsilon_{abcd} E^a \wedge E^b \wedge E^c \wedge E^d L   + \,\cdots \\ + \frac{1}{24} E^\delta_l \wedge E^\gamma_k  \wedge  E_j^\beta \wedge  E_i^\alpha M_{\alpha\beta\gamma\delta}^{ijkl} + \frac{1}{6} E^{\dot{\delta} l}\wedge E^\gamma_k  \wedge  E_j^\beta \wedge  E_i^\alpha M_{\alpha\beta\gamma\dot{\delta} l}^{ijk} \ . \label{AnomalyInvariant} \end{multline} 
The property that this be $d$-closed implies that its pull-back to the bosonic component of superspace $\iota^* \cL[\cF]$ defines a supersymmetric density, because  variation with respect to local supersymmetry in space-time is then ensured to be a total derivative \cite{Voronov,Gates:1997kr,Gates:1997ag}.  The pull-back is simply obtained in components by replacing the supercovariant components $L, \dots , M_{\alpha\beta\gamma\delta}^{ijkl} , M_{\alpha\beta\gamma\dot{\delta} l}^{ijk}$ by their evaluations at $\theta= 0 $, such that 
\begin{multline} \iota^* \cL[\cF] =  \frac{1}{24} \varepsilon_{abcd} e^a \wedge e^b \wedge e^c \wedge e^d L \big|_{\theta=0}   + \cdots \\ + \frac{1}{24} \psi^\delta_l \wedge \psi^\gamma_k  \wedge  \psi_j^\beta \wedge  \psi_i^\alpha M_{\alpha\beta\gamma\delta}^{ijkl}  \big|_{\theta=0}+ \frac{1}{6} \psi^{\dot{\delta} l}\wedge \psi^\gamma_k  \wedge  \psi_j^\beta \wedge  \psi_i^\alpha M_{\alpha\beta\gamma\dot{\delta} l}^{ijk} \big|_{\theta=0}  \label{AnomalyInvariant}\ , \end{multline}
where $\psi^\a_i\ (=E^\a_i |_{\th=0}$ in the Wess-Zumino gauge) is the gravitino one-form. 
The main advantage of the ectoplasm formalism is that all possible invariants admit, by definition, a cocycle representation, irrespectively of whether they can be defined as superspace integrals (also without involving prepotentials in a hypothetical off-shell formulation). Moreover, in order to prove the existence of such an invariant, one simply needs to check the lowest-dimension component of its $d$-closure, as we shall now demonstrate. The differential decomposes into four components \cite{Bonora:1986ix}
\be d =d_0 + d_1 + t_0  +t_1 \ee
of bi-degrees  $(1,0)$, $(0,1)$, $(-1,2)$ and $(2,-1)$ respectively (where the first degree refers to the degree in the even basis forms $E^a$, and the second to the degree in the odd basis forms $(E^\a_i, E^{\adt i})$). The first two components of $d$ are respectively even and odd differential operators while the other two are algebraic, involving the dimension-zero and dimension-three-halves torsions. The latter is not relevant to the present discussion, while the former can be defined in terms of the contraction operator $ \iota_c   E^a \equiv \delta^a_c$  
and  the dimension-zero torsion, given in (2.1), by
\be t_0 L_{p,q} = -E^{\bdt j} \wedge E^\alpha_i \, T^i_{\alpha\bdt j}{}^c \wedge  \iota_c L_{p,q} \ .\ee
Since $d^2=0$ it follows that $t_0$ is nilpotent so that we can define cohomology groups $H_t^{p,q}$ \cite{Bonora:1986ix}. This means that we can analyse superspace cohomology in terms of elements of this group, in any spacetime dimension; see, for example, \cite{Bonora:1986ix,Cederwall:2001bt,Cederwall:2001dx,Howe:2003cy,Berkovits:2008qw}.\footnote{This approach to superspace cohomology is related to pure spinors \cite{Howe:1991mf,Howe:1991bx,Berkovits:2002zk} and also to the cohomology of supersymmetry algebras \cite{Brandt:2009xv,Brandt:2010tz,Movshev:2010mf}.}  A key feature of $D=4$ is that the only non-trivial $t_0$-cohomology groups have $p=0$ for $\cN>2$. Since $t_0$ maps onto lower-dimensional superform components, the lowest-dimensional component of a cocycle must itself necessarily be $t_0$-closed, and one can remove any $t_0$-exact component from it by redefining the cocycle by the addition of a trivial cocycle, $\cL \sim \cL + d \Psi$. In $\cN=4$ supergravity in four dimensions the only non-trivial $t_0$ cohomology classes have bi-degree $(0,4)$ and correspond to irreducible representations of $SU(4)$ for which the Dynkin labels $[r_1,r_2,r_3]$ satisfy $r_1 + 2 r_2 + r_3 = 4$. It is straightforward to check that all the irreducible representations of $SU(4)$ for which the Dynkin labels $[r_1,r_2,r_3]$ satisfy $r_1 + 2 r_2 + r_3 \le 2$ are $t_0$ trivial, because they necessarily contain an $SU(4)$ singlet that can be identified with $E^{\bdt i} \wedge E^\alpha_i$.\footnote{In general, one can decompose a  component of bi-degree $(0,n)$ into the corresponding $(0,p,q)$ components of degree $p$ in $E^\alpha_i$ and $q$ in $E^{\bdt j}$. For the components of degree $(0,n,0)$ and $(0,0,n)$, the symmetric tensor product of the associated representation decomposes into irreducible representations of $SU(4)$ satisfying  $r_1 + 2 r_2 + r_3 =n$ because the antisymmetrisation of three $SL(2,\mathds{C})$ indices vanishes. For any other degree, the property that $r_1 + 2 r_2 + r_3 <n$ implies, for the same reason, that the corresponding form includes a factor $E^{\bdt i} \wedge E^\alpha_i$, and therefore that it is $t_0$-trivial. Indeed, the only way to have lower values of $r_1 + 2 r_2 + r_3$ is either to have such a contraction, or to have three $SU(4)$ indices antisymmetrised.} The first non-trivial cocycle equation therefore states that 
\be  d_1 \cL_{0,4}[\cF] + t_0 \cL_{1,3}[\cF] = 0  \ .\label{3.7}\ee
Using $d^2 =0$ together with $t_0 \cL_{0,4}[\cF]=0$ one then has
\be t_0 \scal{ d_0  \cL_{0,4}[\cF]  + d_1 \cL_{1,3}[\cF]  } = 0 \ ,\ee
and because there is no $t_0$ cohomology class at bi-degree $(1,4)$, one concludes that there exists an $\cL_{2,2}[\cF]$ such that 
\be d_0  \cL_{0,4}[\cF]  + d_1 \cL_{1,3}[\cF] + t_0 \cL_{2,2}[\cF] = 0\ . \ee
Similarly, for the other components one obtains
\bea t_1 \cL_{0,4}[\cF]  + d_0 \cL_{1,3}[\cF] + d_1 \cL_{2,2}[\cF] + t_0  \cL_{3,1}[\cF] &=& 0 \CR
 t_1 \cL_{1,3}[\cF]  + d_0 \cL_{2,2}[\cF] + d_1 \cL_{3,1}[\cF] + t_0  \cL_{4,0}[\cF] &=& 0 \CR
  t_1 \cL_{2,2}[\cF]  + d_0 \cL_{3,1}[\cF] + d_1 \cL_{4,0}[\cF] &=& 0 \ .\eea
Therefore, the property that $  d_1 \cL_{0,4}[\cF] \approx 0 $ in $t_0$-cohomology is a sufficient condition for the existence of the entire cocycle.\footnote{Note that $\cL_{0.4}$ determines an element of the spinorial cohomology group $H_s^{0,4}$ which consists of non-trivial elements of $H_t^{0,4}$ satisfying $  d_1 \cL_{0,4}[\cF] \approx 0 $ in $t_0$-cohomology.}

In $D=4$ we can further split the odd indices into undotted and dotted indices, so $\cL_{0,4}=M_{0,4,0} + M_{0,3,1}+\ldots M_{0,0,4}$. In the case of the `anti-chiral cocycle' $\cL[\cF]$ one can set $M_{0,2,2}=M_{0,1,3}=M_{0,0,4}=0$ and \eq{3.7} becomes\footnote{A true anti-chiral cocycle in flat superspace would have only $M_{0,4,0}$ non-zero.} 
\bea D_{{\eta}}^p M_{\alpha\beta\gamma\delta}^{ijkl} + 2 T_{\eta}^p{}_\alpha^i{}^{\dot{\varsigma}q} M_{\beta\gamma\delta\dot{\varsigma}q}^{jkl} + \  \circlearrowleft \quad     &\approx& \ 0  \label{d11}\\
\bar D_{\dot{\eta}p} M_{\alpha\beta\gamma\delta}^{ijkl}  + 4 D_\alpha^i M_{\beta\gamma\delta\dot{\eta}p}^{jkl}+ \  \circlearrowleft \quad     &\approx& \ 0 \label{d12} \\
2 \bar D_{\dot{\eta}p} M_{\alpha\beta\gamma\dot{\delta}l}^{ijk} + T_{\dot{\eta}p\dot{\delta}l}{}^\varsigma_q M_{\alpha\beta\gamma\varsigma}^{ijkq} + \  \circlearrowleft \quad     &\approx& \ 0 \label{d13}\\
T_{\dot{\eta}p\dot{\gamma}l}{}^\varsigma_q  M_{\alpha\beta\varsigma\dot{\delta}l}^{ijq} + \  \circlearrowleft \quad     &\approx& \ 0 \label{d14}
\eea
where $\approx$ means equal in $t_0$-cohomology, and $+  \circlearrowleft$ denotes the appropriate sum over symmetric permutations of the sets of indices $\{_{\alpha ,}^i{}_{\beta ,}^j{}_{\gamma ,}^k{}_{\delta ,}^l\}$ and $\{_{\dot{\gamma} k , \dot{\delta} l ,\dot{\eta} p }\}$. At bi-degree $(0,5)$, the $t_0$-cohomology consists of all irreducible representations of  $SU(4)$ for which the Dynkin labels $[r_1,r_2,r_3]$ satisfy $r_1 + 2 r_2 + r_3 =5$.

After some work, one verifies that such a cocycle exists for any holomorphic function, $\cF^\ord{0}$, with lowest components
\begin{multline} \label{R2Cocycle} M_{\alpha\beta\gamma\delta}^{ijkl}  =  6 \, \varepsilon_{\alpha\beta} \varepsilon_{\gamma\delta} \Bigl( \cF^\ord{0}(\bar T) \bar M_{\dot{\alpha}\dot{\beta}}^{ij} \bar M^{\dot{\alpha}\dot{\beta} kl} - \bar U^{-2} \bar \partial \cF^\ord{0}(\bar T) \varepsilon^{ijpq} \bar \chi_{\dot{\alpha} p} \bar \chi_{\dot{\beta} q} \bar M^{\dot{\alpha}\dot{\beta} kl}  \Bigr . \\ \Bigl .+ \frac{1}{6} \bar U^{-4} \scal{ \bar \partial - \frac{2 T}{ 1 - T \bar T} } \bar \partial  \cF^\ord{0}(\bar T) \varepsilon^{ijpq} \varepsilon^{klrs} \bar \chi_{\dot{\alpha} p} \bar \chi_{\dot{\beta} q} \bar \chi_r^{\dot{\alpha}}\bar  \chi_s^{\dot{\beta}} \Bigr) \ +   \circlearrowleft  \end{multline} 
and
\be M_{\alpha\beta\gamma\dot{\delta} l}^{ijk}   =  6\,   \varepsilon_{\alpha\beta}   \chi_\gamma^k \bar \chi^{\dot{\eta}}_{\, l} \Bigl( \cF^\ord{0}(\bar T) \bar  M_{\dot{\delta}\dot{\eta}}^{ij} -\frac{1}{3}  \bar U^{-2} \bar \partial \cF^\ord{0}(\bar T) \varepsilon^{ijpq}\bar  \chi_{\dot{\delta} p} \bar \chi_{\dot{\eta} q}  \Bigr)  \ +   \circlearrowleft  \ . \ee
The computation goes as follows: (\ref{d14}) requires that 
\be M_{\alpha\beta\gamma\dot{\delta} l}^{ijk} =  6  \chi_\gamma^k \varepsilon_{\alpha\beta} M^{ij}{}_{\dot{\delta} l}  + \  \circlearrowleft \quad  \ee
for some $M^{ij}{}_{\dot{\delta} k} $ in the ${ \Yboxdim4pt  {\yng(2,1)}}$ of $SU(4)$. Then  (\ref{d13}) requires that 
\be \varepsilon_{\alpha\beta}\bar D_{\dot{\gamma} k} M^{ij}{}_{\dot{\delta} l} \approx \varepsilon_{\alpha\beta} \varepsilon_{\dot{\gamma}\dot{\delta}} N^{ij}{}_{kl} \ee
for some $N^{ij}{}_{kl}$ in the ${ \Yboxdim4pt  {\yng(2,2)}}$ of $SU(4)$. Because of the dimension-one-half component of the torsion, this equation is very constraining; the general solution depends on one anti-holomorphic function 
\be  M^{ij}{}_{\dot{\delta} k} =   \bar \chi^{\dot{\eta}}_{\,  l} \Bigl( \cF^\ord{0}(\bar T) \bar M_{\dot{\delta}\dot{\eta}}^{ij} -\frac{1}{3}  \bar U^{-2} \bar \partial \cF^\ord{0}(\bar T) \varepsilon^{ijpq} \bar \chi_{\dot{\delta} p} \bar \chi_{\dot{\eta} q}  \Bigr) \ . \label{MCC} \ee
Both (\ref{d11}) and (\ref{d13}) separately determine  $M_{\alpha\beta\gamma\delta}^{ijkl} $ as functions of $M_{\alpha\beta\gamma\dot{\delta} l}^{ijk}$ in a consistent way, and (\ref{d12}) is satisfied for both $M_{\alpha\beta\gamma\delta}^{ijkl} $ and $M_{\alpha\beta\gamma\dot{\delta} l}^{ijk}$ individually. 

One easily checks that, for $\cF^\ord{0} = 1$, the  chiral superform is just $R_{\dot{\alpha}\dot{\beta}} \wedge R^{\dot{\alpha}\dot{\beta}}$ so that, in general,  this invariant includes the term 
\be \cL[\cF^\ord{0}] = \cF^\ord{0}(\bar T) R_{\dot{\alpha}\dot{\beta}} \wedge R^{\dot{\alpha}\dot{\beta}} + \dots \ \ .\ee
This invariant is of course complex, and the associated real invariants will be obtained from its real and imaginary parts, which are respectively even and odd with respect to parity. 

This may seem very different from the example of the type IIB superstring $\cE^\ord{0}(T,\bar T) R^4$ invariant, for which the function $\cE^\ord{0}$ is also known to satisfy a Poisson equation \cite{GreenSethi}. However, note that any holomorphic function of this kind necessarily satisfies a Laplace equation
 \be \Delta \cF^\ord{0}(\bar T):=( 1 - T \bar T)^2 \frac{ \partial \, }{\partial \bar T}  \frac{ \partial \, }{\partial  T} \cF^\ord{0}(\bar T)  = 0 \ee
although  this property was not used in checking the cocycle structure. 

The generalisation of these pure $\cN=4$ supergravity cocycles to $\cN=4$ supergravity coupled to $n$ vector multiplets is rather complicated, and requires a discussion of the dimension-four $F^4$ invariant.
%%%%%%%%%
\subsubsection{The $F^4$  invariant} 
\label{F4Inv}
The $F^4$ invariant in  flat $(4,2,2)$ analytic superspace was given in \cite{Drummond:2003ex}. Here we show how to extend this result to the curved case. Because this superspace exists in the curved case there is no problem writing down this invariant in a similar way. However, it is not manifestly invariant under the $SO(6,n)$ duality symmetry which  is rather difficult to show directly. Nevertheless, we can show that this symmetry holds, as it must (because there is a known one-loop divergence), by use of the ectoplasm formalism to deduce that the analytic superspace integral is indeed duality-invariant by uniqueness.

One can easily check that both 
\be U_{34}{}^I \equiv \frac{1}{2} \varepsilon^{\hat{r}\hat{s}}u^i{}_{\hat{r}} u^j{}_{\hat{s}} U_{ij}{}^I \quad \textrm{and} \quad V_{34}{}^{\hat{I}}  \equiv \frac{1}{2} \varepsilon^{\hat{r}\hat{s}}u^i{}_{\hat{r}} u^j{}_{\hat{s}} V_{ij}{}^{\hat{I}} \ee
are G-analytic. Because of (\ref{SO5nConst}), one has
\be \delta_{IJ} U_{34}{}^I U_{34}{}^J = \delta_{\hat{I}\hat{J}} V_{34}{}^{\hat{I}} V_{34}{}^{\hat{J}} \ee
and we obtain the unique $U(4) \times SO(n)$ invariant G-analytic integrand with the correct $U(1)$ weight:
\be S^\ord{1} = \int d\mu_{\scriptscriptstyle (4,2,2)} \Scal{  \delta_{\hat{I}\hat{J}} V_{34}{}^{\hat{I}} V_{34}{}^{\hat{J}}}^2 \ .  \label{HalfHarmo}  \ee
The associated top component then clearly contains a term in $F^4$ 
\be  \cL^\ord{1}  = e \scal{  \delta^{AB}  \delta^{CD}  + 2 \delta^{AC} \delta^{BD} } M_{\alpha\beta A} M^{\alpha\beta}{}_B \bar M_{\dot{\alpha}\dot{\beta} C} \bar M^{\dot{\alpha}\dot{\beta}}{}_D  + \dots \label{F4} \ee
and yields the one-loop counterterm for the matter vector fields. One proves, using superconformal representation theory, that this is the unique $U(4) \times SO(n)$ invariant at this order \cite{Drummond:2003ex}. It follows that this invariant must actually be invariant as well under $SO(6,n)$, despite the fact that the integral formula (\ref{HalfHarmo}) does not display this symmetry manifestly. To prove directly that this is indeed the case, one would need to show that
\be  \int d\mu_{\scriptscriptstyle (4,2,2)} \,  \delta_{\hat{K}\hat{L}} V_{34}{}^{\hat{K}} V_{34}{}^{\hat{L}}  U_{34}{}^I V_{34}{}^{\hat{J}} = 0 \label{SOF4invariance} \ . \ee
In the linearised approximation, the integral of $(W_{34})^4$ over the $(4,2,2)$ measure is indeed invariant under linearised duality transformations, \ie constant shifts, since $(W_{34})^{3}$ is ultra-short. If (\ref{SOF4invariance}) were not zero, it would necessarily start at five points with the linearised invariant part of it defined as the integral of $(W_{34})^{5}$ over the $(4,2,2)$ measure, and there could also be a contribution from the integral of $(W_{34})^{3}$ component over the non-linear $(4,2,2)$ measure; in principle, these might cancel each other, and indeed they do.   However, this would be very complicated to check in practice. Instead, we shall show that there is a duality-invariant cocycle
\begin{multline} \cL^\ord{1} =  \frac{1}{24} \varepsilon_{abcd} E^a \wedge E^b \wedge E^c \wedge E^d L^\ord{1}   + \cdots + \frac{1}{4} E^{\dot{\delta} l}\wedge E^{\dot{\gamma} k}  \wedge  E_j^\beta \wedge  E_i^\alpha M_{\alpha\beta\dot{\gamma} k\dot{\delta} l}^{ij} \\ + \frac{1}{24} E^\delta_l \wedge E^\gamma_k  \wedge  E_j^\beta \wedge  E_i^\alpha M_{\alpha\beta\gamma\delta}^{ijkl} + \frac{1}{6} E^{\dot{\delta} l}\wedge E^\gamma_k  \wedge  E_j^\beta \wedge  E_i^\alpha M_{\alpha\beta\gamma\dot{\delta} l}^{ijk}     + \rm{c.c.} \label{F4Cocycle} \end{multline} 
 associated to this invariant. Within the linearised approximation, one finds that the na\"ive representative of this cocycle obtained from the one-half BPS supermultiplet in the ${ \Yboxdim5pt  {\yng(4,4)}}$ of $SU(4)$ \cite{Howe:1981xy} does depend explicitly on the scalar fields, and so looks non-shift-invariant. For example (with $W^{ij}{}_A = V^{ij}{}_{\hat{I}} U_A{}^{\hat{I}}$), one has
 \begin{multline}  M_{\alpha\beta\gamma\delta}^{ijkl} \sim \varepsilon_{\alpha\beta}   \varepsilon_{\gamma\delta}  \scal{  \delta^{AB}  \delta^{CD}  + 2 \delta^{AC} \delta^{BD} } \\\times \Scal{ W^{ij}{}_ A W^{kl}{}_B M_{\eta\vartheta C} M^{\eta\vartheta}{}_D + W^{ij}{}_A M^{\eta\vartheta}{}_B \varepsilon^{klpq}  \lambda_{\eta p  C} \lambda_{\vartheta q D} + \varepsilon^{ijpq} \varepsilon^{klrs}    \lambda_{\eta p A} \lambda_{\vartheta q B} \lambda^\eta_{r C} \lambda^\vartheta_{s D} } \label{LinearisedCocycle} \end{multline} 
 projected into the ${ \Yboxdim5pt  {\yng(2,2)}}$ of $SU(4)$. However, one can trivialise the components depending explicitly on the scalar fields by adding the exterior derivative of a 3-form with $(0,3,0)$ component 
\be \Psi_{\alpha\beta\gamma}^{ijk} \sim \varepsilon_{\alpha\beta} \scal{  \delta^{AB}  \delta^{CD}  + 2 \delta^{AC} \delta^{BD} }  \Scal{ W^{ij}{}_A W^{kl}{}_B \lambda^\delta_{l C} M_{\gamma\delta D} + W^{kl}{}_A \lambda^\delta_{l B}  \varepsilon^{ijpq} \lambda_{\gamma p C} \lambda_{\delta q D} } \label{LinearisedTrivial} \ee
projected into the ${ \Yboxdim5pt  {\yng(2,1)}}$. Here, formulae (\ref{LinearisedCocycle},\ref{LinearisedTrivial}) are only indicative, and the specific coefficients have not been precisely established. 

To prove that the complete non-linear cocycle can be chosen to be duality-invariant, we will compute  its fermionic components directly. One finds that 
\bea M_\alpha^i{}_\beta^j{}_\gamma^k{}_\delta^l &=& \frac{i}{4} \varepsilon_{\alpha\beta} \varepsilon_{\gamma\delta} \varepsilon^{ijpq} \varepsilon^{klrs} \scal{  \delta^{AB}  \delta^{CD}  + 2 \delta^{AC} \delta^{BD} }   \lambda_{\eta p A} \lambda_{\vartheta q B} \lambda^\eta_{r C} \lambda^\vartheta_{s D} \, + \  \circlearrowleft  \CR
M_\alpha^i{}_\beta^j{}_\gamma^k{}_{\dot{\delta}l} &=& \frac{i}{2} \varepsilon_{\alpha\beta} \varepsilon^{ijpq} \scal{  \delta^{AB}  \delta^{CD}  +  \delta^{AC} \delta^{BD} + \delta^{AD} \delta^{BC} }   \lambda_{\gamma p A} \lambda_{\eta q B} \lambda^\eta_{l C} \bar \lambda^k_{\dot{\delta} D} \, + \  \circlearrowleft  \CR
M_\alpha^i{}_\beta^j{}_{\dot{\gamma} k \dot{\delta}l} &=&  \frac{i}{2} \varepsilon_{\alpha\beta} \scal{  \delta^{AB}  \delta^{CD}  + 2 \delta^{AC} \delta^{BD} }   \lambda^\eta_{\,  k A} \lambda_{\eta l B} \bar  \lambda^{[i}_{\dot{\gamma} C}\bar  \lambda^{j]}_{\dot{\delta} D} \, + \  \circlearrowleft  \CR
&& \  + \frac{i}{2} \varepsilon_{\dot{\gamma}\dot{\delta}} \scal{  \delta^{AB}  \delta^{CD}  + 2 \delta^{AC} \delta^{BD} } \bar   \lambda_{\dot{\eta}  A}^i \bar \lambda^{\dot{\eta} j}_{ B} \lambda_{\alpha [k C} \lambda_{\beta l] D} \, + \  \circlearrowleft 
 \eea
together with their complex conjugates. The cocycle equations are then satisfied, \ie
\bea D_{{\eta}}^p M_{\alpha\beta\gamma\delta}^{ijkl} + 2 T_{\eta}^p{}_\alpha^i{}^{\dot{\varsigma}q} M_{\beta\gamma\delta\dot{\varsigma}q}^{jkl} \, + \  \circlearrowleft \quad     &\approx& \ 0  \label{F11}\\
D_{\dot{\eta}p} M_{\alpha\beta\gamma\delta}^{ijkl}  + 4 D_\alpha^i M_{\beta\gamma\delta\dot{\eta}p}^{jkl}+ 6 T_\alpha^i{}_\beta^j{}^{\dot{\varsigma} q} M_\gamma^k{}_\delta^l{}_{\dot{\eta}p\dot{\varsigma}q}\, +  \  \circlearrowleft \quad     &\approx& \ 0 \label{F12} \\
T_{\dot{\eta}p\dot{\delta}l}{}^\varsigma_q M_{\alpha\beta\gamma\varsigma}^{ijkq} +  2 D_{\dot{\eta}p} M_{\alpha\beta\gamma\dot{\delta}l}^{ijk} + 3 D_\gamma^k M_\alpha^i{}_\beta^j{}_{\dot{\delta} l \dot{\eta} p} + 3 T_\alpha^i{}_\beta^j{}^{\dot{\varsigma} q} M_\gamma^k{}_{\dot{\delta} l \dot{\eta} p \dot{\varsigma} q}\, +  \  \circlearrowleft \quad     &\approx& \ 0 \label{F13} \eea
Equation (\ref{F11}) consists of one single term in $\bar \chi \bar \lambda_A \lambda_A^{\; 3}$ in the $[1;0|1,2,0]$. Equation (\ref{F12}) includes a term in $P_A \lambda_A^{\; 3}$ in the $[0;1|0,2,1]$ and a term in $\bar \chi \bar \lambda_A^{\; 2} \lambda_A^{\; 2}$ in the $[2;1|2,1,1]$. Equation (\ref{F13}) includes two terms in $\lambda_A^{\; 2} \bar \lambda_A P_A$ in the $[1;0|1,2,0]$ and the $[1;2|1,1,2]$, one term in $\chi \lambda_A^{\; 4}$ in the $[1;0|1,2,0]$, and two terms in $\bar \chi \lambda_A \bar \lambda_A^{\; 3}$ in the $[1;0|1,2,0]$ and the $[1;2|1,1,2]$. 

Because the cocycle is unique in cohomology, the property that we can find a duality invariant representative is enough to prove that the integral (\ref{HalfHarmo}) is duality invariant. 

%%%%%%%%%%
\subsubsection{$R^2$ type invariants with vector multiplets}
\label{R2withV}
%%%%%%%%%%
In this subsection we will generalise the cocycle derived in Section \ref{Rdeux} in the presence of vector multiplets. As we shall prove, the cocycle is still anti-chiral in the presence of vector multiplets, but in a weaker sense, \ie only $M_{0,1,3} = M_{0,0,4} = 0 $. The other lowest-dimensional components satisfy  
\bea D_{{\eta}}^p M_{\alpha\beta\gamma\delta}^{ijkl} + 2 T_{\eta}^p{}_\alpha^i{}^{\dot{\varsigma}q} M_{\beta\gamma\delta\dot{\varsigma}q}^{jkl} \, + \  \circlearrowleft \quad     &\approx& \ 0  \label{R11}\\
D_{\dot{\eta}p} M_{\alpha\beta\gamma\delta}^{ijkl}  + 4 D_\alpha^i M_{\beta\gamma\delta\dot{\eta}p}^{jkl}+ 6 T_\alpha^i{}_\beta^j{}^{\dot{\varsigma} q} M_\gamma^k{}_\delta^l{}_{\dot{\eta}p\dot{\varsigma}q}\, +  \  \circlearrowleft \quad     &\approx& \ 0 \label{R12} \\
T_{\dot{\eta}p\dot{\delta}l}{}^\varsigma_q M_{\alpha\beta\gamma\varsigma}^{ijkq} +  2 D_{\dot{\eta}p} M_{\alpha\beta\gamma\dot{\delta}l}^{ijk} + 3 D_\gamma^k M_\alpha^i{}_\beta^j{}_{\dot{\delta} l \dot{\eta} p} \, +  \  \circlearrowleft \quad     &\approx& \ 0 \label{R13} \\
3 T_{\dot{\alpha} i \dot{\beta} j}{}^\varsigma_q M_{\gamma}^k{}_\delta^l{}_{\varsigma}^q{}_{\dot{\eta} p} + 3 D_{\dot{\eta} p} M_{\gamma}^k{}_\delta^l{}_{\dot{\alpha} i \dot{\beta} j }  +  \  \circlearrowleft \quad     &\approx& \ 0 \label{R14} \\
6 T_{\dot{\alpha} i \dot{\beta} j}{}^\varsigma_q M_{\eta}^p{}_\varsigma^q{}_{\dot{\gamma} k \dot{\delta} l }+  \  \circlearrowleft \quad     &\approx& \ 0 \label{R15}
 \eea
The last condition (\ref{R15}) is purely algebraic, and requires that 
\be M_\alpha^i{}_\beta^j{}_{\dot{\gamma} k \dot{\delta} l} \approx  4 \chi_\alpha^i N_{\dot{\gamma}k\dot{\delta} l}{}_\beta^j - 2 \chi_\beta^i N_{\dot{\gamma}k\dot{\delta} l}{}_\alpha^j
\  , \label{T21Cohomology} \ee
for some operator symmetric in exchange of the pairs of indices $\{{\dot{\gamma}k,\dot{\delta} l}\}$. One computes that a $(1,3)$ component satisfying the corresponding $(0,5)$ constraint $T_{-1,2} M_{1,3} \approx 0 $ is necessarily of the form $M_{1,3} \approx T_{-1,2} N_{2,1}$, whereas (\ref{T21Cohomology}) is generally not of the form $T_{-1,2} N_{3,0}$.\footnote{Here, the numerical subscripts denote odd indices and $T_{-1,2}$ is the dimension-one-half torsion with an upper undotted index.} Therefore one can think of (\ref{T21Cohomology}) as a $T_{-1,2}$ cohomology class characteristic of the cocycle. Equation (\ref{R14}) then requires $N_{\dot{\alpha}i\dot{\beta} j}{}_\delta^l $ to satisfy 
\be D_{\dot{\alpha}i}  N_{\dot{\beta} j \dot{\gamma} k }{}_\delta^l  \  +  \  \circlearrowleft \quad   \approx -  \varepsilon_{\adt\bdt} \varepsilon_{ijpq} N_{\dot{\gamma}k}{}^{pq}{}_\delta^l \ +  \  \circlearrowleft \quad \ ,   \ee
so that $M_{3,1}$ has the form
\be M_\alpha^i{}_\beta^j{}_\gamma^k{}_{\ddt l} \approx  6 \varepsilon_{\alpha\beta} \scal{  N_{\dot{\delta}l}{}^{ij}{}_\gamma^k  + \chi_\gamma^k M^{ij}{}_{\ddt l} } \ +  \  \circlearrowleft \quad \  ,  \ee
where the second term defines a degree $(3,1)$ $T_{-1,2}$ cohomology class as in Section \ref{Rdeux}. One finds that the unique solution for $N_{\dot{\alpha}i\dot{\beta} j}{}_\delta^l $ at this dimension is determined by an arbitrary anti-holomorphic function $\mathcal{F}(\bar T)$, 
\be N_{\dot{\alpha}i\dot{\beta} j}{}_\delta^l  =  \mathcal{F}(\bar T) \scal{ 2 \bar \chi_{\bdt i} \lambda_{\delta j}^A \lambda_{\adt A}^l - \bar \chi_{\adt i} \lambda_{\delta j}^A \lambda_{\bdt A}^l } \ +  \  \circlearrowleft \quad \ .   \ee
One then computes that 
\begin{multline}  N_{\dot{\delta}l}{}^{ij}{}_\gamma^k = \mathcal{F}(\bar T)  \Scal{ \frac{1}{2} \bar \lambda^{\dot{\eta} k A} \lambda_{\gamma l A} \bar M_{\ddt \dot{\eta}}^{ij}  - i \bar \chi^{\dot{\eta}}_l P_{(\dot{\delta}|\gamma}^{ij A} \bar \lambda_{\dot{\eta}) A}^k + \frac{1}{8} \varepsilon^{ijpq} \lambda_{\eta l}^{ A} \lambda^{\eta}_{ p A} \lambda_{\ddt q}^{ B} \bar \lambda_{\ddt B}^k }  \\ + \frac{1}{2} \varepsilon^{ijpq} \bar U^{-2} \bar \partial \mathcal{F}(\bar T)  \, \bar \chi_{\ddt l} \bar \chi_{\dot{\eta} p} \bar \lambda^{\dot{\eta} k A} \lambda_{\gamma q A}\ \ .  \end{multline} 
The next equation (\ref{R13}) requires that $M^{ij}{}_{\ddt l}$ is still given by (\ref{MCC}) and implies that 
\begin{multline} \label{R2wthMatterCocycle} M_{\alpha\beta\gamma\delta}^{ijkl}  =  6 \, \varepsilon_{\alpha\beta} \varepsilon_{\gamma\delta} \biggl( \cF(\bar T) \Scal{  \bar M_{\dot{\alpha}\dot{\beta}}^{ij} \bar M^{\dot{\alpha}\dot{\beta} kl} - \varepsilon^{ijpq} \bar \chi_{\adt p} \bar \chi_{\bdt q} \bar \lambda^{\adt k A } \bar \lambda^{\bdt l}_A + \frac{1}{16} \varepsilon^{ijpq} \varepsilon^{klrs} \lambda_{\alpha p}^A \lambda_{r A}^{\alpha} \lambda_{\beta q}^B \lambda_{s B}^\beta }  \biggr . \\
 - \bar U^{-2} \bar \partial \cF(\bar T) \varepsilon^{ijpq} \Scal{ \bar \chi_{\dot{\alpha} p} \bar \chi_{\dot{\beta} q} \bar M^{\dot{\alpha}\dot{\beta} kl}   + \frac{1}{4} \varepsilon^{klrs} \bar \chi_{\adt p} \bar \chi_r^{\adt} \lambda_{\beta q}^A \lambda_{s A}^\beta } \\ \biggl .+ \frac{1}{6} \bar U^{-4} \scal{ \bar \partial - \frac{2 T}{ 1 - T \bar T} } \bar \partial  \cF^\ord{0}(\bar T) \varepsilon^{ijpq} \varepsilon^{klrs} \bar \chi_{\dot{\alpha} p} \bar \chi_{\dot{\beta} q} \bar \chi_r^{\dot{\alpha}} \bar \chi_s^{\dot{\beta}} \biggr) \ +   \circlearrowleft \ .  \end{multline} 
One can finally check that the cocycle built in this way also satisfies (\ref{R11}) and (\ref{R12}). The final expressions for the other components take the form 
\bea M_\alpha^i{}_\beta^j{}_\gamma^k{}_{\ddt l} &\approx& 6 \varepsilon_{\alpha\beta} \biggl( \cF(\bar T) \Bigl( \scal{  \chi_\gamma^k \bar \chi^{\dot{\eta}}_{\, l} + \tfrac{1}{2} \bar \lambda^{\dot{\eta} k A} \lambda_{\gamma l A} } \bar M_{\ddt\dot{\eta}}^{ij} - i \bar \chi^{\dot{\eta}}_l P_{\gamma(\ddt}^{ij A} \bar \lambda_{\dot{\eta})A}^k + \tfrac{1}{8} \varepsilon^{ijpq} \lambda_{\eta l}^A \lambda^{\eta}_{p A} \lambda_{\gamma q}^B \bar \lambda_{\ddt k B} \Bigr) \biggr . \CR
&& \qquad \qquad \biggl . + \; \bar U^{-2} \bar \partial \cF(\bar T) \Bigl( - \tfrac{1}{3} \chi_\gamma^k \bar \chi^{\dot{\eta}}_{\, l} \varepsilon^{ijpq}\bar  \chi_{\dot{\delta} p} \bar \chi_{\dot{\eta} q} + \tfrac{1}{2} \varepsilon^{ijqp} \bar \chi_{\ddt l} \bar \chi_{\dot{\eta} p} \bar \lambda^{\dot{\eta} k A} \lambda_{\gamma q A} \Bigr) \biggr)  \ +   \circlearrowleft \  , \CR
M_\alpha^i{}_\beta^j{}_{\dot{\gamma} k \dot{\delta} l} &\approx&  2 \cF(\bar T) \Bigl( 2  \chi_\alpha^i \scal{ 2 \bar \chi_{\ddt k} \lambda_{\beta l}^A \lambda_{\cdt A}^j - \bar \chi_{\cdt k} \lambda_{\beta l}^A \lambda_{\ddt A}^j }  -  \chi_\beta^i \scal{ 2 \bar \chi_{\ddt k} \lambda_{\alpha l}^A \lambda_{\cdt A}^j - \bar \chi_{\cdt k} \lambda_{\alpha l}^A \lambda_{\ddt A}^j } \Bigr)  \ +   \circlearrowleft \ . \CR
\eea
For $\cF(\bar T)=1$ one must get a supersymmetry invariant that preserves duality symmetry, and indeed, one can recognise that the $(0,4)$  components of $\cL[\cF]$ for $\cF=1$ can be identified with the ones of 
\be \cL[1] \approx R_{\adt\bdt} \wedge R^{\adt\bdt} + \frac{1}{4} R^{AB} \wedge R_{AB} + \frac{i}{2} \cL^\ord{1} - d \Scal{\  \frac{1}{2} E^{\dot{\gamma} k} \wedge E^\beta_j \wedge E^\alpha_i\,  \varepsilon_{\alpha\beta} \bar \chi^{\ddt}_k \bar \lambda_{\cdt i}^A \bar \lambda_{\ddt j A} } \ , \ee
where $\cL^\ord{1}$ is the cocycle (\ref{F4Cocycle}) defining the $F^4$ invariant. The total derivative on the right-hand-side can be disregarded in cohomology, but we display it as it appears in the computation. The second term can be rewritten thanks to (\ref{MaurerCartan}) as
\be R^{AB} \wedge R_{AB} = \frac{1}{2} \varepsilon^{ijkl} P_{ij A} \wedge P_{kl B} \wedge  \frac{1}{2} \varepsilon^{pqrs} P^A_{pq } \wedge P^B_{rs }  \  , \ee 
which is the exterior derivative of a globally defined Wess--Zumino like term. The real part of $\cL[1]$ is therefore locally a total derivative, but its imaginary part includes the $F^4$ type invariant in presence of vector multiplets. We conclude  that the invariant is in general of the form 
\be  \cL[\cF] \approx \cF(\bar T) \Scal{ R_{\adt\bdt} \wedge R^{\adt\bdt} +   \frac{1}{4} R^{AB} \wedge R_{AB} + \frac{i}{2} \cL^\ord{1} } +\dots \  , \ee
where the dots correspond to terms involving the derivatives of the function $\cF$ only. Supersymmetry therefore implies that the function of the complex scalar $\tau$ multiplying the $F^4$ term in the effective action is the same as the one that multiplies the Gauss--Bonnet invariant. Overall, the $F^4$ threshold function is the sum of the real part of a holomorphic function of $\tau$ and a function that only depends on the other moduli parametrising $SO(6,n)/ ( SO(6) \times SO(n))$, and which  itself satisfies other constraints that we shall not display in this paper. 
%%%%%%%%%%
\subsection{Dimension-six invariants: $\partial^2 F^4$ and $R^2 F^2$.}
%%%%%%%%%
\label{Sectiond2F4}
It is well-known that there are no $R^3$ type invariants in supergravity \cite{Grisaru:1976nn}, but in the presence of vector multiplets one can have invariants of types $\partial^2 F^4$ and $R^2 F^2$. These can be constructed using analytic superspace.  For $r=2,3$ one has the $(4,1,1)$ G-analytic superfields 
\be U_{4r}{}^I \equiv u^i{}_4 u^j{}_r U_{ij}{}^I \ , \qquad V_{4r}{}^{\hat{I}} \equiv  u^i{}_4 u^j{}_r V_{ij}{}^{\hat{I}} \ee
and one can define the unique  $U(4) \times SO(n)$ invariant at this order by
\be S^\ord{2} = \int d\mu_{\scriptscriptstyle (4,1,1)} \varepsilon^{rs} \varepsilon^{tu} V_{4r}{}^{\hat{I}} V_{4s}{}^{\hat{J}} V_{4t\hat{I}} V_{4u\hat{J}}\ .  \label{d2F4} \ee
Note that this integrand is antisymmetric in $\hat{I}$ and $\hat{J}$, and hence can only exist if one considers at least two vector multiplets. This invariant clearly includes a $\partial^2 F^4$ term  \cite{Drummond:2003ex}, but there is no {\sl a priori}  reason why it should be fully
 $SO(6,n)$ symmetric. Nevertheless, it could be that it is, as in the case of the $F^4$ invariant.

One can also consider $(4,2,2)$ integrals of dimension-two G-analytic integrands. For a generic function $\cF$ of the complex scalar $\bar T$, one can consider a G-analytic integrand of the form 
\begin{multline}  \cL_{\scriptscriptstyle (4,2,2)}[\cF] = \cF( \bar T) L^\ord{0} + \bar U^{-2} \bar \partial  \cF( \bar T) L^\ord{2} +  \bar U^{-4} \Scal{ \bar \partial^2  \cF( \bar T) - \frac{2T}{1-T\bar T} \bar \partial  \cF( \bar T)  }  L^\ord{4} \\  +  \bar U^{-6} \Scal{ \bar \partial^3  \cF( \bar T) - \frac{6T}{1-T\bar T}  \bar \partial^2 \cF( \bar T) +  \frac{6T^2}{(1-T\bar T)^2}  \bar \partial \cF( \bar T)}  L^\ord{6} \ . \end{multline}
Here the derivatives of the function $\cF$ are covariant K\"{a}hler derivatives. The G-analyticity condition requires that 
\be\begin{split} D_\alpha^r L^\ord{0} &=0 \ , \\
 D_\alpha^r L^\ord{2} &=  2 \chi_\alpha^r L^\ord{4}  \ , \\
  D_\alpha^r L^\ord{4} &=6 \chi_\alpha^r L^\ord{6} \ , \\
   D_\alpha^r L^\ord{6} &=0 \ , \\
   \end{split}\hspace{10mm}\begin{split}
   \bar D_{\adt \hat{r}} L^\ord{0} &=0 \ , \\
 \bar D_{\adt \hat{r}} L^\ord{2} &=  - \bar \chi_{\adt \hat{r}} L^\ord{0}  \ , \\
  \bar D_{\adt \hat{r}} L^\ord{4} &=- \bar \chi_{\adt \hat{r}} L^\ord{2} \ , \\
   \bar D_{\adt \hat{r}} L^\ord{6} &=-\bar \chi_{\adt \hat{r}} L^\ord{4}\ . \label{G422Descent} 
   \end{split}\ee
This sequence will stop at weight 6 provided that the last function satisfies
   \be    \bar \chi_{\adt \hat{r}} L^\ord{6} =0 \ . \label{AlgebraChi} \ee
The existence of the functions $L^\ord{2n}$ is consistent with the supersymmetry algebra thanks to this last condition, because it implies that 
\be \{D_\alpha^r , \bar D_{\dot{\beta} \hat{s}}\} L^\ord{2n} = 2n\,  \chi_\alpha^r \bar \chi_{\dot{\beta} \hat{s}} L^\ord{2n} \ , \ee
which is consistent with (\ref{G422Descent}), but not for $L^\ord{6}$ unless (\ref{AlgebraChi}) applies. It follows that $L^\ord{6}$ must be proportional to $\bar \chi^4$, and because it must also be annihilated by $D_\alpha^r$, the only solution (at this dimension) is 
   \be L^\ord{6} = \frac{1}{12} \delta_{\hat{I}\hat{J}}  V_{34}{}^{\hat{I}} V_{34}{}^{ \hat{J}}  \varepsilon^{\hat{r}\hat{s}} \varepsilon^{\hat{t}\hat{u}} \bar  \chi_{\adt \hat{r}} \bar \chi_{\bdt \hat{s}}  \bar \chi^{\adt}_{ \hat{t}} \bar \chi^{\bdt}_{\hat{u}} \ . \ee
   Because the first factor $(V_{34})^2$ is itself G-analytic, we shall consider the ansatz 
   \be L^\ord{2n} =  \delta_{\hat{I}\hat{J}}  V_{34}{}^{\hat{I}} V_{34}{}^{ \hat{J}} X^\ord{2n} \ , \ee
   where the four functions $X^\ord{2n}$ are required to satisfy  (\ref{G422Descent}). The solution is 
   \bea  \label{Xsolution} 
   X^\ord{6} &=& \frac{1}{12} \varepsilon^{\hat{r}\hat{s}} \varepsilon^{\hat{t}\hat{u}} \bar  \chi_{\adt \hat{r}} \bar \chi_{\bdt \hat{s}}  \bar \chi^{\adt}_{ \hat{t}} \bar \chi^{\bdt}_{\hat{u}}\ ,  \CR
    X^\ord{4} &=& \bar M^{12}_{\adt\bdt}  \bar \chi^{\adt}_3 \bar \chi^{\bdt}_4 + \frac{1}{8}  \varepsilon^{\hat{r}\hat{s}} \varepsilon^{\hat{t}\hat{u}} \bar \chi_{\adt \hat{r}} \bar \chi^{\adt}_{\hat{t}} \lambda_{\beta \hat{s}}^A \lambda^\beta_{\hat{u} A} \ , \CR
      X^\ord{2} &=&\frac{1}{2}  \bar M^{12}_{\adt\bdt}  \bar M^{\adt\bdt 12}   + \frac{1}{4}  \varepsilon^{\hat{r}\hat{s}} \varepsilon_{rs}   \bar \chi^{\adt}_{ \hat{r}} \bar \chi^{\bdt}_{\hat{s}} \bar \lambda^{r A}_{\adt}\bar  \lambda^{s}_{\bdt A} +  i  P_{\alpha\bdt 34}^A \varepsilon^{\hat{r}\hat{s}} \bar \chi^{\bdt}_{\hat{r}} \lambda^\alpha_{\hat{s} A}  - \frac{1}{32}  \varepsilon^{\hat{r}\hat{s}} \varepsilon^{\hat{t}\hat{u}}  \lambda^A_{\alpha\hat{r}} \lambda^\alpha_{\hat{s}A}  \lambda^B_{\beta\hat{t}} \lambda^\beta_{\hat{u}B} \ , \CR
     X^\ord{0} &=&-   P^A_{\alpha\bdt 34} P^{\alpha\bdt}{}_{34 A} -\frac{1}{2}   M_{\alpha\beta 34} \lambda^{\alpha A}_3 \lambda^\beta_{4 A} -\frac{1}{2}  \bar M_{\adt\bdt}^{12} \bar\lambda^{\adt 1 A} \bar\lambda^{\bdt 2}_A  \CR && \hspace{50mm} - \frac{1}{8}\varepsilon^{\hat{r}\hat{s}} \varepsilon_{rs}   \chi_\alpha^r \lambda^{\alpha A}_{\hat{r}} \bar \chi_{\bdt \hat{s}} \bar \lambda^{\bdt s}_A   - \frac{1}{4} \lambda^{[A}_{\alpha 3} \lambda^{\alpha B]}_4 \bar \lambda^1_{\bdt [A} \bar \lambda^{\bdt 2}_{B]}    \ .    \eea

Note that $X^\ord{0}$ is real with respect to harmonic conjugation, as defined in \cite{Galperin:1984av,Hartwell:1994rp}, so that this invariant is real for $\cF=1$.\footnote{This follows from the property that $X^\ord{0} = u^i{}_3 u^k{}_3 u^j{}_4 u^l{}_4 X_{ij,kl}$ for an operator $X_{ij,kl}$ in the real $[0,2,0]$ representation of $SU(4)$, \ie $\bar X^{ij,kl} = \frac{1}{4} \varepsilon^{ijpq} \varepsilon^{klmn} X_{pq,mn}$.} This is the only invariant of this class that is also invariant with respect to $SL(2,\mathds{R})$ duality symmetry, and one can prove that it is in fact (\ref{d2F4}). Indeed, one knows from the linearised analysis that there is only one $SL(2,\mathds{R})$ R-symmetry invariant at this dimension, and one can verify that the two expressions coincide in the linearised approximation. 

For a general monomial function in $\bar T$, $\cF[\bar T] = \tfrac{1}{p+1}\bar T^{p+1}$, the linearised integrand is 
\bea && W_{34}^A W_{34 A} \Bigl( \frac{1}{2}  \bar W^p  \bar M^{12}_{\adt\bdt}  \bar M^{\adt\bdt 12} + p   \bar W^{p-1} \bar M^{12}_{\adt\bdt}  \bar \chi^{\adt}_3 \bar \chi^{\bdt}_4 +  \frac{p(p-1)}{12}  \bar W^{p-2}\varepsilon^{\hat{r}\hat{s}} \varepsilon^{\hat{t}\hat{u}} \bar  \chi_{\adt \hat{r}} \bar \chi_{\bdt \hat{s}}  \bar \chi^{\adt}_{ \hat{t}} \bar \chi^{\bdt}_{\hat{u}} \Bigr) \CR
&=&  \frac{1}{4} \bar D_{\adt 3} \bar D^{\adt}_3 \bar D_{\bdt 4} \bar D^{\bdt}_4 \Scal{ \tfrac{1}{(p+2)(p+1)} W_{34}^A W_{34 A}  \bar W^{p+2} }  \ . \eea
In the linearised approximation, this invariant is therefore associated to the $(4,2,0)$ integral of the  one-quarter BPS integrand $W_{34}^A W_{34 A}  \bar W^{p+2}$ satisfying 
\be D_\alpha^r \Scal{ \tfrac{1}{(p+2)(p+1)} W_{34}^A W_{34 A}  \bar W^{p+2} } = 0 \ . \ee
One straightforwardly computes that the corresponding $(4,2,2)$ integral gives rise to a coupling of the form $ \bar W^n F^2 R^2$ in the linearised approximation, and therefore in general one has
\be  \int d\mu_{\scriptscriptstyle (4,2,2)}  \cL_{\scriptscriptstyle (4,2,2)}[\cF]  \sim \int d^4x e \Scal{ \cF[\bar T] \partial^ 2 F^4 + \bar U^{-2}  \bar \partial \cF[\bar T] M_{\alpha\beta A} M^{\alpha\beta A} \bar C_{\adt\bdt\cdt\ddt} \bar C^{\adt\bdt\cdt\ddt}  + \dots } \ . \label{R2F2I} \ee
Of course this invariant is complex in general, and one must consider separately its real and imaginary parts to get a real invariant. Because the $U(1)$ symmetry is axial, the real part is parity-even while the imaginary part is parity-odd. 

The dimension-six invariants we have discussed exhaust the possible chiral primary operators of $SU(2,2|4)$ that can define invariants with six derivatives in the linearised approximation. We therefore conclude  that the class of invariants depending on a general holomorphic function $\cF(T)$ in this section includes all invariants at this dimension.

\subsection{Dimension-eight invariants} 
%%%%%%%%%
\subsubsection{$R^4$ type invariants}
%%%%%%
The pure supergravity invariants at this dimensions are of $R^4$  type \cite{Deser:1977nt,Deser:1978br,Kallosh:1980fi,Howe:1981xy} and in $D=4$ arise at three loops. At this order, in $\cN=4$, examples of such invariants are given by full-superspace integrals of arbitrary functions of the complex scalar superfield $T$. To analyse these fully, including the case when vector multiplets are present, it will again turn out to be useful to use harmonic-superspacee techniques.

Let us consider first  $(4,1,1)$ harmonic superspace. We can associate four odd normal coordinates with the four involutive odd directions, as in  \cite{Bossard:2011tq,KuzenkoRY}, and use these to relate full-superspace integrals to integrals over the remaining twelve odd coordinates, \ie over $(4,1,1)$ analytic superspace. This programme was carried out in pure $\cN=4$ and $\cN=8$ supergravities in \cite{Bossard:2011tq} to show that the full-superspace integral of the Berezinian of the supervielbein, \ie the volume of superspace, vanishes subject to the classical equations of motion. We refer to  \cite{Bossard:2011tq,KuzenkoRY} for more details; here, we  simply state that the full-superspace integral of an arbitrary function $H(T,\bar T)$ can be rewritten as
\bea   \int d^4 x d^{16} \theta\, E \, H(T,\bar T) &= &\frac{1}{4} \int d\mu_{\scriptscriptstyle (4,1,1)}\, (D^1 D^1) (\bar D_4 \bar D_4) \, H (T,\bar T) \CR
&=&  \frac{1}{4} \int d\mu_{\scriptscriptstyle (4,1,1)}\, (\chi^1\chi^1) ( \bar\chi_4\bar \chi_4) \, (\Delta-2)\Delta H(T,\bar T) \ ,\eea
where, for any spinor $\psi$, $(\psi \psi) =\ve^{\a\b}\psi_\a\psi_\b$, and $(\bar \psi \bar \psi) = \ve^{\adt\bdt} \bar \psi_{\adt} \bar \psi_{\bdt}$, and where we recall that  $\Delta \equiv   ( 1- T \bar T )^2 \partial \bar \partial$
is the $SL(2,\mathds{R})$ invariant Laplace operator on the unit disc. Clearly, this integral vanishes if $H$ is an eigenfunction of the scalar target-space Laplace operator with eigenvalue $0$ or $2$, and in particular if $H(T,\bar T)$ is a constant, or more generally a holomorphic function. Moreover, one straightforwardly computes that
\be (\Delta-2)\Delta \, \Scal{ - \mbox{ln}\left({1-T \bar T} \right) }= -2 \ee
and therefore the full-superspace integral of $K:=\mbox{ln}\left({1-T \bar T}\right)$ is duality-invariant. This is the K\"{a}hler potential for the symmetric space $SU(1,1)/U(1)$, and its full-superspace integral is duality-invariant just as it is in $\cN=1$. This can be seen more easily using the coordinate $\uptau = i \frac{1-T}{1+T}$, in terms of which the K\"{a}hler potential is $K= - \mbox{ln}( \mbox{Im}[\uptau])$. The latter transforms  under  $SL(2,\bbR)$ as
\be - \mbox{ln}( \mbox{Im}[\uptau])\rightarrow - \mbox{ln}\left( \mbox{Im}\Bigl[ \frac{a \uptau + b}{d  + c\,  \uptau} \Bigr]\right) = - \mbox{ln}( \mbox{Im}[\uptau]) +  \mbox{ln}( d  + c\,  \uptau ) +  \mbox{ln}( d  + c\, \bar  \uptau ) \ , \ee
so that the integral of $K$ remains unchanged because the full-superspace integral of any holomorphic function of $\uptau$ vanishes.

For any function $G(T,\bar T)$ we can define a $(4,1,1)$ analytic superspace integral by
\be  S[G]=  \frac{1}{4} \int d\mu_{\scriptscriptstyle (4,1,1)} (\chi^1\chi^1) ( \bar\chi_4\bar \chi_4)  \, G(T,\bar T) \label{R4Pure} \ee
because the integrand is G-analytic, \ie annihilated by $D_\a^1$ and $\bar D_{\adt 4}$. This follows from the properties of $\chi$ under differentiation and from the fact that $D_\a^1 T=\chi_\a^1$, so that differentiating $G$ leads to cubic (and hence vanishing) expressions in $\chi^1$ or $\bar\chi_4$.

A general supersymmetric invariant of this form can always be rewritten as a full-superspace integral of a function $H$ that is a solution of the equation
\be (\Delta-2)\Delta\, H(T,\bar T) = G(T,\bar T) \ . \ee

Given the results of Section 2.2, it is straightforward to check that the consistency requirements for the existence of a normal-coordinate expansion associated to the $(4,1,1)$ measure are still satisfied in the presence of vector multiplets. The computation of the expansion of the Berezinian of the supervielbein can be carried out exactly as in \cite{Bossard:2011tq} in the presence of vector multiplets because the relevant components of the Riemann tensor are still determined by the unique G-analytic vector 
\be B_{\alpha\dot{\beta}} = 2 \chi_\alpha^1 \bar \chi_{\dot{\beta} 4} +\bar  \lambda_{\dot{\beta}}^{1 A} \lambda_{\alpha 4 A} = u^1{}_i u^j{}_4 \scal{ 2 \chi_\alpha^i \bar  \chi_{\dot{\beta} j} + \bar  \lambda_{\dot{\beta}}^{i A} \lambda_{\alpha j A} }\ .\label{Bvect}\ee
We find
\be \,E = \cE \Scal{ 1 - \frac{1}{6} \zeta^\alpha \zeta^{\dot{\beta}} \scal{ 2 \chi_\alpha^1\bar  \chi_{\dot{\beta} 4} +\bar  \lambda_{\dot{\beta}}^{1 A} \lambda_{\alpha 4 A}}} \ ,\ee 
where $(\zeta,\bar\zeta)$ are the normal coordinates, and $\cE$ is the measure on analytic superspace.

The full-superspace integral of an arbitrary function $H(T,\bar T)$ in the presence of vector multiplets can therefore be written
\begin{multline}  \int d^4 x d^{16} \theta \,E \, H(T,\bar T)  =  \frac{1}{4} \int d\mu_{\scriptscriptstyle (4,1,1)} \biggl( (\chi^1 \chi^1) ( \bar\chi_4 \bar \chi_4) \, (\Delta - 2 ) \biggr .   \\
 - \frac{1}{2} (\chi^1 \chi^1) (\l_4^A \l_{4 A} ) \bar U^{-2}  \bar \partial - \frac{1}{2} (\bar\chi_4 \bar \chi_4) (\bar\l^1_A \bar \lambda^{1A} ) U^{-2} \partial 
\\ \biggl .  
+ 2 (\l_{4A} \chi^1)(\bar\l^{1A} \bar\chi_4)+ \frac{1}{4} (\l_{4 A} \l_4^A)  (\bar\l^{1B}\bar \l^1_B) \biggr )   \Delta  \, H(T,\bar T) \ . \end{multline} 
This integral vanishes if $H(T,\bar T)$ is a solution to the Laplace equation, and in particular if it is a constant. Hence, the volume of $D=4$, $\cN=4$ supergravity vanishes even in the presence of vector-multiplet coupling.

A  $(4,1,1)$ analytic superspace integral for a general function $J(T,\bar T)$ can be written
\begin{multline} S[J] =  \frac{1}{4} \int d\mu_{\scriptscriptstyle (4,1,1)} \biggl( (\chi^1 \chi^1) ( \bar\chi_4 \bar \chi_4) \, (\Delta - 2 ) \biggr .   \\
 - \frac{1}{2} (\chi^1 \chi^1) (\l_4^A \l_{4 A} ) \bar U^{-2}  \bar \partial - \frac{1}{2} (\bar\chi_4 \bar \chi_4) (\bar\l^1_A \bar \lambda^{1A} ) U^{-2} \partial 
\\ \biggl .  
+ 2 (\l_{4A} \chi^1)(\bar\l^{1A} \bar\chi_4)+ \frac{1}{4} (\l_{4 A} \l_4^A)  (\bar\l^{1B}\bar \l^1_B) \biggr )   J(T,\bar T) \ . \label{QuaterR4} \end{multline} 
One can also integrate the square of the G-analytic vector $B_{\alpha\dot{\beta}}$ given in (\ref{Bvect}) to get an independent invariant:
\be S_{\scriptscriptstyle 1/4}= \int d\mu_{\scriptscriptstyle (4,1,1)} B_{\a\bdt} B^{\a\bdt}\ .  \label{FakeVolume} \ee
These are the only possible $SO(6,n)$-invariant G-analytic integrands for the $(4,1,1)$ measure at this order.

The duality invariant given in (\ref{QuaterR4}) for $J=1$ can be defined as the full-superspace integral of the K\"{a}hler potential just as in the absence of vector multiplets.  However, the duality invariant (\ref{FakeVolume}) can only be defined as an integral over $(4,1,1)$ harmonic superspace.  

In the linearised approximation, only the scalar field $T$ contains the Riemann tensor, and therefore only the term 
\be  \int d\mu_{\scriptscriptstyle (4,1,1)} (\chi^1\chi^1)(\bar \chi_4\bar\chi_4) \sim    \int d^4 xd^{16} \theta\,  T^2 \bar T^2 \sim  \int d^4 x C_{\alpha\beta\gamma\delta} C^{\alpha\beta\gamma\delta} \bar  C_{\dot{\alpha}\dot{\beta}\dot{\gamma}\dot{\delta}} \bar C^{\dot{\alpha}\dot{\beta}\dot{\gamma}\dot{\delta}}+ \dots \ee
includes an $R^4$ contribution (where $C_{\alpha\beta\gamma\delta}$ is the Weyl tensor).

These invariants are the most general $SO(6,n)$ invariants that can be defined as $(4,1,1)$ harmonic-superspacee integrals. However, this does not imply that they define the most general $R^4$ type invariants preserving  $SO(6,n)$. For instance, the dimensional reduction of the ten-dimensional $R^4$ type invariant admits a coupling of the form $\mbox{Im}(\tau)^2 R^4$, whereas $\Delta \mbox{Im}(\tau)^2  = 2 \mbox{Im}(\tau)^2 $, and this is precisely the eigenfunction that cannot appear in (\ref{QuaterR4}). Note, however, that  one could define a $J = \mbox{Im}(\tau)^2 \mbox{ln}(\mbox{Im}(\tau))$ satisfying 
\be  \Delta  \Scal{ \mbox{Im}(\tau)^2 \mbox{ln}(\mbox{Im}(\tau))  } \equiv  - ( \tau - \bar \tau)^2 \partial \bar \partial \Scal{ \mbox{Im}(\tau)^2 \mbox{ln}(\mbox{Im}(\tau)) } = 2  \Scal{ \mbox{Im}(\tau)^2 \mbox{ln}(\mbox{Im}(\tau))  }  - \mbox{Im}(\tau)^2 \ ; \ee
such an invariant would have terms involving the logarithm of $ \mbox{Im}(\tau)$ in the $\partial^4 F^4$ and $\partial^2 F^2 R^2$ couplings which  is not the case for the invariant obtained by dimensional reduction from $D=10$. We shall see in Section \ref{R4onehalf} that the invariant obtained by such dimensional reduction can only be written as a $(4,2,2)$ superspace integral, so that it is effectively one-half BPS at the non-linear level.

Note that the linearised analysis suggests that one should be able to define an infinite class of invariants associated to the full-superspace integral of the linearised integrand $W^n ( \tfrac{1}{2} \varepsilon^{ijkl} W_{ij} W_{kl} )^2$. One can check that this is indeed possible by solving the equation 
\be - ( \tau - \bar \tau)^2 \partial \bar \partial J(\tau , \bar \tau ) = 2 J(\tau , \bar \tau )  \ . \ee
One can solve this equation as a convergent series in terms of a holomorphic function as
\be J( \tau , \bar \tau ) = \frac{1}{6} \cF(\tau) + \sum_{p=1}^{\infty} (-1)^p \frac{1+p}{(3+p)!} ( \tau - \bar \tau )^p \partial^p \cF(\tau) \ . \ee
Defining the function $J$ as in (\ref{QuaterR4}), one will get by construction an invariant that does not involve any $C^2 \bar C^2$ coupling. Note, however, that the operation defining $J$ from $\cF$ does not preserve any modular property. The only automorphic function satisfying this Poisson equation is the real analytic Eisenstein series $\cE_2(\tau , \bar \tau)$. 
%%%%%%%%
\subsubsection{$\partial^4 F^4 $, $(\partial F)^2 R^2$ and $(\partial^2 T)^2 R^2$ type invariants}
%%%%%%%
The structure that we described in subsection \ref{Sectiond2F4} can also be used to define eight-derivative invariants in $(4,2,2)$ superspace. We consider the same construction as in (3.55) but with 
\be L^\ord{2n} = X^\ord{0} X^\ord{2n} \ . \ee 
Because $X^\ord{0}$ is itself G-analytic, it follows that the corresponding integrand $ \cL_{\scriptscriptstyle (4,2,2)}[\cF] $ is G-analytic. As in the case of the $\partial^2 F^4$ type invariant, for $\cF=1$ the invariant can be rewritten as a $(4,1,1)$ harmonic-superspacee integral 
\bea  \int d\mu_{\scriptscriptstyle (4,2,2)}  ( X^\ord{0} )^2 &\propto& \int  d\mu_{\scriptscriptstyle (4,1,1)} \Scal{ (\l_{4 A}\l_4^A) (\bar\l^1_B\bar\l^{1 B})+2(\l_{4 A} \l_{4 B}) (\bar\l^{1 A}\bar\l^{1 B}) } \ , \label{d4F4411} \eea
but because the right-hand-side is a linear combination of both (\ref{QuaterR4}) for $J(T,\bar T) = 2$ and (\ref{FakeVolume}), it cannot be written as a full-superspace integral. This invariant coincides at the linear level with the $F^4$ type invariant with four more derivatives included, which is completely symmetric (as opposed to the $\partial^2 F^4$ invariant). 

For a non-trivial holomorphic function, the first derivative of $\cF$ multiplies the factor $X^\ord{2}$, which can be identified in the linearised approximation with $\frac{1}{2} \bar M_{\adt\bdt}^{12} \bar M^{\adt\bdt 12}$. For a general monomial function in $\bar T$, $\cF[\bar T] = \tfrac{1}{p+1}\bar T^{p+1}$, the linearised integrand is then 
\bea &&\partial_a W_{34}^A \partial^a W_{34 A} \Bigl( \frac{1}{2}  \bar W^p  \bar M^{12}_{\adt\bdt}  \bar M^{\adt\bdt 12} + p   \bar W^{p-1} \bar M^{12}_{\adt\bdt}  \bar \chi^{\adt}_3 \bar \chi^{\bdt}_4 +  \frac{p(p-1)}{12}  \bar W^{p-2}\varepsilon^{\hat{r}\hat{s}} \varepsilon^{\hat{t}\hat{u}} \bar  \chi_{\adt \hat{r}} \bar \chi_{\bdt \hat{s}}  \bar \chi^{\adt}_{ \hat{t}} \bar \chi^{\bdt}_{\hat{u}} \Bigr) \CR
&=&  \frac{1}{8} D_\alpha^2 D^{\alpha 2} \bar D_{\adt 3} \bar D^{\adt}_3 \bar D_{\bdt 4} \bar D^{\bdt}_4 \Scal{ \tfrac{1}{4 (p+2)(p+1)} \bar \lambda_{\adt}^{1A} \bar \lambda^{\adt 1}_A   \bar W^{p+2} }  \ . \eea
In the linearised approximation, this invariant is therefore  a $(4,1,0)$ integral of the one-eighth BPS integrand $\bar \lambda_{\adt}^{1A} \bar \lambda^{\adt 1}_A   \bar W^{p+2}$ satisfying
\be D_\alpha^1   \Scal{ \tfrac{1}{4 (p+2)(p+1)} \bar \lambda_{\adt}^{1A} \bar \lambda^{\adt 1}_A   \bar W^{p+2} }  = 0 \ , \ee
where the expression in the brackets is related to  the one-eighth BPS chiral primary \eq{TDDW}. This invariant can in fact be defined as a $(4,1,0)$ harmonic-superspacee integral at the non-linear level. The $(4,1,0)$ normal-coordinate expansion of the supervielbein Berezinian is trivial (\ie it stops at zero order), and the corresponding measure is therefore simply obtained by  pulling it back onto the analytic subspace. One can write two G-analytic integrands for the $(4,1,0)$ measure corresponding to this class of invariants, $ \bar \lambda_{\adt}^{1A} \bar \lambda^{\adt 1}_A \bar U^2   \cF^\ord{-2}(\bar T)$ and $V^{1r\hat{I}} V^{1s}{}_{\hat{I}} \bar \chi_{\adt r} \bar \chi^{\adt}_s  \bar U^{-2} \bar \partial \cF(\bar T)$.
The two expressions are equivalent in the linearised approximation, and by uniqueness of the one-eighth BPS chiral primary for a fixed number of fields, we conclude that they must define the same invariant at the non-linear level as well. The first integrand is not manifestly $SL(2,\mathds{R})$ covariant, and the explicit relation between $ \cF^\ord{-2}(\bar T)$ and $\cF(\bar T)$ remains to be worked out. The second integrand is not manifestly $SO(6,n)$ invariant, whereas the corresponding invariant is. 

Note that the integrand $\bar \lambda_{\adt}^{1A} \bar \lambda^{\adt 1}_A   \bar W^{p+2}$ is very similar to the chiral integrand of a chiral superfield dependent Maxwell kinetic term in $\cN=1$ superspace, and one may therefore think that one could write such an invariant as the full-superspace integral of a Chern--Simons like integrand depending on the Maxwell potentials. However, one may check that this is not the case, because of the non-linear terms in the $\cN=4$ algebra.

One computes that the corresponding $(4,1,0)$ integral gives rise to a coupling of the form $ \bar W^n (\partial F)^2 R^2$ in the linearised approximation, and therefore in general one has
\be  \int d\mu_{\scriptscriptstyle (4,2,2)}  \cL_{\scriptscriptstyle (4,2,2)}[\cF]  \sim \int d^4x e \Scal{ \cF[\bar T] (DF)^4 + \bar U^{-2}  \bar \partial \cF[\bar T] D_a M_{\alpha\beta A} D^a M^{\alpha\beta A} \bar C_{\adt\bdt\cdt\ddt} \bar C^{\adt\bdt\cdt\ddt}  + \dots } \ . \label{DF2R2} \ee
Such invariants are the analogues of the invariants discussed in the last section with two more space-time derivatives inserted. As such it is clear that they do not include a $C^2 \bar C^2$ term, and because they are $SL(2,\mathds{R})$ covariant, they cannot be written as $(4,1,1)$ integrals of $SO(6,n)$ invariant integrands (\ref{QuaterR4}). Because these invariants can be written as $(4,1,0)$ harmonic-superspacee integrals, one would have expected  them to be writable as $(4,1,1)$ harmonic-superspacee integrals. This is not necessarily the case at the non-linear level, but one nevertheless expects that there is a corresponding $(4,1,1)$ G-analytic integrand that is not $SO(6,n)$ invariant.

There is another class of invariants that can be defined with eight space-time derivatives in terms of a weight-four holomorphic function: 
\begin{multline}  \cL^\ord{4}_{\scriptscriptstyle (4,2,2)}[\cF] = \bar U^{-4} \cF( \bar T) L^\ord{4} +  \bar U^{-6} \Scal{ \bar \partial  \cF( \bar T) - \frac{4T}{1-T\bar T}   \cF( \bar T)  }  L^\ord{6} \\  +  \bar U^{-8} \Scal{ \bar \partial^2  \cF( \bar T) - \frac{10T}{1-T\bar T}  \bar \partial \cF( \bar T) +  \frac{20T^2}{(1-T\bar T)^2}  \cF( \bar T)}  L^\ord{8} \ .\label{Weight4H} \end{multline}
Here, weight-four indicates that the function transforms  as
\be \cF(\bar \uptau) \rightarrow \frac{\cF\scal{\frac{a \bar \uptau + b}{c \bar \uptau + d}}}{(c \bar \uptau + d)^4} \ , \ee
with respect to $SL(2,\mathds{R})$, because of its coupling to the vector fields and the fermions. The G-analyticity condition requires in this case that 
\be\begin{split} D_\alpha^r L^\ord{4} &= 4 \chi_\alpha^r L^\ord{6}\ , \\
 D_\alpha^r L^\ord{6} &=  10  \chi_\alpha^r L^\ord{8}  \ , \\
   D_\alpha^r L^\ord{8} &=0 \ , \\
   \end{split}\hspace{10mm}\begin{split}
   \bar D_{\adt \hat{r}} L^\ord{4} &=0 \ , \\
 \bar D_{\adt \hat{r}} L^\ord{6} &=  - \bar \chi_{\adt \hat{r}} L^\ord{4}  \ , \\
  \bar D_{\adt \hat{r}} L^\ord{8} &=- \bar \chi_{\adt \hat{r}} L^\ord{6} \ . \label{G422w4Descent} 
   \end{split}\ee
Once again, the supersymmetry algebra is consistent with these conditions provided $L^\ord{8}$ satisfies the algebraic constraint 
\be \bar \chi_{\adt \hat{r}} L^\ord{8} = 0 \ , \ee
which is also required for the descent to stop. Using the result of the last section, one straightforwardly computes the solution to these equations: 
\be L^\ord{4} = ( X^\ord{2})^2 - 2 X^\ord{0} X^\ord{4} \ , \quad L^\ord{6} = X^\ord{2} X^\ord{4} - 3 X^\ord{0} X^\ord{6} \ , \quad L^\ord{8} = \frac{1}{5} ( X^\ord{4})^2 + \frac{3}{5}  X^\ord{2} X^\ord{6} \ .  \ee
 To prove this, we have used the property that the $X^\ord{2n}$ satisfy (\ref{G422Descent}), as well as the property that $X^\ord{4}$ is quadratic in $\bar \chi_{\adt \hat{r}}$ and $X^\ord{6}$ quartic in $\bar \chi_{\adt \hat{r}}$, from which it follows trivially that 
 \be \bar \chi_{\adt \hat{r}} ( X^\ord{4})^2 = 0 \ , \qquad \bar \chi_{\adt \hat{r}} X^\ord{6} = 0 \ , \qquad X^\ord{4} X^\ord{6} = 0 \ . \ee
For a general monomial function in $\bar T$, $\cF[\bar T] = \bar T^{p}$, the linearised integrand is then 
\bea &&\frac{1}{2} \bar M^{12}_{\adt\bdt} \bar M^{12 \adt \bdt}   \Bigl( \frac{1}{2}  \bar W^p  \bar M^{12}_{\adt\bdt}  \bar M^{\adt\bdt 12} + p   \bar W^{p-1} \bar M^{12}_{\adt\bdt}  \bar \chi^{\adt}_3 \bar \chi^{\bdt}_4 +  \frac{p(p-1)}{12}  \bar W^{p-2}\varepsilon^{\hat{r}\hat{s}} \varepsilon^{\hat{t}\hat{u}} \bar  \chi_{\adt \hat{r}} \bar \chi_{\bdt \hat{s}}  \bar \chi^{\adt}_{ \hat{t}} \bar \chi^{\bdt}_{\hat{u}} \Bigr) \CR
&=&  \frac{1}{4}  \bar D_{\adt 3} \bar D^{\adt}_3 \bar D_{\bdt 4} \bar D^{\bdt}_4 \Scal{ \tfrac{1}{2 (p+2)(p+1)} \bar M^{12}_{\adt\bdt} \bar M^{12 \adt \bdt}     \bar W^{p+2} }  \ . \eea
In the linearised approximation, this invariant is therefore given by a $(4,2,0)$ integral of the one-quarter BPS integrand $\bar M^{12}_{\adt\bdt} \bar M^{12 \adt \bdt}     \bar W^{p+2} $ satisfying  the constraint 
\be D_\alpha^r \Scal{ \tfrac{1}{2 (p+2)(p+1)} \bar M^{12}_{\adt\bdt} \bar M^{12 \adt \bdt}     \bar W^{p+2} } = 0 \ ,\qquad r=1,2\ . \ee
One straightforwardly computes that the corresponding integral gives rise to a coupling of structure $ \bar W^p (\partial^2 \bar W)^2 \bar C^2$ in the linearised approximation. This class of invariant defines the type of couplings discussed in \cite{Antoniadis:2006mr}. 
 %%%%%%%
\subsubsection{Protected $R^4$ type invariants}
 \label{R4onehalf}
 We have seen that $R^4$ type invariants determined by a function of the complex scalar $J_\ord{2}$ satisfying  $\Delta J_\ord{2} = 2 J_\ord{2}$ cannot be written as $(4,1,1)$ harmonic-superspacee integrals. We will show in this section that they can, however, be written as $(4,2,2)$ harmonic-superspacee integrals. To do so, we shall show that one can associate a G-analytic integrand to any function $J_\ord{s}(T,\bar T)$ satisfying $\Delta J_\ord{s} = s J_\ord{s}$. They constitute a very restricted set of $R^4$ type invariants which contains precisely the $s=2$ example we could not build as a $(4,1,1)$ harmonic integral. 
 
 We consider a G-analytic integrand of the form 
 \begin{multline}  \cL^\ord{{\rm R}}_{\scriptscriptstyle (4,2,2)}[J_\ord{s}] = U^{-6} \Scal{  \partial^3  J_\ord{s} - \frac{6\bar T}{1-T\bar T}   \partial^2 J_\ord{s} +  \frac{6\bar T^2}{(1-T\bar T)^2}  \partial J_\ord{s}}  L^\ord{-6}_s \\ + U^{-4} \Scal{ \partial^2J_\ord{s}  - \frac{2\bar T}{1-T\bar T} \partial J_\ord{s}   }  L^\ord{-4}_s + U^{-2} \partial J_\ord{s} L^\ord{-2}_s \\ + J_\ord{s} L^\ord{0}_s + \bar U^{-2} \bar \partial J_\ord{s}  L^\ord{2}_s +  \bar U^{-4} \Scal{ \bar \partial^2  J_\ord{s}  - \frac{2T}{1-T\bar T} \bar \partial  J_\ord{s}  }  L^\ord{4}_s \\  +  \bar U^{-6} \Scal{ \bar \partial^3 J_\ord{s}  - \frac{6T}{1-T\bar T}  \bar \partial^2 J_\ord{s} +  \frac{6T^2}{(1-T\bar T)^2}  \bar \partial J_\ord{s} }  L^\ord{6}_s \ . \label{R4422} \end{multline}
Here we assume the function $J_\ord{s}(T,\bar T)$ is real analytic, and $L^\ord{-2n}$ is the complex  conjugate of $L^\ord{2n}$. Because the derivatives are all K\"{a}hler-covariant, one obtains that 
\bea D_{\alpha}^r \biggl( \bar U^{-6} \Bigl(  \bar \partial^3 J_\ord{s}  - \frac{6T}{1-T\bar T}  \bar \partial^2 J_\ord{s} &+&  \frac{6T^2}{(1-T\bar T)^2} \bar \partial J_\ord{s} \Bigr)  \biggr)  = \chi_\alpha^r  \bar U^{-4} \Scal{ \bar \partial^2    - \frac{2T}{1-T\bar T} \bar \partial } \scal{ \Delta J_\ord{s} - 6 } \CR
&=& ( s-6 ) \chi_\alpha^r  \bar U^{-4} \Scal{ \bar \partial^2  J_\ord{s}  - \frac{2T}{1-T\bar T} \bar \partial  J_\ord{s}  }  \ , \CR
D_\alpha^r \biggl(  \bar U^{-4} \Scal{ \bar \partial^2  J_\ord{s}  - \frac{2T}{1-T\bar T} \bar \partial  J_\ord{s}  } \biggr)  &=& ( s-2 )  \chi_\alpha^r \bar U^{-2} \bar \partial J_\ord{s}   \ , \CR
D_\alpha^r \biggl(  \bar U^{-2} \bar \partial  J_\ord{s}  \biggr)   &=& s  \chi_\alpha^r  J_\ord{s}  \ . \eea
Therefore the integrand (\ref{R4422}) is G-analytic, if and only if 
\be\begin{split} D_\alpha^r L^\ord{0}_s &=-s\,  \chi_\alpha^r L^\ord{2}_s \ , \\
 D_\alpha^r L^\ord{2}_s &=  (2-s) \chi_\alpha^r L^\ord{4}_s  \ , \\
  D_\alpha^r L^\ord{4}_s &=(6-s) \chi_\alpha^r L^\ord{6}_s \ , \\
   D_\alpha^r L^\ord{6}_s &=0 \ , \\
   \end{split}\hspace{10mm}\begin{split}
   \bar D_{\adt \hat{r}} L^\ord{0}_s &=- s\,  \bar \chi_{\adt \hat{r}} L_s^\ord{-2}  \ , \\
 \bar D_{\adt \hat{r}} L^\ord{2}_s &=  - \bar \chi_{\adt \hat{r}} L^\ord{0}_s  \ , \\
  \bar D_{\adt \hat{r}} L^\ord{4}_s &=- \bar \chi_{\adt \hat{r}} L^\ord{2}_s \ , \\
   \bar D_{\adt \hat{r}} L^\ord{6}_s &=-\bar \chi_{\adt \hat{r}} L^\ord{4}_s\ , \label{G422DescentReal} 
   \end{split}\ee
 and identically for the complex conjugate. Moreover, $L^\ord{6}_s$ must satisfy $\bar \chi_{\adt \hat{r}} L^\ord{6} = 0$ for consistency. One checks that the supersymmetry algebra does not imply any restriction on the value of $s$. 
 
 We shall not compute the most general solution, but will only consider the cases for which $L^\ord{2n}_s$ can be written as a bilinear in the operators $X^\ord{2n}$ defined in (\ref{Xsolution}). Consistently, we define $X^\ord{-2n}$ as the harmonic conjugate of $X^\ord{2n}$  \cite{Galperin:1984av,Hartwell:1994rp}, \ie   $X^\ord{2n} \equiv u^i{}_3 u^k{}_3 u^j{}_4 u^l{}_4 X^\ord{2n}_{ij,kl}$ with $X^\ord{2n}_{ij,kl}$ in the $[0,2,0]$ of $SU(4)$, and
\be X^\ord{-2n}_{ij,kl} = \frac{1}{4} \varepsilon_{ijpq} \varepsilon_{klmn} \bar X^{\ord{2n} pq,mn} \ . \ee
Note that only $X^\ord{0}$ is real with respect to harmonic conjugation. One finds the unique solution
 \bea L^\ord{0}_s &=& (X^\ord{0})^2 + s X^\ord{2} X^\ord{-2} + s ( s + 2 ) X^\ord{4} X^\ord{-4} + s ( s^2 + 10 s + 12 ) X^\ord{6} X^\ord{-6} \ ,  \CR
 L^\ord{2}_s &=& X^\ord{2} X^\ord{0} + s X^\ord{4} X^\ord{-2} + s ( s + 4 ) X^\ord{6} X^\ord{-4} \ , \CR
 L^\ord{4}_s &=& X^\ord{4} X^\ord{0} + s X^\ord{6} X^\ord{-2} \ , \CR
 L^\ord{6}_s &=& X^\ord{6} X^\ord{0}  . \eea
 Note that for $s=0$ one gets back the invariant of the last section, since then $J_\ord{0}(T,\bar T) = \cF(T) + \cF(\bar T) $ as a solution of the Laplace equation. Also, the invariant  (\ref{Weight4H}) can equivalently be thought of as being defined from a real analytic function $J_\ord{2}$ such that $\cF(\bar T)$ in (\ref{Weight4H}) is defined as
 \be \cF(\bar T) = \bar \partial^2  J_\ord{2}(T,\bar T)   - \frac{2T}{1-T\bar T} \bar \partial  J_\ord{2}(T,\bar T) \  . \ee  
This second integrand together with $ \cL^\ord{{\rm R}}_{\scriptscriptstyle (4,2,2)}[J_\ord{2}]$ define a one-parameter family of $R^4$ type invariant depending on a real analytic function $J_\ord{2}$.

The full-superspace integral of a function $J_\ord{s}(T,\bar T)$ gives rise to an $R^4$ coupling multiplied by the function $J_\ord{s}$ of the complex scalar field, as long as $s \ne 0,2$. One can define the normal-coordinate expansion of such an integrand in terms of the $(4,2,2)$ normal vectors in order to obtain the corresponding integrand for the $(4,2,2)$ measure. We can then compare the corresponding integrand with the ones we have just built. In order to do so, we shall consider the expression of $L^\ord{2n}$ in the linearised approximation, \ie only consider terms quartic in fields. In this approximation $X^\ord{4}$ and $X^\ord{6}$ are negligible, and the expression one gets from the full-superspace integral of $J_\ord{s}(T,\bar T)$ is proportional to  
\bea L^{\ord{0} \prime}_s &=&   (X^\ord{0})^2 + (s-2)  X^\ord{2} X^\ord{-2} + 2 M^{\alpha\beta}_{34} \bar M^{\adt\bdt 12} P_{\alpha\adt 34}^A P_{\beta\bdt 34 A} + \mathcal{O}(5\mbox{pt}) \ ,  \CR
L^{\ord{2} \prime}_s  &=& X^\ord{2} X^\ord{0} + \mathcal{O}(5\mbox{pt})  \ . \eea
Therefore the class of invariants we have defined in this section differs from the invariants that can be written as $(4,1,1)$ superspace integrals in general, for all values of $s$. 

Let us now argue that the same situation should generalise to the full-superspace integral of a general function of both the complex field $T$, and the scalars parametrising the $SO(6,n)/ (( SO(6) \times SO(n))$ symmetric space. The $(4,1,1)$ normal-coordinate expansion of such a general function is much more complicated, and we shall not display it in this paper. One can nonetheless straightforwardly compute that the only term in the integrand that will contribute to the $R^4$ coupling will remain of the same form, \ie $( \Delta - 2 ) \Delta J$. It follows that the $R^4$ coupling vanishes for a function $J$ which is an eigenfunction of the $SL(2,\mathds{R}) / SO(2)$ Laplace operator of eigenvalue $0$ or $2$, independently of its dependence in the other scalars. Similarly, one checks that the most general $(4,1,1)$ G-analytic integrand will contribute to a $R^4$ coupling multiplying $( \Delta - 2 ) J$, independently of the dependence of $J$ in the other scalars. Therefore if $J$ is a solution to $\Delta J = 2 J$, one cannot write such invariants as a $(4,1,1)$ superspace integrals either. Although we shall not prove it in this paper, it seems clear that such invariants can however be defined as $(4,2,2)$ harmonic-superspacee integrals in general.

We conclude that in general, a supersymmetry invariant that is entirely determined by a function of the scalar fields satisfying $\Delta J = 2 J$ with a $J R^4$ coupling can only be written as a $(4,2,2)$ harmonic-superspacee integral, and should therefore be considered as a one-half BPS protected operator. We shall come back to this point in the context of heterotic string theory in Section \ref{Heterotic}. 

%%%%%%%%%%%%%
\section{The $\mathfrak{sl}_2\mathds{R}$ anomaly}\label{sec4}
%%%%%%%%%%%
\subsection{The one-loop  $\mathfrak{sl}_2\mathds{R}$ anomaly}

In $D=4$, $\cN=4$ supergravity the $SL(2,\mathds{R})$ duality invariance acts on the complex  scalar superfield 
\be  \uptau[ T] \equiv i \frac{1 -T }{1 + T } = a + i e^{-2\phi} + \mathcal{O}( \theta) \ee
in the standard way. In order to discuss its anomaly, it is convenient to introduce anticommuting parameters that parametrise the $\mathfrak{sl}_2\mathds{R}$ transformation as
\be \left( \begin{array}{cc} h & \ \ e \\ f &\  -h \end{array} \right) \in \mathfrak{sl}_2 \ . \label{sl2hef} \ee  
One defines the BRST-like operator $\d$ associated to the $\mathfrak{sl}_2\mathds{R}$ symmetry by
\be \delta  \uptau   = e + 2 h \uptau - f \uptau^2\ ,  \qquad \delta f = - 2 h f\ ,  \quad \delta h = e f \ , \quad \delta e = 2 h e \ . \ee 
One then  checks straightforwardly that $f \uptau - h$ is a representative of the unique cohomology class of $\delta$ linear in the anticommuting parameters. Although this term is complex, its real part is $\delta$-exact, \ie $f ( \uptau + \bar \uptau ) - 2 h = - \delta\,  \mbox{ln}[ \uptau - \bar \uptau ]$. Because the invariant $\iota^* \cL [\cF]$ as defined in Sections \ref{Rdeux} and \ref{R2withV} is linear in the function $\cF$, one finds that the anomaly functional is 
\bea \cA^\ord{1} &=& i  \int  \, \iota^* \scal{ \cL[ f \bar \uptau -h] - \bar \cL[f \uptau - h ]}   \\
&=&  \int  \scal{ f e^{-2\phi} R_{ab} \wedge R^{ab} + ( f a - h ) \frac{1}{2} \varepsilon_{abcd} R^{ab} \wedge R^{cd} + \dots } \ . \label{AnomalyInvariant} \nn
 \eea
 Indeed, the anomalous Ward identity implies that the variation of the 1PI generating functional produces a cohomologically non-trivial term $ f e^{-2\phi} $ multiplying $R_{ab} \wedge R^{ab}$ \cite{Bossard:2010dq}, but consistency with supersymmetry then implies that this must occur together with the cohomologically trivial term $f a - h $ multiplying $\frac{1}{2} \varepsilon_{abcd} R^{ab} \wedge R^{cd}$. 
 
 We therefore observe that supersymmetry implies that the anomaly associated to the non-linearly realised generator cannot occur independently of  a breaking of the shift symmetry of the dilaton. In pure supergravity the corresponding expression is nonetheless locally a total derivative, and only the corresponding current conservation is anomalous 
\be d J_h = -\frac{ 1}{32 \pi^2}  \varepsilon_{abcd} R^{ab} \wedge R^{cd}\ . \ee
In the presence of vector multiplets, the corresponding supersymmetry invariant also contains the $F^4$ type invariant described in Section \ref{F4Inv}.  The $F^4$ term is also multiplied by $(fa-h)$ but now the $h$-dependent term is no longer a total derivative so that invariance with respect to a shift of the dilaton is lost.

Writing the Slavnov--Taylor identity for the 1PI generating functional giving the unbroken $\mathfrak{sl}_2\mathds{R}$ Ward identity as
\be \mathcal{S}^{\scriptstyle \mathfrak{sl}_2}[ \Sigma] \equiv  \int d^4 x \sum_\gimel \frac{ \delta^R \Sigma}{\delta \varphi^\gimel(x) }  \frac{ \delta^L \Sigma}{\delta K_\gimel(x) }  - 2 h f \frac{\partial  \Sigma}{\partial  f} + e f  \frac{\partial  \Sigma}{\partial  h} + 2 h e \frac{\partial \Sigma}{\partial e}   = 0 \ ,\ee
one has at the one-loop order 
\be \cS^{\scriptstyle \mathfrak{sl}_2}[\Gamma] = \frac{ 2+n}{32 \pi^2} \cA^\ord{1}  + \mathcal{O}(\kappa^2) \ . \ee
Here $\varphi^\gimel$ stands for all the fields of the theory transforming in a non-trivial way with respect to $SL(2,\mathds{R})$, and $K_\gimel(x)$ stands for the sources coupled to their $\delta$ variations.

The computation of the anomaly coefficient is carried out in terms of the Atiyah--Singer family index theorem according to \cite{Singer,Marcus:1985yy}. The sum of the contributions of the spin $1/2$, $1$ and $3/2$ fields gives
\be ( - 3 \times 4 + 1 \times 4 n ) \, \frac{p_\un}{24} + ( - 2 \times 6 + 2 \times n )\,  \frac{ p_\un}{6} +( 1 \times 4 )  \, \frac{7 p_\un}{8} = ( 2 + n )\,  \frac{p_\un}{2} \ . \label{AnomalyCounting} \ee

Note that the one-loop anomaly is only defined as a cohomology class up to a trivial invariant, and in general one has 
\be \cA^\ord{1}[\mathcal{F}] = i  \int  \, \iota^* \scal{ \cL[ f \bar \uptau -h + \delta \mathcal{F}(\bar \uptau) ] - \bar \cL[f \uptau - h + \delta \mathcal{F}(\uptau ) ]}   \ , \ee
such that the specific choice for $\mathcal{F}$ amounts to a renormalisation prescription. The prescription $\mathcal{F}=0$ is the only one that preserves the shift invariance of the $\cN=4$ supergravity axion. As we shall see, this prescription is the one that permits one to restore the $SL(2,\mathds{Z})$ modular symmetry at the non-perturbative level. It is worth pointing out that this is consistent with the renormalisation prescription of the double copy formulation \cite{Bern:2010ue} used in the computation of the four-graviton amplitude at this order \cite{Bern:2012cd}. This is very important, because we shall need to choose this prescription in order to show that the three-loop divergence must necessarily be associated to a duality invariant in pure $\cN=4$ supergravity.

To explain this agreement with the double-copy formalism, one considers the functional derivative of the broken Slavnov--Taylor identity (\ref{BrokenSTI}) with respect to $g_{\mu\nu}(x)$, $g_{\sigma\rho}(y)$ and $\phi(z)$, and $f$ at the 1-loop order
\bea \frac{ \delta^3 \quad }{\delta g_{\mu\nu}(x) \delta g_{\sigma\rho}(y) \delta \phi(z) }\int  \Bigl( d^4 w \frac{ \delta \Gamma}{\delta a(w)}  \bigl[ ( e^{-4\phi(w)} - a^2(w) ) \cdot \Gamma \bigr]  - \frac{2+n}{32\pi^2}   e^{-2\phi(w)} R^{ab}(w) \wedge R_{ab}(w) \Bigr) \CR
= - 4  \frac{ \delta^3 \Gamma  }{\delta g_{\mu\nu}(x) \delta g_{\sigma\rho}(y) \delta a(z) } +   \frac{2+n}{16 \pi^2} \frac{ \delta \frac{1}{4} \varepsilon^{\kappa\lambda\theta\tau} R_{\kappa\lambda}{}^{ab}(z)  R_{\theta\tau ab}(z) }{\delta g_{\mu\nu}(x) \delta g_{\sigma\rho}(y)  } = 0 \ .   \hspace{30mm} 
 \eea 
 We therefore obtain in momentum space \footnote{We make use here of the notation $ A_{\sigma)(\mu} B_{\nu)(\rho} \equiv \frac{1}{4} \scal{  A_{\sigma\mu} B_{\nu\rho} + A_{\rho\mu} B_{\nu\sigma} + A_{\sigma\nu} B_{\mu\rho}+ A_{\rho\nu} B_{\mu\sigma} } $.}
 \be \big\langle g_{\mu\nu}(p_1) g_{\sigma\rho}(p_2) a(-p_1-p_2) \big\rangle = \frac{2+n}{64\pi^2} \varepsilon_{\sigma)(\mu}{}^{\kappa\lambda} \scal{ \eta_{\nu)(\rho} p_1 \cdot p_2 - p_{2 \nu)} p_{1 (\rho} } p_{1 \kappa} p_{2 \lambda} \ . \ee
One similarly obtains that 
\bea \frac{ \delta^3 \quad }{\delta g_{\mu\nu}(x) \delta g_{\sigma\rho}(y) \delta a(z) }\int  \Bigl( d^4 w \frac{ \delta \Gamma}{\delta \phi(w)}  \bigl[ a(w) \cdot \Gamma \bigr]  - \frac{2+n}{64\pi^2}  \varepsilon_{abcd} a(w) R^{ab}(w) \wedge R^{cd}(w) \Bigr) \CR
=   \frac{ \delta^3 \Gamma  }{\delta g_{\mu\nu}(x) \delta g_{\sigma\rho}(y) \delta \phi(z) } -   \frac{2+n}{64 \pi^2} \frac{ \delta \frac{1}{4} \varepsilon^{\kappa\lambda\theta\tau} \varepsilon_{abcd} R_{\kappa\lambda}{}^{ab}(z)  R_{\theta\tau}{}^{cd}(z) }{\delta g_{\mu\nu}(x) \delta g_{\sigma\rho}(y)  } = 0 \ ,   \hspace{30mm} 
 \eea 
 which gives
 \be \big\langle g_{\mu\nu}(p_1) g_{\sigma\rho}(p_2) \phi(-p_1-p_2) \big\rangle =- \frac{2+n}{32\pi^2} \varepsilon_{\sigma)(\mu}{}^{\kappa\lambda} \varepsilon_{\nu)(\rho}{}^{\theta\tau}p_{1 \kappa} p_{2 \lambda}  p_{1 \theta} p_{2 \tau}   \ . \ee
 We note that the corresponding on-shell 3-point amplitudes vanish due to the constraints on the kinematics. However, one can get a non-trivial on-shell amplitude if one allows oneself to consider complex external momenta. As $a(x)$ is twice the canonically normalised axion, the corresponding 3-point amplitude coincides with the one obtained from the double copy formalism \cite{LanceZvi}
 \be M_{3}( |+,+\rangle_1,   |+,+\rangle_2 ,  |-,+\rangle_3 ) =   \frac{i [12]^4}{16 \pi^2 } \ , \ee 
 in the absence of vector multiplets. The anomaly implies that several (rigid) $U(1)$ violating on-shell amplitudes do not vanish, as described in detail in \cite{Carrasco:2013ypa}. It would be interesting to compare the renormalisation prescription used there with ours, which preserves the shift symmetry of the axion, at higher number of points. However, it would then be necessary to include the one-loop correction to the insertion of the non-linear transformation of the fields, so that one cannot compare directly to the effective action of \cite{Carrasco:2013ypa}.

Note that if one quantises the theory by  gauge-fixing the local $U(1)$ symmetry within the BRST formalism (rather than quantising the theory by choosing fixed coordinates on the scalar manifold), one must also take into account the $U(1)$  `anomaly' associated with the fermion zero modes \cite{diVecchia:1984jh}. This `anomaly' is not a true anomaly since it can be eliminated by introducing a non-invariant counterterm in the bare action \cite{deWit}, which reads \footnote{The coefficient is obtained by considering only the fermion contribution in (\ref{AnomalyCounting}).}
\be \Sigma^\ord{1}_{\scriptscriptstyle \rm{dWG}} = \frac{18+n}{48 \pi^2}   \int  \, \iota^*  \mbox{Im}\Scal{   \bar \cL\bigl[\mbox{ln}\scal{ U ( 1+T) }\bigr]}    \  . \label{WitGrisaru}\ee
The transformation of the function $U (1+T) $ with respect to $U(1)\times SL(2,\mathds{R})$ 
\be U (1+T)  \ \rightarrow  e^{-2i\alpha}\, U (1+T) \,  ( c \tau + d ) \ , \ee
is precisely such that its $U(1)$ transformation cancels the one-loop `anomaly', and so its $\mathfrak{sl}_2$ transformation gives the fermion field contribution to the rigid $\mathfrak{sl}_2$ anomaly in (\ref{AnomalyCounting}). Here $\alpha$ is a general real superfield whereas $c$ and $d$ are constants parametrising an $SL(2,\mathds{R})$ matrix altogether with $a$ and $b$ (with $a d -bc  =1$), so that infinitesimally $c = f + \mathcal{O}(2)$ and $d=1-h+\mathcal{O}(2)$ of (\ref{sl2hef}). However, despite the fact that $U$ is a chiral superfield, this counterterm is not supersymmetric (\ie $d \bar \cL[\mbox{ln}( U)] \ne 0$). Because of the structure of the supersymmetry algebra, there is in fact no supersymmetry invariant that is not invariant with respect to the local  $U(1)$ symmetry. Note that within the formulation in which only physical scalars are considered, the supersymmetry transformations  preserve, by construction, a chosen $U(1)$ gauge (\eg $U =\Utau$).  They are therefore defined as the $U(1)$-covariant supersymmetry transformations plus field-dependent $U(1)$ gauge transformations. Because the supersymmetry Ward identities are preserved in this formulation, they must be broken by the fermion triangle diagram in the $U(1)$-covariant formulation in which the $U(1)$ symmetry is gauge-fixed within the BRST formalism. It therefore follows that the counterterm  (\ref{WitGrisaru}) restores both the $U(1)$ gauge symmetry and supersymmetry itself in the latter formulation.\footnote{Such entanglements of gauge and supersymmetry anomalies are familiar from the anomaly analysis of Ho\v{r}ava-Witten type brane constructions \cite{Pugh:2010ii}.}

\subsection{Renormalisation of the anomaly at higher orders}

At higher orders in perturbation theory the broken Slavnov--Taylor identity is still valid, although one must consider the one-loop anomaly $\cA^\ord{1}$ as an insertion into the 1PI generating functional \cite{PS} 
\be \cS^{\scriptstyle \mathfrak{sl}_2}[\Gamma] = \frac{ 2+n}{32 \pi^2}[  \cA  \cdot \Gamma] \label{BrokenSTI}\ .   \ee
One must also consider the possibility that $\cA^\ord{1}$ gets finite corrections compatible with supersymmetry and the Wess--Zumino condition as given below in \eq{WZ},
\be \cA = \cA^\ord{1} + \sum_k a_k \kappa^{2k} \cA^\ord{1+k}  \ .  \ee
Defining the linearised Slavnov--Taylor functional operator 
\be \mathcal{S}^{\scriptstyle \mathfrak{sl}_2}_{| \Sigma} \equiv  \int d^4 x \sum_\gimel \left( \frac{ \delta^R \Sigma}{\delta \varphi^\gimel(x) }  \frac{ \delta^L \ }{\delta K_\gimel(x) } -  \frac{ \delta^R \Sigma}{\delta K_\gimel(x) }  \frac{ \delta^L \ }{\delta \varphi^\gimel (x) } \right)   - 2 h f \frac{\partial \ }{\partial f} + e f  \frac{\partial \ }{\partial  h} + 2 h e \frac{\partial  \ }{\partial  e}  \ ,  \ee
one finds that the functional identity $\mathcal{S}^{\scriptstyle \mathfrak{sl}_2}_{| \Gamma}\cS^{\scriptstyle \mathfrak{sl}_2}[\Gamma]=0$ implies the Wess--Zumino condition
\be \mathcal{S}^{\scriptstyle \mathfrak{sl}_2}_{| \Gamma}[  \cA  \cdot \Gamma]  = 0 \ . \label{WZ}\ee
Each new finite correction must be a solution to the tree-level Wess-Zumino consistency condition $\mathcal{S}^{\scriptstyle \mathfrak{sl}_2}_{| \Sigma} \cA^\ord{1+k}=0$, and if it can be reexpressed  as $\cA^\ord{1+k} = \mathcal{S}^{\scriptstyle \mathfrak{sl}_2}_{| \Sigma} \Sigma^\ord{k+1}$, it can be eliminated by a  redefinition of the bare action. Any non-trivial finite correction to the anomalous Ward identity is therefore itself associated to an algebraic anomaly. Such an anomaly can only be a supersymmetry invariant obtained from the function $f \tau - h$, or its imaginary part. Moreover such an anomaly must preserve parity (where $f$ is parity odd). 

In pure supergravity, there is no supersymmetry invariant available at two loops. At three loops, one could consider a full-superspace integral invariant $S[G(\uptau,\bar\uptau)]$ defined as a $(4,1,1)$ harmonic-superspacee integral (\ref{QuaterR4}), or the $(4,2,2)$ harmonic-superspacee integral defined from a weight four holomorphic function (\ref{Weight4H}). As we have noted, the latter cannot define a non-trivial anomaly, and one can check that the invariant $S[\mbox{Im}(\tau)]$ is parity even. We conclude that the anomalous Ward identity (\ref{BrokenSTI}) is preserved until the three-loop order in pure supergravity. Therefore the parabolic subgroup symmetry associated to the parameters $h$ and $e$ is preserved at three loops.

In the presence of vector multiplets the set of possible invariants is greatly expanded. We have not been able to determine the $SO(6,n)$ transformation property of the two-loop invariants studied in subsection \ref{Sectiond2F4}. If they proved to be $SO(6,n)$ invariants, the imaginary part of the invariant  (\ref{R2F2I}) for $\cF = f \uptau - h $ would then define a consistent correction to the anomaly. At the three-loop order one can consider the invariant (\ref{DF2R2}) with the same holomorphic function, which would then be manifestly $SO(6,n)$ invariant. Note that these potentially consistent corrections to the anomaly are all defined as chiral superspace integrals in the linearised approximation -- respectively $(4,2,0)$ for  $(\ref{R2F2I})$ and $(4,1,0)$ for $(\ref{DF2R2})$. In the presence of vector multiplets, the shift symmetry of the dilaton is already broken at the one-loop order, and these higher-order corrections will only modify the specific form of the insertion in the anomalous Ward identity.

The broken Slavnov--Taylor identity (\ref{BrokenSTI}) is a functional equation satisfied by the renormalised 1PI generating functional, and one must also take care of the renormalisation of the local operator defining the anomaly. We are going to see that the triangular matrix of anomalous dimensions of the local operator defining the anomaly is directly related to the triangular matrix of beta functions associated to non-duality invariant logarithmic divergences. If we consider the invariant $\cA$ as a local insertion, {\sl a priori} one would expect it to be renormalised by other local operators $\cA_k$ of higher dimension. Note that, while the finite invariants $\cA^\ord{k+1}$ associated to finite corrections to the broken Slavnov--Taylor identity can be removed whenever they are $\delta$-exact, the operators $\cA_k$ renormalising $\cA$ give rise to non-trivial contributions to the Callan--Symanzik equations even when they are $\delta$-exact. In fact, we shall see that they are always $\delta$-exact. In the presence of logarithmic divergences in the amplitudes, the classical action needs to be renormalised by supersymmetry invariants $S_k$. This gives rise to the functional Callan--Symanzik equation 
\be \Scal{ \mu \frac{ \partial \, }{\partial \mu } - \kappa \frac{ \partial \, }{\partial \kappa} + \sum_{k \ge 1} \beta_k \kappa^{2k-2} \frac{ \partial \; }{\partial z_k} - \sum_{k\ge 1} \gamma_k \kappa^{2k} u \frac{ \partial \; }{\partial u_k} + \dots } \Gamma = 0 \label{csym}\ee
where $u$ is the constant source for the anomaly operator $\cA$, the $u_k$ are constant sources for the operators $\cA_k$ that renormalise $\cA$, and the $z_k$ are constant sources for the invariants $S_k$. The dots in (\ref{csym}) stand for terms that vanish when the $u_k$ and the $z_k$ are set to zero, and also for BRST exact terms associated to the renormalisations of fields and of the gauge-fixing term. If some of the invariants $S_k$ associated to logarithmic divergences do not satisfy the $\mathfrak{sl}_2\mathds{R}$ Slavnov--Taylor identity, one must introduce other sources $u^\prime_k$ for their variations $\delta S_k$. The broken Slavnov--Taylor identity (\ref{BrokenSTI}) can then be recast into the functional form
\be   \cS^{\scriptstyle \mathfrak{sl}_2}[ \Gamma]  + \Scal{ \sum_{k\ge 1} z_k \frac{ \partial\, }{\partial u^\prime_k} - \frac{2+n}{32\pi^2}\frac{ \partial \; }{\partial u} } \Gamma = 0 \ .  \ee
In order to be compatible, these two functional operators must commute:
\be \Scal{ \beta_1 \frac{ \partial \, }{\partial u^\prime_1} + \sum_{k\ge 1} \kappa^{2k} \Scal{ \beta_{k+1} \frac{ \partial \, }{\partial u^\prime_{k+1}}  - \frac{2+n}{32\pi^2} \gamma_k \frac{ \partial \; }{\partial u_k} } + \dots } \Gamma = 0\ ,  \ee
where the dots stand for terms that cannot mix with the ones explicitly written. We directly deduce that a one-loop logarithmic divergence must necessarily be associated with a duality-invariant counterterm, as expected. In general, the variation of an operator may acquire finite corrections in $\kappa^2$, but one can always normalise the sources such that
\be \frac{ \partial \Gamma}{\partial u_k} = \frac{ \partial \Gamma}{\partial u^\prime_{k+1} }  + \sum_{l=1}^{k} b_{k,l} \beta_{k+1-l}  \kappa^{2l} \frac{ \partial \Gamma}{\partial u^\prime_{k+1-l}}    \ , \ee
for some coefficients $b_{k,l}$. If the source $z_{k+1}$ is associated to a first logarithmic divergence ($\beta_l = 0$, $\forall l \le k$), then $u^\prime_{k+1}= u_k$;  moreover,  the $\mathfrak{sl}_2\mathds{R}$ variation of a local operator associated to a logarithmic divergence is directly related to a logarithmic divergence of the local operator $\cA$ considered as an insertion.

\subsection{Recovering $SL(2,\mathds{Z})$ symmetry}
\label{SL2Z} 
Composing an infinitesimal transformation with a finite $SL(2,\mathds{R})$ transformation
\be g \equiv  \left( \begin{array}{cc} a & \ \ b \\ c &\  \ d \end{array} \right)   \in SL(2,\mathds{R}) \ ,\ee  
one obtains that 
\be \rho_{(1 + \delta g ) g}  \Gamma = \rho_g \Gamma  + \frac{2+n}{16\pi^2}  \int  \, \iota^* \mbox{Im} \Scal{ \bar \cL\Bigl[ f \frac{a \uptau +b}{c \uptau +d }  -h\Bigr]}  \ , \ee
where $\delta g = \left( \begin{array}{cc} h & \ \ e \\ f &\  -h \end{array} \right)$.  Integrating this equation, one computes that 
\be \rho_g \Gamma = \Gamma  +  \frac{2+n}{16\pi^2}  \int  \, \iota^* \mbox{Im} \Scal{ \bar \cL\bigl[ \mbox{ln}( c \uptau + d )\bigr]}  \ . \ee
The question is then, can one find a holomorphic function such that one can reabsorb this anomalous transformation just for elements of the modular group $SL(2,\mathds{Z} )$? The answer to that question is well-known to string theorists: one must consider the corrected effective action 
\be \Gamma^{\scriptscriptstyle {\rm S}} = \Gamma -   \frac{2+n}{8\pi^2}  \int  \, \iota^* \mbox{Im} \Scal{ \bar \cL\bigl[ \mbox{ln}( \eta(\uptau) )\bigr]} \ , \label{ModularCT} \ee
where $\eta$ is the Dedekind eta function. Then the transformation of the effective action with respect to $SL(2,\mathds{Z} )$ is 
\bea \rho_g \Gamma^{\scriptscriptstyle {\rm S}}  &=& \Gamma^{\scriptscriptstyle {\rm S}}  -  \frac{2+n}{8\pi^2} \frac{\pi \tilde{b}}{12}  \int  \, \iota^* \mbox{Re} \Scal{ \bar \cL[1]} \CR
&=& \Gamma^{\scriptscriptstyle {\rm S}}  +2\pi    \frac{2+n}{2} \tilde{b} \frac{-1}{24} \int \frac{1}{8\pi^2 } R^{ab}\wedge R_{ab} \CR
&=& \Gamma^{\scriptscriptstyle {\rm S}}  + 2\pi  ( 2 + n ) \tilde{b} \frac{\hat{A}}{2} \ ,\eea
where $\tilde{b} = b$ if $c=0$ and is an integer for any $SL(2,\mathds{Z})$ element. By Rokhlin's theorem, $\hat{A}$ is an even number on any smooth manifold admitting a spin structure, and the theory is therefore $SL(2,\mathds{Z})$ invariant for any number of vector multiplets. This is in perfect agreement with the topological string amplitude computation carried out in \cite{Harvey:1996ir} in type IIA compactified on $K3 \times T^2$ for $n=22$.

One computes in a specific $U(1)$ gauge that
\bea U^{-2} \partial \mbox{ln}\left( \eta\Bigl( i \frac{1 -T }{1 + T }  \Bigr)\right)  &=& - \frac{\pi}{6} \mbox{Im}[\uptau] \, \hat{E}_2(\uptau,\bar\uptau) - \frac{1}{2} \CR
U^{-4} \Scal{ \partial^2 - \frac{2 \bar{T}}{1-T \bar T} \partial } \mbox{ln} \left( \eta\Bigl( i \frac{1 -T }{1 + T }  \Bigr)\right) &=& \frac{\pi^2}{18} \mbox{Im}[\uptau]^2 \, \scal{ \hat{E}_2^{\; 2}(\uptau,\bar\uptau)-E_4(\uptau)} - \frac{1}{2} \ , \eea 
where 
\be \hat{E}_2(\tau,\bar \tau) = E_2(\tau)  - \frac{3}{\pi \mbox{Im}[\tau]} \ , \ee
is modular of weight 2, as opposed to the Eisenstein series $E_2$ itself. By construction, the $w$th power of the K\"{a}hler derivative of the holomorphic function $\mbox{ln}\scal{ \eta(\tau)} $ gives a modular function of weight $w$, minus $\frac{(w-1)!}{2}$. The additional $\frac{(w-1)!}{2}$ constant breaks modular invariance exactly in the right way, such that the 1-loop integration of the massless modes compensates for it. The overall string-theory amplitude including the non-analytic part is $SL(2,\mathds{Z})$ invariant and will not include such a term. For instance, the real part of  $\mbox{ln}\scal{ \eta(\tau)} $ gets a non-holomorphic correction in the amplitude 
\be \mbox{ln}\scal{ \eta(\tau)}  + \mbox{ln}\scal{ \eta(\bar \tau)} \rightarrow  \frac{1}{2}  \mbox{ln}\scal{  \mbox{Im}(\tau) |\eta(\tau)|^4}\ee
and the $\frac{(w-1)!}{2}$ factor corresponds precisely to the $w$-th power of the covariant K\"{a}hler derivative acting on the non-holomorphic term 
\be \prod_{k=0}^{w-1} \scal{ ( \tau - \bar \tau ) \partial + k }  \frac{1}{2}  \mbox{ln}\scal{ \mbox{Im}(\tau)}  = \frac{(w-1)!}{2} \ . \ee
For example, the four-graviphoton amplitude will admit a threshold function $\frac{2+n}{144}  \scal{ \hat{E}_2-E_4} $, multiplying $U_{ij}{}^I F_{I} + V_{ij \hat{I}} F^{\hat{I}} - \tau \star \scal{U_{ij}{}^I F_{I} + V_{ij \hat{I}} F^{\hat{I}} }$ to the fourth, such that it is $SL(2,\mathds{Z})$ invariant. 

The non-linear analysis of the invariant also reveals that the matter four-photon amplitude is determined up to its $\tau$ independent part as $ (2+n) \mbox{ln}( \mbox{Im}(\tau) |\eta(\tau)|^4)$. It follows that the $F^4$ threshold function in heterotic string theory compactified on $T^6$ (for which $n=22$) is the sum of the one-loop contribution, which only depends on the vector-multiplet scalars, and the non-perturbative correction associated to NS 5-branes wrapping $T^6$, which is identical to the $R^2$ threshold function \cite{Harvey:1996ir}. The invariant expressing the former coupling is the $(4,2,2)$ harmonic-superspacee integral of a homogenous function of degree four in the G-analytic superfields $U_{34}{}^I,V_{34}{}^{\hat{I}}$, while that for the latter is the $F^4$ component of the invariant appearing as a counterterm in (\ref{ModularCT}). In more general heterotic string-theory compactifications, the $F^4$ coupling will still decompose in the same way (although the quasi-modular function will be different), such that the quasi-modular function of $\tau$ is determined by supersymmetry from the exact $R^2$ correction \cite{Gregori:1997hi}. This is consistent with the property that the contribution to the $F^4$ type invariant is one-loop exact in heterotic string theory in ten dimensions \cite{Yasuda:1988fi}.

The transformation properties of the one-loop contribution to the effective action with respect to $O(6,22,\mathds{Z})$ are themselves extremely constrained by supersymmetry, and one should thus be able to predict the contributions of NS 5-branes wrapping $K3\times T^2$ to the $F^4$ coupling in type IIA string theory  \cite{Kiritsis:2000zi} on the basis of supersymmetry. 

At weak coupling in heterotic string theory, \ie $\mbox{Im}(\tau) \rightarrow +  \infty$, the $R^2$ coupling gets a universal threshold function \cite{Gregori:1997hi} of the form 
\be  \frac{2+n}{32\pi^2} \Scal{  \mbox{ln}( \mbox{Im}(\tau) ) + \pi n_0     \mbox{Im}(\tau) + \mathcal{O}( e^{-{\rm Im}(\tau)}) } \ , \ee
and because of the structure of the supersymmetry invariant, the $F^4$ coupling will also possess the same logarithmic term in Einstein frame. As explained in \cite{Green:2010sp}, a logarithm of the string coupling constant cannot appear in the string amplitude in string frame, while its appearance in Einstein frame comes from a logarithm of the Mandelstam variable $s$. It follows that the matter four-photon amplitude at the one-loop level in heterotic string theory admits a logarithm of $s$ that is associated to a logarithmic divergence in field theory, as has been argued in \cite{Green:2010sp}. This conclusion is indeed consistent with the one-loop divergence computed in field theory \cite{Fischler:1979yk,Fradkin}.

In order to be able to discuss higher-order corrections in the string-theory effective action, it is important to study first of all the higher-order corrections required to promote the counterterm (\ref{ModularCT}) to a supersymmetric effective action. In order to do so using superspace techniques, it is necessary to compute the corresponding correction to the superspace geometry. To do this, one can use the off-shell version of $\cN=4$ (conformal) supergravity as given in \cite{Howe:1981gz} and then introduce field strengths for the $6+n$ vectors as well as for the scalars in the vector-multiplet sector. Solving the Bianchi identities for these superspace field strengths order by order in $\k^2$ in terms of the physical fields will then also enable one to determine the dimension-one functions  appearing in the torsion and curvature in terms of them. A similar procedure was implemented in \cite{Howe:2010nu} in the context of the supersymmetric Yang--Mills theory in ten dimensions. In that case, the authors were able to use the known $\alpha^\prime{}^2$ $F^4$ type correction to the superspace field strength \cite{Bergshoeff:1986jm} in order to compute the induced $\alpha^\prime{}^4$ $\partial^4 F^4$ type correction to the string-theory effective action, thereby reproducing the bosonic components computed in \cite{Koerber:2002zb}.  It would be particularly interesting to compute the higher-order corrections induced by the $R^2$ type correction to the effective action in  $\cN=4$ supergravity. Such an example would, in principle, allow a determination of the relation between the function of the complex scalar field $\tau$ multiplying the $R^4$ correction and the quasi-modular function $\mbox{ln}( \eta(\uptau) )$ (although in this case one would need to go two orders higher). Such an explicit example could also shed light on the conjectured behaviour of the $\partial^6 R^4$ threshold function as compared to the one multiplying $R^4$ in maximal supergravity \cite{Green:2005ba}.

\section{Implications for perturbative quantum field theory}

\subsection{Consequences of the anomaly}
\label{ImplicationUV}
Let us first consider pure $\cN=4$ supergravity. In this case, we know that there is no on-shell duality-invariant candidate counterterm before the three-loop order, and there is no on-shell supersymmetry invariant candidate at the two-loop order. At the three-loop order, one finds two classes of invariants: the $(4,1,1)$ integral invariants  (\ref{R4Pure}) depending on a general real function $G^\ord{3}(T,\bar T)$, and the $(4,2,2)$ integral invariants (\ref{Weight4H}) depending on a general weight-four holomorphic function $\cF^\ord{3}(\bar T)$. The theory is therefore clearly finite up to the three-loop order, and the Callan--Symanzik equation states that 
\be \Scal{ \mu \frac{ \partial \, }{\partial \mu }  -  \kappa \frac{ \partial \, }{\partial \kappa}}  \Gamma \approx -  \beta_3 \kappa^4 \bigl[ S[G^\ord{3},\cF^\ord{3}] \cdot \Gamma \bigr]   + \cO(\kappa^6) \ee
for an invariant defined as 
\be S[G^\ord{3},\cF^\ord{3}] = \frac{1}{4} \int d\mu_{\scriptscriptstyle (4,1,1)} (\chi^1\chi^1) (\bar\chi_4\bar\chi_4) \, G^\ord{3}(T,\bar T) + \mbox{Re}\Bigl[ \int d\mu_{\scriptscriptstyle (4,2,2)}  \cL_{\scriptscriptstyle (4,2,2)}^\ord{4}[ \cF^\ord{3}]  \Bigr] \ .  \ee  
Here $\approx$ means equal in BRST cohomology, \ie up to terms associated to field redefinitions or the renormalisation of the gauge-fixing action. Compatibility with the broken Slavnov--Taylor identity then requires  
\be  \Scal{  \mu \frac{ \partial \, }{\partial \mu }  -  \kappa \frac{ \partial \, }{\partial \kappa}}  [ \cA \cdot \Gamma ] \approx  - \gamma_2 \kappa^4 [ \cA_2 \cdot \Gamma]   + \cO(\kappa^6)\approx - 16 \pi^2 \beta_3 \kappa^4  \cS^{\scriptstyle \mathfrak{sl}_2}_{|\Gamma}  \bigl[ S[G^\ord{3},\cF^\ord{3}] \cdot \Gamma \bigr]    + \cO(\kappa^6) \ee
or, equivalently, 
\be  \gamma_2  \cA_2 =  16 \pi^2  \beta_3 \, \delta S[ G^\ord{3},\cF^\ord{3}] \ . \ee
Because the anomaly $\cA$ only depends non-trivially on $f$ at this order, it follows that $S[G^\ord{3},\cF^\ord{3}]$ must still be invariant with respect to the parabolic subgroup associated to the parameters $h$ and $e$. Because $\delta S[G^\ord{3},\cF^\ord{3}] = S[\delta G^\ord{3},\delta\cF^\ord{3}]$ and  $S[\delta G^\ord{3},\delta\cF^\ord{3}]$ is non-zero for any non-zero function $\delta G^\ord{3}$ or $\delta \cF^\ord{3}$, both functions $G^\ord{3}(\uptau, \bar \uptau)$ and $\cF^\ord{3}(T)$ must themselves be invariant with respect to the action of the parabolic subgroup, \ie 
\be  ( \partial + \bar \partial ) G^\ord{3}(\uptau ,\bar \uptau) = 0 \ , \qquad ( \uptau \partial + \bar \uptau \bar \partial )   G^\ord{3}(\uptau ,\bar \uptau)= 0\ ,   \ee
and
\be \partial \cF^\ord{3}(\uptau)  = 0 \ , \qquad ( \uptau \partial + 2 ) \cF^\ord{3}(\uptau) = 0   \ . \ee
One can easily check that the only solution, up to a rescaling, is $G^\ord{3}=1$ and $\cF^\ord{3}=0$.  Thus, at this order, shift and scaling invariance together with local supersymmetry are enough to require that a counterterm be duality invariant,\footnote{Note that in components, $\tau = a + i e^{-2\phi}$, these equations just imply  that the counterterm can only depend on scalars through contractions with the vector fields and $e^{2\phi} \partial_\mu a$ and $\partial_\mu \phi$ which are not necessarily duality invariant.} and  we therefore conclude that the anomaly operator cannot get renormalised at the two-loop order, and that the only available counterterm consistent with all required symmetries at this order is the $R^4$ type duality invariant 
\be S_3 = \frac{1}{4} \int d\mu_{\scriptscriptstyle (4,1,1)} \varepsilon^{\alpha\beta} \varepsilon^{\dot{\alpha} \dot{\beta}}  \chi_\alpha^1 \chi_\beta^1  \chi_{\dot{\alpha}4} \chi_{\dot{\beta}4} \ . \ee
This invariant can also be written as the full-superspace integral of the K\"{a}hler potential, as we have seen in the last section.

For $\cN=4$ supergravity coupled to $n$ vector multiplets the discussion becomes more complicated because the theory is known to diverge already at the 1-loop order \cite{Fischler:1979yk,Fradkin},
\be \Scal{  \mu \frac{ \partial \, }{\partial \mu }  -  \kappa \frac{ \partial \, }{\partial \kappa}}  \Gamma \approx - \frac{n+2}{ 32^2  \pi^2 }  [ S^\ord{1}  \cdot \Gamma ]   + \mathcal{O}(\kappa^2) \ , \ee
where $S^\ord{1}$ is the duality invariant (\ref{HalfHarmo}). It is striking that the coefficient of the divergence is the same as that of the anomaly, and, moreover, the duality invariant associated to it is precisely the one defining the anomaly in the shift symmetry of the dilaton. This suggests that one can interpret the latter as being an order $\epsilon$ breaking of this symmetry in a hypothetical supersymmetry invariant regularisation scheme. This is also in agreement with the string-theory interpretation discussed in Section \ref{SL2Z}. 

A consequence of the anomaly is that the candidate counterterm for a three-loop divergence  in the presence of vector-multiplet coupling is not necessarily duality-invariant, because its duality variation is associated to a divergence of the $F^4$ type invariant $S^\ord{1}$ taken as an insertion. The $F^4$ invariant is itself duality invariant, and the operators that can renormalise it at the two-loop level must also preserve duality invariance, despite the appearance of the anomaly at one loop. There are only two such invariants available: the full-superspace integral of the K\"{a}hler potential, and the $\partial^4 F^4$ type $(4,1,1)$ harmonic integral (\ref{d4F4411}). The latter does not generalise, however, to an invariant defined in terms of a generic function of the complex scalar, but only to the class of invariants (\ref{DF2R2}) defined in terms of a holomorphic function $\cF(\bar T)$. Such a two-loop divergence in the insertion of $F^4$ would imply the appearance of a three-loop divergence  in an invariant (\ref{DF2R2}) for $\cF(\bar T) \sim \mbox{ln}\scal{ \frac{1-\bar T}{1+\bar T}}$ which is inconsistent with the shift symmetry of the axion because 
\be \delta  \mbox{ln}( \tau / i  ) = 2 h + \frac{e}{\tau} - f \tau \ . \ee
It follows that the only candidate counterterms consistent with the anomalous $\mathfrak{sl}_2$ Ward identities at the three-loop level are the full-superspace integral of a function
\be H^\ord{3}(\uptau,\bar \uptau)  = - c_\un\,  \mbox{ln}( \mbox{Im}(\uptau) )  + \tfrac{1}{2} c_\deux \, \mbox{ln}^2( \mbox{Im}(\uptau) ) \ , \ee
and the duality-invariant $(4,1,1)$ harmonic integral (\ref{d4F4411}). The coefficient $c_\un$ multiplies the duality-invariant counterterm written as (\ref{QuaterR4}) for $J=1$, whereas $c_\deux$ multiples the counterterm that would be associated to a 2-loop divergence of the $F^4$ type invariant $S^\ord{1}$ taken as an insertion into the duality-invariant counterterm (\ref{QuaterR4}) for $J=1$. The corresponding $R^4$ coupling is
\be - ( 2 c_\un + c_\deux ) + 2 c_\deux   \mbox{ln}( \mbox{Im}(\uptau) ) \ . \ee
Following the string-theory amplitude argument of \cite{Green:2010sp}, a logarithmic divergence in $ \mbox{ln}( \mbox{Im}(\uptau) )$ must be associated to a duality-invariant logarithm-squared divergence. This is perfectly consistent with the present field-theory interpretation that such a term could only appear if the $F^4$ supersymmetric form-factor with four gravitons is divergent at the two-loop order, because one would  then expect a $\mbox{ln}^2(s)$ behaviour in the three-loop four-graviton amplitude originating from the $\mbox{ln}(s)$ behaviour of the four-photon one-loop amplitude. 

\subsection{Superspace non-renormalisation theorems}

Now we return to the structural requirement for ultraviolet counterterms in an $\cN=4$ duality-invariant supergravity theory. Superspace non-renormalisation theorems in super Yang--Mills and supergravity theories are based on a set of rules that permit one to determine the allowed ultraviolet divergences. In order to obtain the strongest results, it is important to know the maximal degree of supersymmetry that can be realised linearly ``off-shell'' (\ie without use of the classical equations of motion to achieve closure of the supersymmetry algebra). The precise answer to this off-shell question for $D=4$, $\cN=4$ supergravity is unknown, but one may draw hints from  $D=10$, $\cN=1$ supergravity, which does have an off-shell formulation at the linearised level \cite{Howe:1982mt}. Dimensionally reducing this off-shell theory to $D=4$ yields a system containing $\cN=4$ supergravity coupled to six $\cN=4$ Maxwell multiplets. 

Decoupling the Maxwell multiplets without destroying the off-shell structure is of course a problem, and there are arguments \cite{Siegel:1981dx, Rivelles:1982gn} against the existence of off-shell multiplets for numbers of $\cN=4$ Maxwell multiplets different from $6$ modulo $8$ with finite numbers of auxiliary fields. The dimensions of the spinor representations required to write the kinetic term of the physical fermions must be  integral multiples of one-half the dimension of the smallest irreducible representation of the $\cN=4$ super Poincar\'e group \cite{Rivelles:1982gn}, \ie  $128$. This implies that only $\cN=4$ supergravity coupled to 6 modulo 8 vector multiplets can possibly admit an off-shell realisation involving finitely many auxiliary fields. However, there is no argument against the existence of an off-shell formulation of $\cN=4$ supergravity coupled to any number $n$ of vector multiplets in harmonic superspace. For the present discussion, we will simply assume that an off-shell harmonic-superspacee formulation exists for the various couplings of $\cN=4$ supergravity to even numbers of $\cN=4$ Maxwell multiplets.
It has so far proved impossible to find an off-shell version of $\cN=4$ SYM \cite{SokZup}, but one aspect of the problem, the self-duality constraint on the scalars, can presumably be avoided if we have an even number of vector multiplets.

Another element in the derivation of non-renormalisation theorems is the use of the background-field method (a review is given in \cite{Howe:1988qz}). The geometric formulations of super Yang--Mills and supergravity theories in superspace begin with  superspace gauge connections $A_M(x,\theta)$ and superspace vielbeins $E_M{}^A(x,\theta)$. In an off-shell theory, however, these are not the entities in terms of which one quantises, as they are subjected to various types of constraints \cite{Gates:1979wg, Stelle:1980uw} that are needed to allow for the construction of an off-shell superspace action. These constraints on the geometrical superfields can be solved in terms of superspace prepotentials analogous to the general scalar prepotential of $D=4$, $\cN=1$ super Maxwell theory, or to the vector superfield prepotential of $D=4$, $\cN=1$ supergravity \cite{Siegel:1978mj}.

The key structural feature shown by use of the background field method is that although one definitely needs to solve the superspace constraints in terms of prepotentials for the {\em quantum} superfields on the internal lines of superspace Feynman diagrams, this is not necessary for the {\em background} fields that occur on external Feynman diagram lines \cite{Grisaru:1981xm,Grisaru:1982zh}. Moreover, when the quantum fields are expressed in terms of prepotentials, all terms in the expansion of the quantum action that are actually used in the derivation of superspace Feynman rules are written as full-superspace integrals. This obtains even though the background fields occur only through the geometrical superspace connections and supervielbeins.

The consequences for the ultraviolet divergence structure of the theory are then immediate. Counterterms must be written purely in terms of the background superspace connections and supervielbeins, and they must also be written as full-superspace integrals. The simplest instance of such a non-renormalisation theorem is in the non-gauge context of the $D=4$, $\cN=1$ Wess--Zumino model, where chiral superspace integrals over superpotentials are disallowed as counterterms: all counterterms must be writable in terms of full-superspace integrals without the use of prepotentials.

There are exceptions to the above rule at the one-loop level in gauge and supergravity theories, owing to problems in decoupling ghost superfields \cite{Grisaru:1981xm,Grisaru:1982zh,Howe:1983sr}. At the one-loop level, and only at this level, it turns out to be necessary to introduce prepotentials also for the background fields in order to achieve a decoupling of the infinite series of ``ghosts for ghosts'' occurring in extended gauge and supergravity theories. This makes for a one-loop exception to the non-renormalisation theorem, but for two loops and higher, the theorem becomes fully valid.

In the present discussion of divergences in half-maximal supergravity theories we shall suppose that there exists a full sixteen-supercharge off-shell formulation such that the action defining the Feynman rules is also invariant with respect to duality symmetry. This second requirement is necessary because we shall assume that the counterterms must be both writable as full-superspace integrals of the covariant superfields and also duality-invariant. Moreover we shall assume that this off-shell formulation satisfies the property that one can decouple the infinite tower of ghosts-for-ghosts by introducing a quadratic action for the second generation of ghosts and the Nielsen--Kallosh ghost, depending on the background prepotentials only, and not on the quantum prepotentials \cite{Grisaru:1981xm,Grisaru:1982zh,Howe:1983sr}.

\subsection{Descent equations for co-forms}
We shall consider the classical Lagrangian density in superspace as a local operator. Because it is a density, its variation with respect to an infinitesimal  super-diffeomorphism generated by a vector $\Xi^M$ is
\be s \cL^\ord{0} = \partial_M \Scal{ (-1)^M  \Xi^M \cL^\ord{0} } \ . \ee
If we define the action of the BRST operator in a similar way, but taking $\Xi^M$ as a super-diffeomorphism ghost with its own transformation, 
\be s \Xi^M =  \Xi^N \partial_N \Xi^M  \ , \ee
we obtain 
\bea s  \Scal{ (-1)^M  \Xi^M \cL^\ord{0} } &=& \partial_N \Scal{ (-1)^{M+N} \Xi^N \Xi^M \cL^\ord{0}} \ , \CR
s  \Scal{ (-1)^{M+N} \Xi^N \Xi^M \cL^\ord{0}} &=& \partial_P \Scal{ (-1)^{M+N+P} \Xi^P \Xi^N \Xi^M \cL^\ord{0}} \ , \CR
\dots && 
\eea
and so on indefinitely. This is the superspace generalisation of the BRST construction defined in \cite{Baulieu:1985gy}. Therefore if one wants to consider the Lagrange density as a local operator in a BRST invariant way, one must consider the infinite series of local operators with their associated sources
\be \int d^4x d^{16} \theta \Scal{  \cL^\ord{0} u +   \Xi^M \cL^\ord{0} u_M  + \tfrac{1}{2}    \Xi^N \Xi^M \cL^\ord{0} u_{MN} + \dots } \ee
such that these sources transform as an extended cocycle, \ie 
\bea s u &=& 0 \CR
s u_M &=& - \partial_M u \CR
s u_{MN} &=& - \partial_M u_N + (-1)^{MN} \partial_N u_M \CR
\dots &&
\eea
or,  equivalently, 
\be ( d + s ) \Scal{ u + dz^M u_M + \frac{1}{2} dz^N \wedge dz^M u_{MN} + \frac{1}{6}dz^P \wedge  dz^N \wedge dz^M u_{MNP} + \dots } = 0 \ . \ee  
This construction is consistent because the transformations of the sources are linear.

The complete BRST invariance of such a theory would also involve other ghost superfields associated to the other gauge invariances of the theory, as well as pre-gauge invariances associated to the introduction of prepotentials for the corresponding superfields $E_M{}^A,\, T,\, V_{ij}{}^{\hat{I}}$, etc... Without an explicit fully supersymmetric formulation to hand, it is not possible to go into more detail about these other symmetries.

One will be led to the same kind of construction when considering the duality invariance of a Lagrange density, because the latter is not itself strictly invariant with respect to duality transformations either. We will illustrate this shortly with the example of the K\"{a}hler gauge invariance of an $\cN=(2,2), D=2$ supersymmetric non-linear sigma model, for which the superspace action's integrand (the K\"{a}hler potential) is not itself ``duality'' (\ie K\"ahler gauge) invariant.  

In principle, one should consider descent equations in two directions: with operators of increasing ghost number and also with increasing numbers of anticommuting constant parameters for the duality transformations. For simplicity, we will only consider the duality Ward identities, and will disregard the supersymmetry BRST Ward identities, which we do not know in detail for the $\cN=4$ theory. In a similar way, one will have a chain of co-form operators associated to the duality variation of the Lagrange density 
\bea \delta \cL^\ord{0} &=& (-1)^M \partial_M   \cL^{\ord{0} \, M} \ , \CR
 \delta \cL^{\ord{0} \, M}  &=&(-1)^{N}  \partial_N   \cL^{\ord{0} \, NM} \ , \CR
 \delta \cL^{\ord{0} \, NM}  &=&(-1)^P  \partial_P   \cL^{\ord{0} \, PNM} \ , \CR
  \dots &&\label{DualityDescent} \eea
where we define a co-form of degree $n$ as an object transforming as the tensor product of a density with the graded antisymmetric tensor product of $n$ vectors. Of course, such an object would be equivalent to a $(d-n)$-form on a Riemannian $d$-dimensional manifold via contraction with the Levi--Civita tensor, but in superspace they are distinct objects. Note that a co-form or a form can admit an arbitrary high degree in superspace, and there is correspondingly no notion of a top form. 

Note that if some of the quantum fields satisfy constraints, such as chiral superfields in $\cN=1$ theories, it may happen that the descent becomes more complicated, because one may then get more solutions to the consistency conditions. Here we shall assume that the theory is quantised in terms of unconstrained superfields for simplicity. However, because our argument for a non-renormalisation theorem will only involve the first term in the descent, this assumption will not have untoward consequences.

Exactly as one does in enforcing supersymmetry BRST invariance, one can ensure duality invariance by introducing a source for each of these co-forms,
\be \int d^4x d^{16} \theta  \Scal{  \cL^\ord{0} u + \cL^{\ord{0} \, M}  u_M + \tfrac{1}{2} \cL^{\ord{0} \, NM} u_{MN}   + \dots }\ , \ee
such that the sources transform with respect to duality as an extended cocycle 
\be ( d + \delta  ) \Scal{ u + dz^M u_M + \frac{1}{2} dz^N \wedge dz^M u_{MN} + \frac{1}{6}dz^P \wedge  dz^N \wedge dz^M u_{MNP} + \dots } = 0 \ . \ee  
The extended cocycle is a cohomology class of the extended exterior derivative $d + \delta $, so one can consider the chain of co-forms as defining a cohomology class. Since a density Lagrangian $\cL$ does not depend on the anticommuting duality parameters, it accordingly cannot be $\delta$-exact. However, a Lagrange density is only defined up to a total divergence 
\be \cL \approx \cL + (-1)^M \partial_M \uppsi^M \ . \ee
Similarly the whole chain is only defined up to an ambiguous exact extended co-form
\bea \cL^M &\approx&  \delta \uppsi^M + (-1)^N \partial_N \uppsi^{NM} \CR
 \cL^{NM} &\approx&  \delta \uppsi^{NM} +(-1)^P \partial_P \uppsi^{PNM} \CR 
 \cL^{PNM} &\approx&  \delta \uppsi^{PNM} + (-1)^Q \partial_Q \uppsi^{QPNM} \CR 
\dots &&
\eea
As in the component formulation, the modification of the extended co-form by the addition of a trivial extended co-form amounts to a $\delta$-exact modification of the source term 
\be \delta \int  d^4x d^{16} \theta \Scal{   \uppsi^{ M} u_M + \tfrac{1}{2}    \uppsi^{NM}  u_{MN}+ \tfrac{1}{6}   \uppsi^{PNM} u_{MNP}  \dots } \ee
and is trivial in $\delta$-cohomology. It follows that, as long as the duality Ward identities are satisfied, the whole chain of co-forms must be renormalised consistently as a single cohomology class.

This construction is a superspace generalisation of the one developed in \cite{PS,Sorella1,Sorella2} in the framework of algebraic renormalisation. One must in principle consider the possibility that the associated Ward identity may be anomalous. As in the former construction, such a Ward identity is only anomalous if the original symmetry is. In our case, this means that the only potential anomaly to this Ward identity is associated to the one-loop anomaly (\ref{BrokenSTI}). A potential anomaly would define a chain of co-forms whose first component would define the density for an anomaly, as for example (\ref{AnomalyInvariant})
\be \cA^\ord{1} = \int d^4 x d^{16} \theta  \cW \ , \ee
such that 
\bea \delta \cW &=&(-1)^M  \partial_M   \cW^{M} \ , \CR
 \delta\cW^{ M}  &=&(-1)^N  \partial_N  \cW^{NM} \ , \CR
 \delta\cW^{ NM}  &=& (-1)^P \partial_P  \cW^{PNM} \ , \CR
  \dots &&\label{AnomalyDescent} \eea
If the associated density turned out to vanish, it would  imply that the lowest-degree non-vanishing co-form would be divergence-free. Within an off-shell formulation, there is no such non-trivial divergence-free $k$ co-form available that would not itself be the divergence of a $k+1$ co-form. We conclude that the only anomaly that can occur for this Ward identity is the one associated with the anomaly (\ref{BrokenSTI}). Therefore it will only affect the non-linearly realised generator of $\mathfrak{sl}_2\mathds{R}$, and the Ward identity will remain valid for the parabolic subgroup in the absence of vector multiplets. 

\subsection{The $\cN=(2,2)$ non-linear sigma model}

To illustrate what one can learn from such Ward identities, let us revisit the example of an $\cN=(2,2)$ non-linear sigma model in two dimensions \cite{{Howe:1986ys}}. In this case, the action is obtained from a K\"{a}hler potential $K(T,\bar T)$ as the superspace integral
\be S =  \int d^2 x d^4 \theta K(T,\bar T) \ . \ee
By power counting, the theory can only be renormalised through a modification of the K\"{a}hler potential at each order in perturbation theory. Accordingly, one can define the beta function as a function of the complex variables $t^a$ parametrising the K\"{a}hler space, \ie 
\be \mu \frac{ d K(t,\bar t) }{d\mu} = \beta(t,\bar t) \ . \ee
We stress that $t^a$ are coordinates and not fields or superfields, and we just use them for parametrising functions (or potentials) defined on the corresponding K\"{a}hler space. 

Using the formal path integral formulation 
\be \exp\bigl[iW[J,\bar J]\bigr] = \int \cD T\cD\bar T \exp\Bigl[ i  \int d^2 x d^4 \theta K(T,\bar T) + i \int  d^2 x d^2 \theta  J_a T^a + i \int  d^2 x d^2 \bar \theta  \bar J_{\bar a} \bar T^{\bar a}  \Bigr] \ , \ee
one deduces that 
\begin{multline}  \int d^n t d^n \bar t \; F(t,\bar t) \frac{ \delta \,}{\delta K(t,\bar t)}   \exp\bigl[iW[J,\bar J]\bigr] \\ =  \int \cD T\cD\bar T  \Scal{ i \int d^2 x d^4 \theta  F(T,\bar T)} \exp\Bigl[ i  \int d^2 x d^4 \theta K(T,\bar T) + i \int  d^2 x d^2 \theta  J_a T^a + i \int  d^2 x d^2 \bar \theta  \bar J_{\bar a} \bar T^{\bar a}  \Bigr] \ . \end{multline}
This equation formally leads to the identity
\be \int d^n t d^n \bar t \; F(t,\bar t) \frac{ \delta \,}{\delta K(t,\bar t)} \Gamma =  \int d^2 x d^4 \theta  \Bigl[    F[T(x,\theta),\bar T(x,\theta)]  \cdot \Gamma \Bigr] \ , \label{ExactScheme} \ee
where the right-hand-side is the insertion of $F(T,\bar T)$ as a local operator in the 1PI generating functional $\Gamma$. In general, this equation can acquire corrections, and the functions $F$ on both sides of the identity could differ by higher-order corrections in the coupling constants. Nevertheless, we shall assume that it is satisfied in some appropriate renormalisation scheme, for the sake of simplicity. The consistency of this identity with the Callan--Symanzik equation implies
\be \Bigl[\mu  \frac{ d \, }{d\mu} , \int  d^n t d^n \bar t \; K(t,\bar t) \frac{ \delta \,}{\delta K(t,\bar t)} \Bigr] \Gamma = \mu  \frac{ d \, }{d\mu} \;  \int  d^n t d^n \bar t \; K(t,\bar t) \frac{ \delta \,}{\delta K(t,\bar t)}  \Gamma \ . \ee
One then obtains using (\ref{ExactScheme}) the identity 
\be    -  \int d^n w  d^n \bar w \; K(w,\bar w) \frac{ \delta  \beta(t,\bar t)}{\delta K(w,\bar w)} = \gamma(t,\bar t) \ , \label{BetaGamma} \ee
where $\gamma(t,\bar t)$ is the anomalous dimension of the local operator $K(t,\bar t)$ 
\be \mu \frac{ d\ }{d\mu} \Bigl[ K(T,\bar T) \cdot \Gamma \Bigr] = \Bigl[ \gamma(T,\bar T) \cdot \Gamma \Bigr] \ . \ee
Equation (\ref{BetaGamma}) implies that the beta function is directly related to the anomalous dimension of the Lagrange density considered as a local insertion. Using the homogeneity of the Feynman rules in the K\"{a}hler potential at $
\ell$ loops, one concludes that the $\ell$-loop beta function $\beta^\ord{\ell}$ is related to the $\ell$-loop gamma function $\gamma^\ord{\ell}$ as
\be ( \ell-1) \beta^\ord{\ell}(t,\bar t) = \gamma^\ord{\ell}(t,\bar t)  \ . \ee 
This equation is very useful, because it relates the beta function to the anomalous dimension of a local operator, which is itself constrained by the Ward identities to be renormalised consistently as a whole cohomology class.

Within the background-field method, one defines a prepotential $V^a$ for the quantum fluctuations of $T^a$ as
\be T^a = \bar D^\alpha \bar D_\alpha X^a[V,T,\bar T] \ , \ee
such that $X^a=X^a(1)$ for the solution $X^a(s)$ to the equation \cite{Howe:1986ys}
\be \frac{ d^2 X^a}{ds^2 } + \Gamma^a_{bc}(T,\bar T) \frac{ d X^b}{ds } \frac{ d T^c}{ds } = - \frac{1}{3} R^a{}_{bc\bar d}(T,\bar T) \frac{ d X^b}{ds } \frac{ d T^c}{ds } \frac{ d\bar  T^{\bar d}}{ds } \ , \ee
where $T^a(s) =  \bar D^\alpha \bar D_\alpha X^a(s) $, taken with the initial conditions 
\be X^a(0) = X^a_{\scriptscriptstyle \rm B}  \ , \qquad \frac{ d X^a}{ds }(0) = V^a \ . \ee
One then computes 
\be X^a[V,T,\bar T] = X^a_{\scriptscriptstyle \rm B}  + V^a  - \frac{1}{2} \Gamma^a_{bc}(T,\bar T)  V^b \bar D^\alpha \bar D_\alpha V^c + \cO(V^3) \label{Xexpension}  \ , \ee
and the action reduces to (with $T_{\scriptscriptstyle \rm B} =  \bar D^\alpha \bar D_\alpha X_{\scriptscriptstyle \rm B} $) \cite{Howe:1986ys}
\begin{multline} S =  \int d^2 x d^4 \theta \Bigl(  K(T_{\scriptscriptstyle \rm B},\bar T_{\scriptscriptstyle \rm B}) + g_{a\bar b}(T_{\scriptscriptstyle \rm B},\bar T_{\scriptscriptstyle \rm B}) \bar D^2 V^a D^2 \bar V^{\bar b} \Bigr . \\ \Bigl . + \tfrac{1}{6} R_{a\bar b c\bar d}(T_{\scriptscriptstyle \rm B},\bar T_{\scriptscriptstyle \rm B}) \bar D^2 V^a \bar D^2 V^c D^2 \bar V^{\bar b}D^2 \bar V^{\bar d} + \cO(D_\alpha  T_{\scriptscriptstyle \rm B} ) \Bigr)   \ . \label{ActionBF} \end{multline}
We see that the action does not depend explicitly on the background prepotentials $X^a_{\scriptscriptstyle \rm B}$, and that although the latter will be involved explicitly in the gauge-fixing at the one-loop order, the involvement of the background prepotentials will not extend to higher loop orders.

Now let us assume that the K\"{a}hler space of the theory admits some isometries, such that 
\be \delta K(T,\bar T) = c^i F_i(T) + c^i F_i(\bar T) \ee
for some variation 
\be \delta T^a = c^i f^a_i(T) \, , \ee
of the fields. The variation of the prepotential $X^a$ is obtained accordingly using functions $f^a_{ib}(T)$ satisfying 
\be f^a_{ib}(t) t^b = f^a_i(t) \ , \ee
as
\be \delta X^a =  c^i f^a_{ib}(T) X^b \ . \ee 
 One can deal with a constant term in $f^a_i(t)$ provided there is a nowhere-vanishing homogenous function $\cF^\ord{r}(t)$ of degree $r$ on the K\"{a}hler manifold, such that 
\be \frac{ \partial_a \cF^\ord{r}(t)}{r \cF^\ord{r}(t)} t^a = 1 \ . \ee
One then defines  $f^a_{ib}(t)$ as a Taylor series
\be f^a_{ib}(t) = \frac{ \partial_a \cF^\ord{n}(t)}{r \cF^\ord{n}(t)} f^a_i(0) + \partial_b  f^a_i(0) + \frac{1}{2} \partial_b \partial_c  f^a_i(0) t^c + \frac{1}{6} \partial_b \partial_c  \partial_d f^a_i(0) t^c t^d + \dots \ . \ee
According to (\ref{DualityDescent}) we define 
\be \delta K(T,\bar T)  = \bar D^\alpha  \scal{ c_i F_{ia}(T) \bar D_\alpha X^a } + D^\alpha  \scal{ c_i F_{i\bar a}(\bar T)  D_\alpha \bar X^{\bar a} } \ , \ee
where 
\be F_{ia}(t) t^a = F_i(t) \ . \ee
One can always choose the representative $K_{i \alpha}$ of $F_{ia}(T) \bar D_\alpha X^a$ such that it only depends on the background prepotential through its classical component 
\begin{multline} K_{i\alpha}[T_{\scriptscriptstyle \rm B},V]  =    F_{ia}(T_{\scriptscriptstyle \rm B}) \bar D_\alpha X^a_{\scriptscriptstyle \rm B} + \partial_a F_i(T_{\scriptscriptstyle \rm B}) \bar D_\alpha ( X^a - X^a_{\scriptscriptstyle \rm B} )  \\  + \frac{1}{2} \partial_a \partial_b  F_i(T_{\scriptscriptstyle \rm B}) \bar D_\alpha ( X^a - X^a_{\scriptscriptstyle \rm B} ) ( T^b - T^b_{\scriptscriptstyle \rm B}) + \dots   \ . \end{multline}
By the structure of the duality symmetry algebra, the chain will stop here. We conclude that the Feynman rules of the theory coupled to 
\be \int d^2 x d^4 \theta  \Scal{ u K[T_{\scriptscriptstyle \rm B},V] + u^\alpha c^i K_{i \alpha}[T_{\scriptscriptstyle \rm B},V]  +\bar u^{\alpha}c^iK_{i \alpha}[\bar T_{\scriptscriptstyle \rm B},\bar V]}  \ee
only involve the background field $T_{\scriptscriptstyle \rm B}$ and not its prepotential. Therefore the associated gamma function must be a function of the scalar fields $T$, and not the associated prepotential. Indeed note that neither the ghosts-for-ghosts nor the Nielsen--Kallosh ghost will contribute to the insertion of such a local operator in the 1PI generating function. Moreover, $K_\alpha$ is an operator of negative dimension, and therefore must be protected. It follows that the gamma-function $\gamma(t,\bar t)$ must be fully invariant with respect to the isometries. Using equation (\ref{BetaGamma}) we conclude that the $n\ge 1$ loop order contribution to the beta function or the counter-Lagrangian must be a fully duality-invariant function of the scalars. 

This is consistent with the analysis carried out in \cite{Howe:1986ys}, because starting from the action (\ref{ActionBF}), it was  concluded that contributions to the beta functions beyond the one-loop order must be scalar functions of the Riemann tensor, which are themselves invariant with respect to the isometries of the K\"{a}hler manifold. 

If we discuss the simplest case of an $SU(1,1)/U(1)$ K\"{a}hler space in this framework, the complex modulus field $\uptau$ is itself a nowhere-vanishing homogenous function on the target-space manifold, and one can define 
\be \delta X = e \frac{X_{\scriptscriptstyle \rm B}}{\uptau_{\scriptscriptstyle \rm B}} + 2 h X  - f \scal{  \uptau_{\scriptscriptstyle \rm B} X + \uptau ( X - X_{\scriptscriptstyle \rm B}) } \ . \ee
The variation splits into 
\bea \delta X_{\scriptscriptstyle \rm B} &=&  e \frac{X_{\scriptscriptstyle \rm B}}{\uptau_{\scriptscriptstyle \rm B}} + 2 h X_{\scriptscriptstyle \rm B}  - f   \uptau_{\scriptscriptstyle \rm B} X_{\scriptscriptstyle \rm B} \ , \CR 
\delta ( X - X_{\scriptscriptstyle \rm B}  ) &=&  2 h ( X - X_{\scriptscriptstyle \rm B}  )   - f (   \uptau_{\scriptscriptstyle \rm B} +  \uptau )  ( X - X_{\scriptscriptstyle \rm B})  \ .
\eea
One then computes 
\bea \delta K(\uptau,\bar\uptau) &=& - 2 h + f ( \uptau + \bar \uptau) \CR
&=& \bar D^\alpha \Scal{ -\frac{h}{\uptau_{\scriptscriptstyle \rm B}} \bar D_\alpha X_{\scriptscriptstyle \rm B} + f \bar D_\alpha X } +  D^\alpha \Scal{ -\frac{h}{\bar \uptau_{\scriptscriptstyle \rm B}} D_\alpha \bar X_{\scriptscriptstyle \rm B} + f D_\alpha \bar X } \ , 
 \eea
and 
\be K_\alpha = -\frac{h}{\uptau_{\scriptscriptstyle \rm B}} \bar D_\alpha X_{\scriptscriptstyle \rm B} + f \bar D_\alpha X \ , \ee
satisfies $\delta K_\alpha = 0 $, showing that the counter-Lagrangian terms at loops $L\ge1$ must therefore be invariant under the isometric ``duality'' symmetries.
 
\subsection{Non-renormalisation in $\cN=4$ supergravity}

Let us now return to $\cN=4$ supergravity without vector multiplets. In order to apply the above reasoning, one must not only assume that there is an off-shell formulation of the theory in superspace with all supercharges realised linearly, but also that the action defining the Feynman rules is itself duality invariant. Such a formulation has not been constructed, even for supergravity theories with fewer supersymmetries, such as $\cN=2$. It is clear from the structure of such a theory in components \cite{Henneaux:1988gg,Schwarz:1993vs,Hillmann:2009zf}, that a corresponding superspace formulation could not be manifestly Lorentz invariant. Therefore the only way to construct it would seem to require the introduction of Lorentz harmonics. Such a formulation was introduced  in \cite{Sokatchev:1985tc}, and was later used to give an off-shell version of $\cN=2$ (maximal) super-Yang--Mills in five dimensions with sixteen supercharges realised linearly \cite{Sokatchev:1988qr}. However such a formulation suffers from several technical complications that have not been fully clarified so far. One problem is due to the use of harmonic variables parametrising a non-compact coset space, although this can be cured by considering the quotient of the Lorentz group by a maximal parabolic subgroup \cite{Galperin:1991gk}. 

The general construction that we have described in this section has not taken into account the possible complications unavoidably associated to such a putative Lorentz harmonic-superspacee formulation of $\cN=4$ supergravity. Instead, we have extrapolated the more conventional argument one would carry out within a more standard superspace theory with finitely many auxiliary fields. Therefore, we should stress that this argument is by no means an actual proof of the three-loop non-renormalisation theorem, but rather is an attempt to motivate a three-loop non-renormalisation theorem using reasonable generalisations of the standard tools available in supersymmetric field theories.

Let us suppose that the reasoning proposed in this section does indeed apply to $\cN=4$ supergravity. The unique duality invariant counterterm  that can be written as a full-superspace integral is the integral of the K\"{a}hler potential. The associated density is not itself duality invariant, but satisfies instead 
\be \delta \Scal{ \,E (-1) \mbox{ln}\Scal{ \tfrac{-i}{2} ( \uptau - \bar \uptau )}} = - 2 h \,E  + f \,E ( \uptau + \bar\uptau ) \ . \label{deltaE}\ee
The right-hand-side of \eqref{deltaE} is the sum of a holomorphic function of $\uptau$ plus an anti-holomorphic function, and we have seen in the preceding section that the associated integral over full superspace vanishes subject to the classical equations of motion. Within an off-shell formulation, one can always carry out a perturbative redefinition of the variables such that this integral will also vanish. It then follows that it must be a total derivative in harmonic superspace. However, the scalar field $\uptau$ and the supervielbein Berezinian can only be the total derivative of a function depending explicitly on the hypothetical prepotentials of the theory.   

To be more explicit, we could reasonably assume that the chiral scalar superfields satisfy something like
\be \uptau \sim \bar D^{8} X[V ] \ , \label{PropoPrepo}  \ee
for a functional $X$ of the prepotentials of the theory, written collectively as $V$. Then it would follow that 
\be \,E  \scal{ - h  + f \uptau } \sim D_{\dot{\alpha} i} \Scal{ \,E  \Scal{   -\frac{h}{\uptau_{\scriptscriptstyle \rm B}}  + f } \bar D^{7 {\dot{\alpha} i}} X_{\scriptscriptstyle  B} + \dots }  \ . \ee
According to the discussion of the last section, such an operator $K^M$, depending explicitly on the prepotentials, could not possibly renormalise the local operator $
 \cL^{\ord{1}\, M}$, and so neither could $ \,E (-1) \mbox{ln}\scal{ \tfrac{-i}{2} ( \uptau - \bar \uptau )} $ renormalise the classical Lagrange density $\cL^\ord{0}$ as a local operator.
 
 Note that the appearance of the one-loop anomaly (\ref{BrokenSTI}) implies that, although one cannot rely on the Ward identity for the non-linearly realised generator, it is enough to consider the Ward identity associated to the parabolic subgroup in order to conclude that this operator cannot renormalise the classical Lagrange density $\cL^\ord{0}$ as a local operator. Using then
 \be (\ell - 1 )  \beta_\ell = \gamma_\ell \ , \ee
it would follow that the full-superspace integral of the K\"{a}hler potential cannot correspond to a logarithmic divergence of the theory. 
\section{Supergeometry in five dimensions}
In this section and the next we shall discuss the status of logarithmic divergences in pure $\cN=2$ supergravity in five dimensions (\ie half-maximal, but note that the $D=5$ half-maximal theory has more degrees of freedom than does $D=4$, $\cN=4$). In order to discuss the invariants we shall need some details on the corresponding supergeometry which are not yet available in the literature. For this purpose it is more efficient to compute the supergeometry in the maximal $D=5$, $\cN=4$ supergravity, and then truncate the results to $\cN=2$. 
\subsection{Maximal ($\cN=4$) supergravity}
 In $\cN$-extended supergravity in five dimensions, the structure group is $Sp(1,1) \times Sp(\csN)$, and the spinors satisfy a symplectic-Majorana condition of the form $\bar\psi^{\a i}=\O^{\a\b}\O^{ij} \psi_{\b j}$, where $\Omega_{ij}$ is the $Sp(\csN)$ symplectic form satisfying $\Omega^{ik} \Omega_{jk} = \delta^i_j$, and {\it idem} for the $Sp(1,1)$ symplectic form $\Omega_{\alpha\beta}$. We first discuss  the on-shell maximal theory which has forty-two scalars in the coset $E_{6(6)} / Sp_{\scriptscriptstyle \rm c}(4)$, forty-eight totally antisymmetric and symplectic traceless dimension-one-half fermions, $\chi_\a^{ijk}$ (satisfying $\bar \chi^\a_{ijk} = \Omega^{\a\b} \Omega_{il} \Omega_{jp} \Omega_{kq} \chi_\b^{lpq}$), twenty-seven vector fields $F^{ij}$ (satisfying $\bar F_{ij} =  \Omega_{ik} \Omega_{jl}F^{kl}$), again antisymmetric and symplectic traceless, as well as eight gravitini and the graviton. We can determine the geometry from dimensional analysis, using the Bianchi identities to fix unknown coefficients. The dimension-zero torsion is
 \be T_\alpha^i{}_\beta^j{}^{\, a} = - i \Omega^{ij} \gamma^a{}_{\alpha\beta} \ , \ee
 At dimension one-half the Bianchi identity implies that the only non-vanishing torsion is 
 \be  T_\alpha^i{}_\beta^j{}_\gamma^k  =  \Omega_{\gamma[\alpha} \chi_{\beta]}^{ijk} + \Omega_{\alpha\beta} \chi_\gamma^{ijk} \ .  \ee
Note that there is no sign ambiguity in raising the last indices in the torsion since one uses two symplectic forms 
 \be T_\alpha^i{}_\beta^j{}^\gamma_k = T_\alpha^i{}_\beta^j{}_\delta^l \Omega_{lk} \Omega^{\delta\gamma} \ .  \ee
One understands that 
 \be \gamma^a_{\alpha\beta} = \Omega_{\alpha\gamma} \gamma^{a\, \gamma}{}_\beta \ , \ee
 where 
 \be  \gamma^{a\, \alpha}{}_\gamma   \gamma^{b\, \gamma}{}_\beta + \gamma^{b\, \alpha}{}_\gamma   \gamma^{a\, \gamma}{}_\beta = 2 \eta^{ab} \delta^\alpha_\beta \ .  \ee
 Moreover, we chose conventions for the gamma matrices such that 
 \be \gamma^{abc}_{\alpha\beta} =  - \frac{i}{2} \varepsilon^{abc}{}_{de} \gamma^{de}_{\alpha\beta} \ee
 where $\gamma^{abc} = \gamma^{[a} \gamma^b \gamma^{c]} $ and $\gamma^{ab} = \gamma^{[a} \gamma^{b]}$. One has then identities like 
 \bea \varepsilon^{ab}{}_{cde} \gamma^e_{\alpha\beta} \gamma^{cd}_{\gamma\delta} &=& - 2 i \scal{ \Omega_{\alpha\beta} \gamma^{ab}_{\gamma\delta} + 4 \Omega_{\gamma)[\alpha} \gamma^{ab}_{\beta](\delta} } \ , \CR
 \gamma^{[a}_{\alpha\beta} \gamma^{b]}_{\beta\delta} &=& 2 \Omega_{\delta][\alpha} \gamma^{ab}_{\beta][\gamma} \ .  \eea

 At dimension one, the independent quantities are defined by the derivatives of the scalar fields, the twenty-seven Maxwell fields strengths and bilinears in the fermion superfields. We therefore consider $P_a^{ijkl}$ defined from the $E_{6(6)}$ matrix $V^{ij}{}_{IJ}$ as
 \be D_a V^{ij}{}_{IJ} = P_a^{ijkl} V_{klIJ} \ , \ee 
and $M_{ab}^{ij}$ (which also includes a bilinear term in the fermions) and note that the bilinear $\chi^2$ decomposes into irreducible representations according to 
 \begin{multline} 
  \Omega_{pq} \chi^{ijp}_\alpha \chi^{klq}_\beta = \frac{1}{4} \Omega_{\alpha\beta} \Scal{ N^{[i,j]kl} - \frac{2}{3} \Omega^{k][i} N^{j],[l}} + \frac{1}{4} \gamma^a_{\alpha\beta} \Scal{ N^{[i,j]kl}_a - \frac{2}{3} \Omega^{k][i} N_a^{j],[l}} \\  - \frac{1}{8} \gamma^{ab}_{\alpha\beta} \Bigl(  N_{ab}^{ij,kl} -\frac{1}{10} \scal{ \Omega^{ij} N_{ab}^{kl} + \Omega^{kl} N_{ab}^{ij} + 8 \Omega^{k][i} N_{ab}^{j][l} } \Bigr . \\ \Bigl . - \frac{1}{216} \scal{ \Omega^{ij} \Omega^{kl} + 8 \Omega^{k][i} \Omega^{j][l}} N_{ab} \Bigr) \end{multline} 
where sets of adjacent indices are understood to be antisymmetrised, while the commas separate different columns of the corresponding Young tableaux.  All expressions are assumed to be symplectic traceless. $N^{i,jkl}$ is in the symplectic traceless  ${ \Yboxdim5pt  {\yng(2,1,1)}}$, $N^{i,j}$ in the  ${ \Yboxdim6pt  {\yng(2)}}$, $N^{ij,kl}$ in the symplectic traceless  ${ \Yboxdim6pt  {\yng(2,2)}}$, and $N^{ij}$ in the symplectic traceless ${ \Yboxdim6pt  {\yng(1,1)}}$. The bilinears in $\chi^{ijk}_\alpha$ also include $N^{ijk,lpq}$ and $N_a^{ijk,lpq}$ in the  ${ \Yboxdim4pt  {\yng(2,2,2)}}$, and $N_{ab}^{ij,klpq}$ in the  ${ \Yboxdim4pt  {\yng(2,2,1,1)}}$, but they will not appear in the dimension-one components of the torsion and the Riemann tensor.  The notation $\Omega^{k][i} N^{j][l}$ is also used. It is defined by
\be 4 \Omega^{k][i} N^{j][l}  = \Omega^{ki} N^{jl} - \Omega^{kj} N^{il}- \Omega^{li} N^{jk} + \Omega^{lj} N^{ik} \ .\ee
We will consider the standard constraint $T_{ab}{}^c = 0 $.
 
 First we use the Bianchi identity 
\be R_\alpha^i{}_\beta^j{}_c{}^d   = T_c{}_\alpha^i{}_\eta^k \Omega_{kl} \Omega^{\eta\varsigma} T_\varsigma^l{}_\beta^j{}^d + T_c{}_\beta^j{}_\eta^k \Omega_{kl} \Omega^{\eta\varsigma} T_\varsigma^l{}_\alpha^i{}^d \ ,  \ee
from which one can determine the components of $ R_\alpha^i{}_\beta^j{}_c{}^d $ in terms of those of $T_a{}_\beta^j{}_\gamma^k$.  Then we use
\be R_\alpha^i{}_\beta^j{}^{kl} \Omega_{\gamma\delta} + \frac{1}{4} R_\alpha^i{}_\beta^j{}_{cd} \Omega^{kl} \gamma^{cd}_{\gamma\delta} \ +   \circlearrowleft \ = D_\alpha^i T_\beta^j{}_\gamma^k{}_\delta^l + T_\alpha^i{}_\beta^j{}^a T_a{}_\gamma^k{}_\delta^l + T_\alpha^i{}_\beta^j{}_\eta^p \Omega_{pq} \Omega^{\eta\varsigma} T_\varsigma^q{}_\gamma^k{}_\delta^l \ +   \circlearrowleft \ee
where $+ \circlearrowleft $ indicates that one should add cyclic permutations of the three first pairs of indices {\tiny $\left(\begin{array}{ccc} i&j&k \\\alpha & \beta&\gamma\end{array}\right)$}. This allows us to kill some {\sl a priori}  allowed terms such as the component involving
\be \gamma^{ab}_{\alpha\beta} \Omega^{k)(i} M^{j)(l}_{ab}\ ,   \ee
of $ R_\alpha^i{}_\beta^j{}^{kl}$, the components involving 
\be \gamma_{a \beta\gamma} N^{j,k} \ ,  \qquad \Omega^{jk} \varepsilon_a{}^{bcde} N_{bc} \gamma_{de \beta\gamma}  \ \ee
of  $T_a{}_\beta^j{}_\gamma^k$ and the related components involving 
\be - 2 i  \gamma_{cd \alpha\beta} N^{i,j}\ ,  \qquad 4 i \varepsilon_{cd}{}^{abe} \Omega^{ij} \gamma_{e\alpha\beta} N_{ab} \ \ee
of  $ R_\alpha^i{}_\beta^j{}_{cd} $. This also determines all but two coefficients for all the other representations. The latter can then be determined by requiring that the superfield $\chi_\alpha^{ijk}$ defines the covariant derivative of the scalar superfield $V^{ij}{}_{IJ}$, and by requiring that the algebra closes accordingly on it. 

After some work one obtains  
\bea  \label{ScalarMomenta5} 
D_\alpha^k V^{ij}{}_{IJ} &=& \Scal{ 2 \Omega^{k[i} \chi_\alpha^{jpq]} + \frac{3}{2} \Omega^{[ij} \chi_\alpha^{pq]k}} V_{pqIJ} \ , \CR 
D_\alpha^i \chi_\beta^{jkl} &=& - 2 i \gamma^a_{\alpha\beta} P_a^{ijkl} + \gamma^{ab}_{\alpha\beta} \Scal{ 3 \Omega^{i[j} M_{ab}^{kl]} + M^{i[j}_{ab} \Omega^{kl]} } - \frac{1}{8} \gamma^a_{\alpha\beta} N_a^{i,jkl} + \frac{3}{8} \Omega_{\alpha\beta} N^{i,jkl}\ ,  \CR
\eea
and finds that the torsion and curvature components are 
\bea T_a{}_\beta^j{}_\gamma^k &=& \frac{4 i}{3} \gamma^b_{\beta\gamma} M_{ab}^{jk} + \frac{1}{6} \varepsilon_{abc}{}^{de} \gamma^{bc}_{\beta\gamma} \Scal{ M_{de}^{jk} - \frac{3}{32} N_{de}^{jk} } \CR
&& \qquad  -  \frac{i}{48} \Omega_{\beta\gamma} N_a^{j,k} + \frac{i}{24} \gamma_a{}^b_{\, \beta\gamma} N_b^{j,k} + \frac{i}{192} \Omega^{jk} N_{ab} \gamma^b_{\beta\gamma} \ ,  \CR
 R_\alpha^i{}_\beta^j{}_{cd} &=&  \frac{8}{3} \Omega_{\alpha\beta} M^{ij}_{cd} + \frac{2i}{3} \varepsilon_{cde}{}^{ab} \gamma^e_{\alpha\beta} \Scal{ M_{ab}^{ij} - \frac{3}{32} N_{ab}^{ij}} + \frac{i}{24} \varepsilon_{cdab}{}^e \gamma^{ab}_{\alpha\beta} N_e^{i,j} + \frac{1}{96} \Omega^{ij} \Omega_{\alpha\beta} N_{cd} \ ,  \CR
 R_\alpha^i{}_\beta^j{}^{kl} &=& \frac{1}{4}  \Omega^{ij} N^{k,l}_{\alpha\beta} - \frac{1}{2} \Omega^{i](k} N^{l),[j}_{\alpha\beta} + \frac{3}{4} N_{\alpha\beta} ^{(k,l)ij} \CR
 && \qquad + \frac{1}{16} \gamma^{ab}_{\alpha\beta} \Scal{\, \frac{7}{72} \Omega^{i)(k}\Omega^{l)(j} N_{ab} - \frac{9}{5}  \Omega^{i)(k} N_{ab}^{l)(j} +  N_{ab}^{i)(k,l)(j} } \ , 
\eea
where $N_{\alpha\beta}^{i,jkl} = \frac{1}{4} \Omega_{\alpha\beta} N^{i,jkl} + \frac{1}{4} \gamma^a_{\alpha\beta} N_a^{i,jkl}$ and $N_{\alpha\beta}^{i,j} = \frac{1}{4} \Omega_{\alpha\beta} N^{i,j} + \frac{1}{4} \gamma^a_{\alpha\beta} N_a^{i,j}$ are bilinears in the reducible antisymmetric representation of $Sp(1,1)$. The constant multiplying the $N_a^{i,j}$ term in $T_a{}_{\beta\gamma}^{jk}$ requires one to check the existence of the 27 $d$-closed superfield strengths $F^{IJ} = V^{-1 IJ}{}_{ij} F^{ij} $  with 
\be F_{ab}{}^{kl} = M_{ab}^{kl} - \frac{1}{8} N_{ab}^{kl} \ , \quad F_a{}_\beta^j{}^{kl} = - \frac{i}{4} \gamma_{a\beta}{}^\gamma \chi_\gamma^{ijk} \ , \quad F_{\alpha\beta}^{ij}{}^{kl} = \Omega_{\alpha\beta}\Scal{ \Omega^{k][i} \Omega^{j][l} + \frac{1}{8} \Omega^{ij} \Omega^{kl}} \ . \ee
We have checked that the supersymmetry algebra of $\{D_\a^i , D_\b^j\}$ is satisfied on the scalar fields for the term involving the derivative of $\chi_\alpha^{ijk}$ that matches the dimension 1/2 torsion, and that the terms involved in the product of two derivatives of the scalar in $N^{ij,kl}_{ab}$ reproduce the correct term in the $R_\alpha^i{}_\beta^j{}^{kl}$ component acting on the scalars. The former permits one to fix the coefficient of the $\Omega_{\alpha\beta} N^{i,jkl}$ term in the derivative of $\chi_\alpha^{ijk}$, whereas the second permits one to fix the overall coefficient of the derivative of the scalars. Checking the dimension-zero torsion relation, one then fixes the coefficient of $P_a^{ijkl}$ in the derivative of $\chi_\alpha^{ijk}$. 
 \subsection{Pure $\cN=2$ supergravity}
 We now consider the truncation to pure $\cN=2$ supergravity.\footnote{There is an off-shell version of the $\cN=2$ theory which describes a multiplet that is dual to the $\cN=2, D=5$ supercurrent \cite{Howe:1981nz}. This could be used as an alternative starting point, but we do not consider it here.} For this purpose, we decompose $Sp(4) \supset Sp(2) \times Sp(2)^\prime$ such that the truncated fermion superfield reduces to 
 \be   \chi_\alpha^{ijk} = - 3 \Omega^{[ij} \chi_\alpha^{k]} \ , \quad \chi_\alpha^{i\hat{\jmath}\hat{k}} = \frac{1}{2} \Omega^{\hat{\jmath}\hat{k}} \chi_\alpha^i \ee
 where the unhatted indices now define the $\cN=2$ internal $Sp(2)$ indices, whereas the hatted indices are for the complementary $Sp(2)^\prime$ in $Sp(4)$. The second term is determined as a function of the former by imposing symplectic tracelessness with respect to $Sp(4)$. The vector field-strength tensor decomposes into the symplectic traceless component and the singlet according to
 \be F^{ij} = F^{\prime ij} - \frac{1}{4} \Omega^{ij} F \ , \quad F^{\hat{\imath}\hat{\jmath}} =  \frac{1}{4} \Omega^{\hat{\imath}\hat{\jmath}} F \label{Fdecomposes} \ee
 where $F^{\prime ij}$ is symplectic traceless. The $Sp(2)$ components of the scalar superfield reduce to
 \be V^{ij}{}_{KL} = \frac{1}{8} \Omega^{ij} \Omega_{KL} e^{\Phi} + e^{-\frac{1}{2}\Phi } \Scal{ \delta^{ij}_{KL} - \frac{1}{4} \Omega^{ij} \Omega_{KL}} ,\, V^{ij}{}_{\hat{K}\hat{L}} = - \frac{1}{8} \Omega^{ij} \Omega_{\hat{K}\hat{L}} e^{\Phi} \label{ScalarDecompose} \ee
 where $IJ$ and $\hat{K}\hat{L}$ are respectively $Sp(2)$ and $Sp(2)^\prime$ rigid indices. It is consistent to identify rigid and local indices, because $R_{\alpha\beta}^{i\,\,  j}{}^{kl} = 0 $ by construction in the absence of matter. The bilinears in the fermion reduce to
 \bea N_{\alpha\beta}^{i,jkl} &=& - 3 \Omega^{[kl} \chi_{[\alpha}^{j]} \chi_{\beta]}^i \ , \quad  N_{\alpha\beta}^{i,j\hat{k}\hat{l}} = \frac{1}{2} \Omega^{\hat{\jmath}\hat{k}} \chi_{[\alpha}^i \chi_{\beta]}^j\ , \quad N_{\alpha\beta}^{i,j} = 3 \chi^i_{[\alpha} \chi^j_{\beta]} \CR
 N^{i)(k,l)(j)}_{ab}  &=& \frac{1}{10} \gamma_{ab}^{\alpha\beta} \Scal{  - 18 \Omega^{i)(k} \chi^{l)}_\alpha \chi_\beta^{(j} + 7 \Omega^{i)(k} \Omega^{l)(j} \Omega_{pq} \chi^p_\alpha \chi^q_\beta } \CR
 N^{ij}_{ab} &=&  \gamma_{ab}^{\alpha\beta} \Scal{ - \chi^i_\alpha \chi^j_\beta + \frac{7}{8} \Omega^{ij} \Omega_{kl} \chi^k_\alpha \chi^l_\beta} \ , \quad N_{ab} = 9  \gamma_{ab}^{\alpha\beta} \Omega_{ij} \chi_\alpha^i \chi_\beta^j \ .
 \eea
It then follows  from (\ref{ScalarMomenta5}) that 
 \bea D_\alpha^i \Phi &=& -\chi_\alpha^i \CR
 D^i_\alpha \chi_\beta^j &=& \frac{i}{2} \Omega^{ij} \gamma^a{}_{\alpha\beta} D_a \Phi + \frac{1}{3} \gamma^{ab}{}_{\alpha\beta} \scal{ 2 M^{\prime ij}_{ab} + \Omega^{ij} M_{ab} }  - \frac{1}{2} \scal{ \chi^i_{[\alpha} \chi^j_{\beta]} - \Omega_{\alpha\beta} \Omega^{\gamma\delta} \chi_\gamma^i \chi_\delta^j }  \label{Dchi5} 
 \eea
 where $M_{ab}^{\prime ij}$ and $M_{ab}$ are defined as in (\ref{Fdecomposes}). The scalar superfield is real, and satisfies the second-derivative  constraints 
 \be \gamma^{a \alpha\beta} D_\alpha^{(i} D_\beta^{j)} e^{-\frac{1}{2} \Phi} = 0 \ , \quad \Omega^{\alpha\beta} D_\alpha^{(i} D_\beta^{j)} e^{\frac{3}{2} \Phi} = 0\ .  \ee  
 
 \subsection{Harmonic superspace} 
 Harmonic superspaces in five dimensions are of type $(2\cN,p)$, meaning that a G-analytic field of this type will be annihilated by $p$ four-component spinor derivatives that mutually anti-commute.  We write $2\cN$ because  a G-analytic field will depend on $(2\cN-p)$ four-component odd coordinates, with $\cN$ being the maximal possible value of $p$.\footnote{In other words, a G-analytic field of type $(2\cN,p)$ is $p/{2\cN}$ BPS.} In $\cN=4$ supergravity, only the $(8,1)$ harmonic structure associated to the coset  $Sp(4)/( U(1) \times Sp(3)) $ is consistent with the dimension-one-half torsion component because 
 \be u^1{}_i u^1{}_j  T_\alpha^i{}_\beta^j{}_\gamma^k = 0 \label{TorsionG}\ ,  \ee
 where the harmonic variables are defined such that 
 \be u^1{}_i u^{\bar 1}{}_j \Omega^{ij} = 1\ ,  \qquad u^r{}_i u^s{}_j \Omega^{ij} = \Omega^{rs} \ , \ee
 the other contractions are null and $\Omega^{rs}$ is the $Sp(3)$ symplectic matrix. 
 
 The Riemann tensor components are also consistent with the torsion because 
 \be u^1{}_i u^1{}_j  R_\alpha^i{}_\beta^j{}^{kl}  = \frac{1}{16} u^1{}_i u^1{}_j \gamma^{ab}_{\alpha\beta} \Scal{ N_{ab}^{i(k,l)j} - \frac{9}{5}  \Omega^{i(k} N_{ab}^{l)j} + \frac{7}{72}  \Omega^{i(k}\Omega^{l)j} N_{ab} } \ ,  \ee
clearly satisfies
 \be u^1{}_i u^1{}_j u^1{}_k R_\alpha^i{}_\beta^j{}^{kl} = 0 \label{GAC} \ . \ee
It turns out that the Lorentz curvature  \be u^1{}_i u^1{}_j R_\alpha^i{}_\beta^j{}_{cd} = \frac{i}{24} \varepsilon_{cdab}{}^e \gamma^{ab}_{\alpha\beta}  u^1{}_i u^1{}_j N_e^{i,j} \ee
is expressible in terms of the G-analytic vector 
\be B_a \equiv u^1{}_i u^1{}_j N_a^{i,j} \ . \ee
To show that $B_a$ is indeed G-analytic, let us rewrite $B_a$ in terms of 
\be \chi^{1rs}_\alpha \equiv u^1{}_i u^r{}_j u^s{}_k \chi_\alpha^{ijk} \ , \ee
 as
\be B_a = \frac{1}{4} \gamma_a^{\alpha\beta} \Omega_{rt} \Omega_{su} \chi_\alpha^{1rs} \chi_\beta^{1tu} \label{GaB} \ . \ee 
We have
\bea D_\alpha^1 \chi_\beta^{1rs} &=& - \frac{1}{8} u^1{}_i u^1{}_j u^r{}_k u^s{}_l \scal{ \gamma^a_{\alpha\beta} N_a^{i,jkl} - 3 \Omega_{\alpha\beta}  N^{i,jkl} } \CR
&=&  \scal{ \delta_{\alpha\beta}^{\gamma\delta}  - \Omega_{\alpha\beta} \Omega^{\gamma\delta} } \Omega_{tu} \, \chi_\alpha^{1rt} \chi_\beta^{1su} \ , 
\eea
 so that 
 \bea D_\alpha^1 B_a &=& \frac{1}{2} \gamma_a^{\beta\gamma} \scal{ \delta_{\alpha\beta}^{\eta\varsigma} - \Omega_{\alpha\beta}\Omega^{\eta\varsigma} } \,  \Omega_{vw} \Omega_{rt} \Omega_{su} \, \chi_\eta^{1rv} \chi_\varsigma^{1sw} \chi_\gamma^{1tu} \CR
  &=& \frac{1}{2} \gamma_a^{\beta\gamma} \scal{ \delta_{\alpha\beta}^{\eta\varsigma} - \Omega_{\alpha\beta}\Omega^{\eta\varsigma} } \delta_{\gamma}^\zeta \Omega_{vw} \Omega_{rt} \Omega_{su} \, \chi_{[\eta}^{1rv} \chi_\varsigma^{1sw} \chi_{\zeta]}^{1tu} \CR
  &=& \frac{1}{4}   \gamma_a^{\beta\gamma} \Omega_{\beta\gamma} \Omega_{\alpha\delta} \Omega^{[\eta\varsigma} \Omega^{\zeta\delta]}  \Omega_{vw} \Omega_{rt} \Omega_{su} \, \chi_{[\eta}^{1rv} \chi_\varsigma^{1sw} \chi_{\zeta]}^{1tu} \CR
  &=& 0 \ .
 \eea
 Note that this is the unique G-analytic vector (up to an overall G-analytic function) since the obstruction 
 \be \{ D_\alpha^1 , D_\beta^1 \} B_a = \frac{i}{24} \varepsilon_{abc}{}^{de} \gamma^{bc}_{\alpha\beta} B_d B_e = 0 \ ,  \ee 
 only vanishes because the curvature is proportional to $B_a$ itself.

At the linearised level, one can define the $\partial^6 R^4$ invariant starting with a G-analytic integrand quadratic in the linearised G-analytic superfield $W^{1rst} \sim V^{1[r}{}_{IJ}  V^{st]}{}_{KL} \Omega^{IK} \Omega^{JL} $, but this does not extend to the non-linear level. Indeed, its na\"{i}ve generalisation $V^{1r}{}_{ IJ} \equiv u_i u^r_j V^{ij}{}_{IJ}$ is not G-analytic since
\be D_\alpha^1 V^{1r}{}_{IJ} =  - \frac{1}{2} \Omega_{st} \chi_\alpha^{1rs} V^{1t}{}_{IJ} \ .\ee
There is therefore no dimension-zero G-analytic superfield of the correct $U(1)$ weight, and the $\partial^6 R^4$ supersymmetry invariant cannot be defined as a harmonic-superspacee integral at the non-linear level.

The $\cN=2$ supergravity theory admits a  $(4,1)$ harmonic-superspacee structure associated with  the coset $Sp(2) / ( U(1) \times Sp(1) )$. The harmonic variables in this case satisfy
  \be u^1{}_i u^{\bar 1}{}_j \Omega^{ij} = 1\  , \qquad u^r{}_i u^s{}_j \Omega^{ij} = \varepsilon^{rs} \ , \ee
where $\varepsilon^{rs}$ is the $Sp(1)$ antisymmetric tensor  and the other contractions are null. However, since the torsion supertrace does not vanish in this case, \ie
 \be (-1)^A T^i_{\alpha A}{}^A = \chi_\alpha^i \ , \ee
 (where the index $A$ runs over all bosonic and fermionic indices), the G-analyticity condition involves the flat $U(1)$  connection $u^1{}_i \chi^i_\alpha$ 
 \be u^1{}_i  D_\alpha^i  \, u^1{}_j  \chi_\beta^j  + u^1{}_j  D_\beta^j \,  u^1{}_i \chi_\alpha^i  = 0 \label{FlatCon} \ , \ee
 and reads 
 \be u^1{}_i \scal{  D_\alpha^i + \chi_\alpha^i } \cF = 0 \ . \label{GASection} \ee 
We will prove this in the next section by consistency with the normal-coordinate expansion of a generic scalar superfield together with Stokes' theorem in superspace.

\section{Invariants in five dimensions} 

As in four dimensions, the first duality invariants will be of the same dimension as the full-superspace integral of a function of the $D=5$ scalars. In $\cN=4$ supergravity, this property was already discussed in \cite{Bossard:2010bd}. In pure $D=5$, $\cN=2$ supergravity, the first $Sp(2)$ invariant can be derived from the exact 6-superform 
\be e^{-\Phi}  F \wedge R^{ab} \wedge R_{ab} \ee
where $d ( e^{-\Phi}  F  )= 0 $. However, this invariant includes the terms
\be \cL  \sim  \frac{1}{2} \varepsilon_{abcde} e^{-\phi} e^a \wedge R^{bc} \wedge R^{de} + A \wedge R^{ab} \wedge R_{ab} + \dots \ee
so it is clearly not invariant with respect to a shift of the dilaton superfield. In any case, it is not relevant in perturbation theory in five dimensions because of its mass dimension. 

In this section we will show that the integral of the Berezinian of the supervielbein  does not vanish in maximal supergravity in five dimensions, but it does vanish in the half-maximal theory. However, the volume of $\cN=2$ superspace vanishes, so one can still write the shift-invariant $R^4$-type invariant as a full-superspace integral of the dilaton superfield.  In order to do this, we shall compute the normal-coordinate expansion of the  supervielbein Berezinian.

\subsection{Normal-coordinate expansion of $\,E$} 
 One checks, in a similar fashion to \cite{Bossard:2011tq}, that all the requirements for the existence of complex  normal coordinates
 \be \zeta^{\hat{A}} \equiv  \{ \zeta^\alpha \equiv u^i{}_1 \delta_\mu^\alpha \theta^\mu_i  \, , \ z^{\bar 1 r} , z^{\bar 1 \bar 1} \}\ ,  \ee 
 are satisfied, because the associated tangent vectors satisfy the involutive algebra
 \bea \{ \tilde E_\alpha^1 , \tilde E_\beta^1 \} &=& 2 \Omega^1_{(\alpha\beta)}{}^\gamma \tilde E_\gamma^1 + 2 \Omega^1_{(\alpha}{}^1{}_1 \tilde E^1_{\beta)}  +  \scal{ N_{\alpha\beta} ^{1\bar 1,1r} - \tfrac{9}{10} N_{\alpha\beta}^{1r} } d_{\bar 1 r}  + \frac{1}{2}  \scal{ N_{\alpha\beta}^{1\bar 1,1\bar 1} - \tfrac{9}{5} N_{\alpha\beta}^{1\bar 1}- \tfrac{7}{72} N_{\alpha\beta}  }  d_{\bar 1 \bar 1} \ , \CR
 \left[ d_{\bar 1 r} , d_{\bar 1 s}\right] &=& -\Omega_{rs}  d_{\bar 1 \bar 1} \ ,  \eea 
 with all other graded commutators vanishing. Here $d_{\bar 1 r} $ and $d_{\bar 1 \bar 1}$ are the vectors on the harmonic coset space that act on the harmonic variables by
 \be d_{\bar 1 r} u^{\bar 1}{}_i = \Omega_{rs} u^s{}_i \ , \qquad d_{\bar 1 r} u^s{}_i =  \delta^s_r u^1{}_i \ , \qquad d_{\bar 1 \bar 1} u^{\bar 1}_i = 2 u^1{}_i \ , \ee
 and trivially on the others. The complex coordinates $z^{\bar 1 r} , z^{\bar 1 \bar 1}$ are the complex normal coordinates on the coset space associated to these vectors. The vectors $\tilde E_\a^1$ are the horizontal lifts of the basis vectors $E_\a^1$ to the harmonic superspace (\ie they contain connection terms).
 
 As in four dimensions, one can check that the normal-coordinate expansion of the supervielbein Berezinian multiplying any scalar superfield factorises into the normal-coordinate expansion of the harmonic measure and the normal-coordinate expansion of the supervielbein Berezinian together with the scalar superfield given in terms of the fermionic coordinates expansion alone. One can therefore forget about the complex normal-coordinate expansion of the harmonic measure, and simply consider the normal-coordinate expansion in terms of the fermionic coordinates $\zeta^\alpha$ as in \cite{KuzenkoRY}. 
 
 Before discussing the specific examples of maximal and half-maximal supergravities in five dimensions, we shall rederive the formula for the normal-coordinate expansion of the supervielbein Berezinian in an alternative way. 

We will start quite generally and consider a supergravity theory that admits four fermionic normal coordinates $\zeta^\alpha$ (possibly together with harmonic ones which can be disregarded), with a possibly non-zero torsion supertrace 
 \be (-1)^A T_{\alpha A}{}^A \equiv \chi_\alpha \label{TraceTorsion} \ .\ee
 Note that we use $\alpha, \beta \dots $ as indices for the 4 normal coordinates, which would be the $Sp(1,1)$ indices in five dimensions, and which would stand for both fundamental and complex conjugate $SL(2,\mathds{C})$ indices together with the associated $U(1)$ weights in four dimensions.  By Stokes' theorem, the integral of a total derivative over superspace must vanish 
 \be \label{Stokes} 0 = \int d^d x d^{4k}\theta   \partial_M \scal{ \,E E_\alpha{}^M \Xi^\alpha} = \int d^d x d^{4k}\theta \,E \scal{ D_\alpha \Xi^\alpha + \chi_\alpha \Xi^\alpha } \ .\ee
If one assumes the existence of normal coordinates, with a general normal-coordinate expansion of the supervielbein Berezinian  given by
 \be \,E  = \cE \Scal{ 1 + \zeta^\alpha e_\alpha + \zeta^\beta \zeta^\alpha e_{\alpha\beta} + \zeta^\gamma \zeta^\beta \zeta^\alpha e_{\alpha\beta\gamma} +\zeta^\delta  \zeta^\gamma \zeta^\beta \zeta^\alpha e_{\alpha\beta\gamma\delta} }  \ , \label{ExpansionBerE} \ee
 then Equation (\ref{Stokes}) implies that 
 \bea &&  \int d^4 \zeta \Scal{  1 + \zeta^\alpha e_\alpha + \zeta^\beta \zeta^\alpha e_{\alpha\beta} + \zeta^\gamma \zeta^\beta \zeta^\alpha e_{\alpha\beta\gamma} +\zeta^\delta  \zeta^\gamma \zeta^\beta \zeta^\alpha e_{\alpha\beta\gamma\delta} } \exp[ \zeta^\eta D_\eta ] ( D_\varsigma + \chi_\varsigma ) \Xi^\varsigma\big| \CR
 &=& \frac{1}{24} \varepsilon^{\alpha\beta\gamma\delta} \Scal{ D_\alpha D_\beta D_\gamma D_\delta + 4 e_\alpha D_\beta D_\gamma D_\delta + 12 e_{\alpha\beta} D_\gamma D_\delta + 24 e_{\alpha\beta\gamma} D_\delta + 24 e_{\alpha\beta\gamma\delta} } ( D_\eta + \chi_\eta) \Xi^\eta\big| \CR
 &=& 0 \eea
 for any tensor superfield $\Xi^\alpha$, where the notation $|$ to the right of a superfield indicates that the latter is evaluated at $\zeta^\alpha = 0$, as in \cite{KuzenkoRY}.  In more geometrical terms, $X|$ is the pullback of the superfield $X$ to the analytic superspace from which the normal-coordinate expansion is defined. Note that for a general vector field, $\Xi^A$, this expression would only be required to be a total derivative in the analytic superspace, but for a vector, $\Xi^\alpha$, normal to the analytic superspace,  this expression must vanish. The various terms in $D^{5-n} \Xi$ then determine uniquely the components $e_n$ of the normal-coordinate expansion of $ \,E$. To carry out this computation explicitly, we will consider theories for which one has
 \be T_{\alpha\beta}{}^\gamma = 0 \ , \quad D_\eta R_{\alpha\beta\gamma}{}^\delta = 0 \ , \quad D_\alpha \chi_\beta + D_\beta \chi_\alpha = 0  \label{ConstraintTorCurv} \ , \ee
 as well as
 \be \varepsilon^{\alpha\beta\gamma\delta} R_{\varsigma\alpha\beta}{}^\varsigma R_{\iota\gamma \delta}{}^\eta = \frac{1}{4} \delta_\iota^\eta  \varepsilon^{\alpha\beta\gamma\delta} R_{\varsigma\alpha\beta}{}^\varsigma R_{\vartheta\gamma \delta}{}^\vartheta  \ , \quad  \varepsilon^{\alpha\beta\gamma\delta} R_{\iota\alpha\beta}{}^\varsigma R_{\varsigma\gamma \delta}{}^\eta = \frac{1}{4} \delta_\iota^\eta  \varepsilon^{\alpha\beta\gamma\delta} R_{\vartheta\alpha\beta}{}^\varsigma R_{\varsigma\gamma \delta}{}^\vartheta \label{CurvatureTrace} \ . \ee
 These equations are indeed satisfied in four dimensions \cite{Bossard:2011tq}, and they are also satisfied in supergravity in five dimensions. We have checked that the torsion (\ref{TorsionG}) satisfies this criterion, that the corresponding curvature (\ref{GAC}) is G-analytic, and that $\chi_\alpha$ defines a flat $U(1)$ connection (\ref{FlatCon}).  Because the curvature (\ref{GAC}) only depends on an $Sp(1,1)$ vector, it follows from representation theory that the constraint (\ref{CurvatureTrace}) must be satisfied. 

Using these equations, one computes that 
\bea e_\alpha &=& \chi_\alpha \big| \CR
 e_{\alpha\beta}&=& \frac{1}{6} R_{\gamma\alpha\beta}{}^\gamma \big| + \frac{1}{2} D_\alpha \chi_\beta \big| + \frac{1}{2} \chi_\alpha   \chi_\beta \big|\CR
  e_{\alpha\beta\gamma}&=& \frac{1}{6} \chi_{[\alpha} R_{\delta|\beta\gamma]}{}^\delta \big| + \frac{1}{6} D_{[\alpha} D_\beta \chi_{\gamma]} \big| + \frac{1}{2} \chi_{[\alpha}  D_\beta \chi_{\gamma]} \big| + \frac{1}{6} \chi_\alpha   \chi_\beta \chi_\gamma \big|\CR
   e_{\alpha\beta\gamma\delta} &=& \frac{1}{180} R_{\eta[\alpha\beta}{}^\varsigma R_{\varsigma|\gamma\delta]}{}^\eta\big| + \frac{1}{72}R_{\eta[\alpha\beta}{}^\eta R_{\varsigma|\gamma\delta]}{}^\varsigma \big|+   \frac{1}{12} R_{\eta[\alpha\beta}{}^\eta ( D_\gamma \chi_{\delta]} + \chi_\gamma\chi_{\delta]}) \big| \CR
   && \quad + \frac{1}{24} D_{[\alpha} D_\beta D_\gamma \chi_{\delta]} \big| + \frac{1}{8} D_{[\alpha} \chi_\beta D_\gamma \chi_{\delta]} \big| + \frac{1}{6} \chi_{[\alpha} D_\beta D_\gamma \chi_{\delta]} \big|  \CR
   && \qquad \quad+ \frac{1}{4} \chi_{[\alpha} \chi_\beta D_\gamma \chi_{\delta]} \big| + \frac{1}{24} \chi_\alpha   \chi_\beta \chi_\gamma \chi_\delta  \big| \ .
\eea
One checks that these expressions are indeed compatible with the formula given in \cite{KuzenkoRY}, reproducing the result
\bea \zeta^\alpha D_\alpha \mbox{ln}[\,E] &=& \zeta^\alpha \chi_\alpha  + \frac{1}{3} \zeta^\beta \zeta^\alpha R_{\gamma\alpha\beta}{}^\gamma \big|  + \frac{1}{45} \zeta^\delta \zeta^\gamma \zeta^\beta \zeta^\alpha  R_{\eta\alpha\beta}{}^\varsigma R_{\varsigma\gamma\delta}{}^\eta\big|  \label{LogExp} \CR
&=& (-1)^A T_{\alpha A}{}^A - \Omega_{\beta\alpha}{}^\beta \zeta^\alpha - \delta_\mu^\alpha ( E_\alpha{}^\mu - \delta_\mu^\alpha) \ ,\eea
where, by definition, $\chi_\alpha = \exp[\zeta^\beta D_\beta ] \chi_\alpha \big|$ \footnote{Or more explicitly $\chi_\alpha = \chi_\alpha \big| + \zeta^\beta \scal{ D_\beta \chi_\alpha}\big|   + \frac{1}{2} \zeta^\gamma \zeta^\beta \scal{D_\beta D_\gamma \chi_\alpha} \big| +   \frac{1}{6}  \zeta^\delta \zeta^\gamma \zeta^\beta \scal{D_\beta D_\gamma D_\delta \chi_\alpha} \big| $.} and where  (\ref{TraceTorsion}) and (\ref{ConstraintTorCurv}) have been used to show that the normal-coordinate expansions of $ \Omega_{\beta\alpha}{}^\beta$ and $E_\alpha{}^\mu$ take the same form as given in \cite{KuzenkoRY}. 

Let us now consider the full-superspace integral of a scalar superfield $K$. Using the normal-coordinate expansion, one computes that 
\be  \int d^d x d^{4k}\theta \, \,E \, K = \int d\mu_{\scriptscriptstyle (k,1)} \, \cF[K] \ , \ee
where $d\mu_{\scriptscriptstyle (k,1)}$ is the measure over $(k,1)$ harmonic superspace\,\footnote{That would be $(k,1,1)$ with $k = \cssN$ in four dimensions and $(k,1)$ with $k = 2\cssN$ in five dimensions.} $d\mu_{\scriptscriptstyle (k,1)} = dx^d d^{4(k-1)}\theta du\,  \cE$ with $du$ being the Haar measure over the harmonic coset manifold, and $\cE$ is the measure defined in (\ref{ExpansionBerE}), and where the function $\cF[K]$ is equal to
\be  \frac{1}{24} \varepsilon^{\alpha\beta\gamma\delta} \Scal{ D_\alpha D_\beta D_\gamma D_\delta + 4 e_\alpha D_\beta D_\gamma D_\delta + 12 e_{\alpha\beta} D_\gamma D_\delta + 24 e_{\alpha\beta\gamma} D_\delta + 24 e_{\alpha\beta\gamma\delta} }K \big|\ .  \ee
Using this expression and using (\ref{ConstraintTorCurv}) and (\ref{CurvatureTrace}), one computes that $\cF[K]$ satisfies 
\be ( D_\alpha + \chi_\alpha ) \cF[K] = 0  \ , \ee
\ie $\cF[K]$ is only G-analytic on a line bundle with a flat connection. Therefore we conclude in general that one can integrate any $G$-analytic section on this line bundle over $(k,1)$ harmonic superspace. 

\subsection{$R^4$ type invariants}
Before considering the full-superspace integral of a general function of the dilaton superfield $\Phi$ in $\cN=2$ supergravity in five dimensions, let us discuss the simpler $\cN=4$ example. In this case,   the torsion supertrace vanishes, and the only relevant component of the Riemann tensor is 
\be R_{\alpha\beta\gamma}{}^\delta = \frac{i}{96} \varepsilon_{abcd}{}^e \gamma^{ab}{}_{\alpha\beta} \gamma^{cd}{}_\gamma{}^\delta \, B_e \ ,  \ee
where $B_a$ is the unique G-analytic vector (\ref{GaB}). One straightforwardly computes that 
\be R_{\eta[\alpha\beta}{}^\varsigma R_{\varsigma|\gamma\delta]}{}^\eta = 0 \ ,  \ee
and therefore 
\be \,E = \cE \Scal{ 1 - \frac{1}{24} \zeta^\beta \zeta^\alpha \gamma^a_{\alpha\beta} B_a \big| - \frac{1}{24} \varepsilon_{\alpha\beta\gamma\delta} \zeta^\alpha \zeta^\beta \zeta^\gamma \zeta^\delta \frac{1}{144} \eta^{ab} B_a B_b\big| } \ , \ee
and so the volume
\be \int d^5 x d^{32} \theta \,E  = - \frac{1}{144} \int d\mu_{\scriptscriptstyle (8,1)} \,   \eta^{ab} B_a B_b\big|  \ee
is not zero. Note, however, that this result is irrelevant for the ultraviolet behaviour of the theory, because the $\partial^8 R^4$ type invariant does not have the correct power counting to be a candidate counterterm in five dimensions.

In $\cN=2$ supergravity, the Riemann tensor satisfies the same equation, and is determined by the torsion supertrace $\chi_\alpha \equiv u^1{}_i \chi_\alpha^i$ as
\be R_{\alpha\beta\gamma}{}^\delta =  - \delta^\delta_{(\alpha} \chi_{\beta)} \chi_\gamma - \Omega_{\gamma(\alpha} \chi_{\beta)} \chi_\eta \Omega^{\eta\delta} + \frac{1}{2} \delta^\delta_{(\alpha} \Omega_{\beta)\gamma} \Omega^{\eta\varsigma} \chi_\eta \chi_\varsigma \ .\ee
One straightforwardly computes that 
\begin{gather}
D_\alpha \chi_\beta = - \frac{1}{2} \scal{ \chi_\alpha\chi_\beta - \Omega_{\alpha\beta} \Omega^{\gamma\delta} \chi_\gamma \chi_\delta} \  , \CR
D_{[\alpha} D_\beta \chi_{\gamma]} = - \frac{1}{2} \chi_\alpha \chi_\beta \chi_\gamma \ , \qquad D_{[\alpha} D_\beta D_\gamma \chi_{\delta]} = \frac{1}{4}  \chi_\alpha \chi_\beta \chi_\gamma \chi_\delta \ , 
\end{gather}
and 
\be R_{\gamma\alpha\beta}{}^\gamma = - 3 \scal{ \chi_\alpha\chi_\beta -\tfrac{1}{4}  \Omega_{\alpha\beta} \Omega^{\gamma\delta} \chi_\gamma \chi_\delta} \ , \ee
so that 
\be\begin{split} e_\alpha &= \chi_\alpha \big| \ , \\
e_{\alpha\beta\gamma} &= - \frac{5}{12} \chi_\alpha \chi_\beta \chi_\gamma \big| \ , 
\end{split}\hspace{10mm}\begin{split}
e_{\alpha\beta} &= - \frac{1}{4} \chi_\alpha\chi_\beta \big| + \frac{3}{8} \Omega_{\alpha\beta} \Omega^{\gamma\delta} \chi_\gamma \chi_\delta \big| \ , \\
e_{\alpha\beta\gamma\delta} &= 0 \ . 
\end{split}\ee
As a consistency check, one computes using this formula that 
\be \zeta^\alpha D_\alpha\mbox{ln}[ \,E] = \zeta^\alpha \chi_\alpha \big| - \frac{3}{2} ( \zeta^\alpha \chi_\alpha)^2 \big| + \frac{3}{4} \zeta^\beta \zeta^\alpha \Omega_{\alpha\beta} \Omega^{\gamma\delta} \chi_\gamma \chi_\delta \big| - \frac{1}{4} ( \zeta^\alpha \chi_\alpha)^3 \big| + \frac{1}{24} ( \zeta^\alpha \chi_\alpha)^4 \big| \ee
indeed coincides with formula (\ref{LogExp}). This may be a surprise: the volume of superspace does not vanish on-shell  in maximal supergravity, but it does in the half-maximal theory as a consequence of the contribution of the non-trivial torsion supertrace.

Using this formula, one can now integrate an arbitrary function $K$ of the dilaton superfield $\Phi$ over full superspace to obtain
\be \int d^5 x d^{16} \theta \,E K(\Phi) = \frac{1}{24} \int d\mu_{\scriptscriptstyle (4,1)} \varepsilon^{\alpha\beta\gamma\delta}  \chi_\alpha \chi_\beta \chi_\gamma \chi_\delta ( \partial - 3 ) ( \partial - \tfrac{3}{2} ) \partial ( \partial+ \tfrac{3}{2}) K(\Phi) \ .  \ee
Since the integrand starts at 4-points, one can use the linearised analysis to compute that such an invariant contains a term in $R^4$ of the form
\bea &&  \int d^{16} \theta \,E K(\Phi) \CR
&\sim& \varepsilon^{\alpha\beta\gamma\delta} \varepsilon^{\eta\varsigma\vartheta\iota} \varepsilon^{\upsilon\kappa\xi\zeta} \varepsilon^{\varpi\lambda\mu\nu} C_{\alpha\eta\upsilon\varpi} C_{\beta\varsigma\kappa\lambda} C_{\gamma\vartheta\xi\mu} C_{\delta\iota\zeta\nu}( \partial - 3 ) ( \partial - \tfrac{3}{2} ) \partial ( \partial+ \tfrac{3}{2})  K(\Phi)  + \dots \quad \quad
\eea
where $C_{\alpha\beta\gamma\delta} \equiv \frac{1}{16} \gamma^{ab}_{(\alpha\beta} \gamma^{cd}_{\gamma\delta)} R_{abcd} $ is the Weyl tensor. Clearly this integral vanishes if $K = e^{3\Phi},\  e^{\pm\frac{3}{2} \Phi}$ or if it is a constant. Therefore the integral of the $\cN=2$ supervielbein Berezinian vanishes. However, one can still write the only invariant that preserves the dilaton shift symmetry as the full-superspace integral
\be \int d^5 x d^{16} \theta \,E \Phi = \frac{9}{32} \int d\mu_{\scriptscriptstyle (4,1)} \varepsilon^{\alpha\beta\gamma\delta}  \chi_\alpha \chi_\beta \chi_\gamma \chi_\delta\ .  \ee
This invariant satisfies all the required symmetries of the theory and is a full-superspace integral, so that there is no obvious non-renormalisation associated with it. However, we shall argue in the last section that the same reasoning that we sketched within supergravity in four dimensions would in principle imply that such an invariant would be forbidden by a non-renormalisation theorem within a hypothetical formulation of the theory in superspace with all 16 supercharges realised linearly.

The situation remains very similar in the presence of vector multiplets. We shall not display the explicit components of the Riemann tensor and the torsion in the presence of vector multiplets in this paper. The latter can be straightforwardly  extracted from the $\cN=4$ ones for $n=5$ vector multiplets, and, using the property that the vector multiplets only carry  $SO(5)$ vector indices contracted using the corresponding metric $\delta_{AB}$, one can then straightforwardly generalise all formulae to any number of vector multiplets. Here, we shall only explain how to extract the quantities that are important for the computation of the superspace volume. In this case the bilinears in the Dirac fermions also include the matter-field Dirac fermions $\lambda_\alpha^A$ and one has 
\be N_{\alpha\beta}^{i,j} = 3 \chi_{[\alpha}^i \chi_{\beta]}^j + \lambda_{[\alpha}^{Ai} \lambda_{\beta]A}^j \ , \quad N_{\alpha\beta}^{i,jkl} = - 3 \Omega^{[kl} \chi_{[\alpha}^{j]} \chi_{\beta]}^i+  \Omega^{[kl} \lambda_{[\alpha}^{Aj]} \lambda_{\beta]A}^i \ .\ee
Using these expressions one computes that $\chi_\alpha \equiv u^1{}_i \chi_\alpha^i$ and $\lambda_\alpha^A \equiv u^1{}_i \lambda^{Ai}_\alpha$ satisfy  
\bea D_\alpha \chi_\beta &=&- \frac{1}{2} \scal{ \chi_\alpha\chi_\beta - \Omega_{\alpha\beta} \Omega^{\gamma\delta} \chi_\gamma \chi_\delta} + \frac{1}{6} \scal{ \lambda^A_\alpha\lambda_{\beta A} - \Omega_{\alpha\beta} \Omega^{\gamma\delta} \lambda^A_\gamma \lambda_{\delta A} } \ ,  \CR
D_\alpha \lambda_\beta^A &=& \chi_{[\alpha} \lambda^A_{\beta]}  - \Omega_{\alpha\beta} \Omega^{\gamma\delta} \chi_\gamma \lambda^A_\delta \ ,  \eea
and therefore
\bea D_{[\alpha} D_\beta \chi_{\gamma]} &=&- \frac{1}{2} \chi_\alpha \chi_\beta \chi_\gamma  - \frac{1}{2} \chi_{[\alpha} \lambda_\beta^A \lambda_{\gamma]A} + \frac{1}{2} \Omega_{[\alpha\beta} \chi_{\gamma]} \Omega^{\delta\eta} \lambda_\delta^A \lambda_{\eta A} \ , \\
D_{[\alpha} D_\beta D_\gamma \chi_{\delta]} &=& \frac{1}{4} \chi_\alpha \chi_\beta \chi_\gamma \chi_\delta + \frac{3}{2}  \chi_{[\alpha} \chi_\beta  \lambda_\gamma^A \lambda_{\delta]A} - \frac{1}{12} \lambda_{[\alpha}^A \lambda_{\beta A} \lambda^B_\gamma \lambda_{\delta] B} - \frac{1}{6} \Omega_{[\alpha\beta} \lambda_\gamma^A \lambda_{\delta]A} \Omega^{\eta\varsigma} \lambda_\eta^B \lambda_{\varsigma B}\nn \ . \eea
Together with 
\be R_{\gamma\alpha\beta}{}^\gamma = - 3 \chi_\alpha \chi_\beta - \lambda_\alpha^A \lambda_{\beta A} + \frac{1}{4} \Omega_{\alpha\beta} \Omega^{\gamma\delta}  \scal{  3 \chi_\gamma \chi_\delta + \lambda_\gamma^A \lambda_{\delta A}} \ , \ee
this gives 
\bea e_\alpha &=& \chi_\alpha \big| \ , \CR
e_{\alpha\beta} &=& - \frac{1}{4} \chi_\alpha \chi_\beta\big| + \frac{3}{8} \Omega_{\alpha\beta} \Omega^{\gamma\delta} \chi_\gamma \chi_\delta\big| - \frac{1}{12} \lambda_\alpha^A \lambda_{\beta A} \big|- \frac{1}{24} \Omega_{\alpha\beta} \Omega^{\gamma\delta} \lambda_\gamma^A \lambda_{\delta A} \big|\ , \CR
e_{\alpha\beta\gamma} &=& - \frac{5}{12} \chi_\alpha \chi_\beta \chi_\gamma \big| - \frac{1}{6} \chi_{[\alpha} \lambda_\beta^A \lambda_{\gamma]A}\big| + \frac{1}{24} \Omega_{[\alpha\beta} \chi_{\gamma]} \Omega^{\delta \eta} \lambda_\delta^A \lambda_{\eta A}\big|\ ,  \CR
e_{\alpha\beta\gamma\delta} &=& 0\ . \eea
The volume therefore still vanishes on-shell  in the presence of vector multiplets. The full-superspace integral of a function of the dilaton $K$ is expressible as a $(4,1)$ harmonic-superspacee integral by
\begin{multline} \int d^5 x d^{16} \theta \,E K(\Phi) = \frac{1}{8} \int d\mu_{\scriptscriptstyle (4,1)}\Omega^{\alpha\beta} \Omega^{\gamma\delta} \Bigl(   \chi_\alpha \chi_\beta \chi_\gamma \chi_\delta ( \partial - 3 ) ( \partial - \tfrac{3}{2} )  ( \partial + \tfrac{3}{2}) \Bigr . \\ \Bigl . 
- \chi_{[\alpha} \chi_\beta \lambda_{\gamma}^A \lambda_{\delta] A} ( \partial - \tfrac{3}{2} ) ( \partial + 3 ) + \frac{2}{3}  \chi_\alpha \chi_{\beta} \lambda_\gamma^A \lambda_{\delta A} ( \partial - \tfrac{3}{2} ) \partial \Bigr . \\ \Bigl .  +\frac{1}{12} \lambda_{[\alpha}^A \lambda_{\beta A} \lambda^B_\gamma \lambda_{\delta]B} ( \partial + 3) + \frac{1}{9} \lambda^A_\alpha \lambda_{\beta A} \lambda^B_\gamma \lambda_{\delta B} ( \partial - \tfrac{3}{2}) \Bigr) \partial K(\Phi) \ . \end{multline} 
In particular, the integral of the dilaton is 
\begin{multline} \int d^5 x d^{16} \theta \,E \Phi = \frac{1}{16} \int d\mu_{\scriptscriptstyle (4,1)}\Omega^{\alpha\beta} \Omega^{\gamma\delta} \Bigl(   \frac{27}{2}  \chi_\alpha \chi_\beta \chi_\gamma \chi_\delta + 9 \chi_{[\alpha} \chi_\beta \lambda_{\gamma}^A \lambda_{\delta] A}  \Bigr . \\ \Bigl .  + \frac{1}{2} \lambda_{[\alpha}^A \lambda_{\beta A} \lambda^B_\gamma \lambda_{\delta]B} - \frac{1}{3} \lambda^A_\alpha \lambda_{\beta A} \lambda^B_\gamma \lambda_{\delta B} \Bigr) \ .\label{5DR4invariant} \end{multline} 
In the presence of vector multiplets, the most general $SO(5,n)$ invariant $G$-analytic integrand of mass dimension two is 
\begin{multline} \Omega^{\alpha\beta} \Omega^{\gamma\delta} \Bigl(   \chi_\alpha \chi_\beta \chi_\gamma \chi_\delta ( \partial - 3 ) ( \partial - \tfrac{3}{2} )  ( \partial + \tfrac{3}{2}) \Bigr . \\ \Bigl . 
- \chi_{[\alpha} \chi_\beta \lambda_{\gamma}^A \lambda_{\delta] A} ( \partial - \tfrac{3}{2} ) ( \partial + 3 ) + \frac{2}{3}  \chi_\alpha \chi_{\beta} \lambda_\gamma^A \lambda_{\delta A} ( \partial - \tfrac{3}{2} ) \partial \Bigr . \\ \Bigl .  +\frac{1}{12} \lambda_{[\alpha}^A \lambda_{\beta A} \lambda^B_\gamma \lambda_{\delta]B} ( \partial + 3) + \frac{1}{9} \lambda^A_\alpha \lambda_{\beta A} \lambda^B_\gamma \lambda_{\delta B} ( \partial - \tfrac{3}{2}) \Bigr) G(\Phi) \, ,  \end{multline}
so that all such $(4,1)$ integrals can be expressed as full-superspace integrals of a primitive $K(\Phi)$ of $G(\Phi)$.

Note that the field $B_a$ defining the component $R_{\alpha\beta\gamma}{}^\delta$ of the Riemann tensor is the unique G-analytic vector, as in four dimensions, but its square is a G-analytic function whereas the integrand of the harmonic measure must be a G-analytic section (\ref{GASection}) because of the non-vanishing torsion supertrace. Therefore there is no associated duality invariant in five dimensions. In fact one checks that for $G(\Phi) = e^{\frac{3}{2} \Phi}$ one obtains an invariant that only depends on the matter fields in the quartic approximation. Through dimensional reduction, the latter gives rise to an $SL(2,\mathds{R})$ invariant in four dimensions that depends non-trivially on the matter scalar fields. We conclude that the $(4,1,1)$ superspace integral (\ref{d4F4411}) does not lift to five dimensions.

In conclusion, $\cN=2$ supergravity coupled to $n$ vector multiplets in five dimensions admits only one invariant candidate that is invariant with respect to a shift of the dilaton superfield and that can be written as a $(4,1)$ harmonic-superspacee integral. It can also be written as the full-superspace integral of the dilaton superfield itself. 

\subsection{Protected invariants}
We have discussed invariants that can be written as $(4,1)$ harmonic-superspacee integrals in the last section, but they do not exhaust all the possible $R^4$ type invariants one can write in five dimensions. Similarly to four dimensions, some $R^4$ invariants can only be written as $(4,2)$ harmonic-superspacee integrals. The main difference in five dimensions is that one can already distinguish these invariants by the structure of the four-graviton amplitude, because there are two distinguished $Sp(1,1)$ quartic invariants in the Weyl tensor that define supersymmetry invariants. 

The $(4,2)$ harmonic variables $u^r{}_i, \, u_{r i}$ parametrise the symmetric space $Sp(2) / U(2)$, and satisfy
\be u^r{}_i u^s{}_j \Omega^{ij} = 0 \ , \qquad u^r{}_i u_{s j} \Omega^{ij} = \delta^r_s \ ,\ee
and the reality condition $\bar u^{i}_r = \Omega^{ij}  u_{r j}$. As for the $(4,1)$ harmonic superspace, the non-vanishing supertrace of the torsion defines a flat connection (where we use (\ref{Dchi5}))
 \be u^r{}_i  D_\alpha^i  \, u^s{}_j  \chi_\beta^j  + u^s{}_j  D_\beta^j \,  u^r{}_i \chi_\alpha^i  = 0 \label{FlatCon2} \ , \ee
 and the appropriate G-analyticity condition for an integrand $\cF_{\scriptscriptstyle (4,2)}$ of the $(4,2)$ measure is that it satisfies
 \be u^r{}_i \scal{  D_\alpha^i + \chi_\alpha^i } \cF_{\scriptscriptstyle (4,2)} = 0 \ . \label{GASection2} \ee

 In the linearised approximation, one can define the G-analytic superfield 
 \be M_{\alpha\beta}^{12} = \frac{1}{4} \gamma^{ab}{}_{\alpha\beta} u^1{}_i u^2{}_j M^{\prime ij}_{ab} \ , \ee
 where we chose to define it in terms of the $Sp(1,1)$ spinor indices for convenience. Any quartic polynomial in $M^{12}_{\alpha\beta}$ therefore defines a G-analytic integrand for the $(4,2)$ measure in the linearised approximation. One straightforwardly checks that 
 \bea && \int d\mu_{\scriptscriptstyle (4,1)} \, \varepsilon^{\alpha\beta\gamma\delta} \varepsilon^{\eta\varsigma\vartheta\iota} M_{\alpha\eta}^{12} M_{\beta\varsigma}^{12} M_{\gamma\vartheta}^{12} M^{12}_{\delta\iota} \;  \propto\;  \int d^5 x d^{16} \theta \, \Phi^4 \CR
 &\propto&  \int d^5 x\; \scal{  \varepsilon^{\alpha\beta\gamma\delta} \varepsilon^{\eta\varsigma\vartheta\iota} \varepsilon^{\upsilon\kappa\xi\zeta} \varepsilon^{\varpi\lambda\mu\nu} C_{\alpha\eta\upsilon\varpi} C_{\beta\varsigma\kappa\lambda} C_{\gamma\vartheta\xi\mu} C_{\delta\iota\zeta\nu} + \dots } \ .  \eea
This class of invariant clearly corresponds to full-superspace integrals, and must admit a non-linear form for an arbitrary function of the dilaton that is not $e^{\pm\frac{3}{2} \Phi}$, $e^{3 \Phi}$ or $1$.   
 
One has also the additional linearised invariant 
\bea &&  \int d\mu_{\scriptscriptstyle (4,1)} \, \scal{ \Omega^{\alpha\gamma} \Omega^{\beta\delta} M_{\alpha\beta}^{12} M_{\gamma\delta}^{12} }^2 \CR
&\propto& \int d^5 x\; \scal{  \varepsilon^{\alpha\beta\gamma\delta} \varepsilon^{\eta\varsigma\vartheta\iota} ( \Omega^{\upsilon\kappa} \Omega^{\varpi\lambda}  C_{\alpha\eta\upsilon\varpi} C_{\beta\varsigma\kappa\lambda} ) ( \Omega^{\xi\zeta} \Omega^{\mu\nu} C_{\gamma\vartheta\xi\mu} C_{\delta\iota\zeta\nu} )  + \dots } \ ,  \eea
which cannot be written as a full-superspace integral, even at the linearised level. We will not prove the existence of such a G-analytic integrand in this paper, but one can infer, from the existence of the two ten-dimensional Chern--Simons type invariants associated to the gauge anomaly, that it exists at least for a power of the dilation of $e^{-\frac{3}{2} \Phi}$.

However, it is not clear if one can define an independent duality invariant with this structure. Equivalently, one may wonder if one can define independent invariants involving the matter fields, which could only be written as $(4,2)$ superspace integrals. We will not answer this question in this paper. Nonetheless we note that such an invariant, even if it existed, would be more constrained by non-renormalisation theorems. 

\subsection{Consequences for logarithmic divergences}
The situation for divergences in five dimensions is very similar to the one in four. There is only one available duality-invariant counterterm at two loops (\ref{5DR4invariant}) which can be written as a $(4,1)$ harmonic-superspacee integral. It can also be written as a full-superspace integral, but not of a duality-invariant integrand. It has been recently computed \cite{Bern:2012gh} that the associated UV divergence is indeed absent in $\cN=2$ supergravity, and there are hints from string theory suggesting that this result should apply independently of the number of vector multiplets.\footnote{We are grateful to Pierre Vanhove for this comment.} 

Before arguing that the result of the computation \cite{Bern:2012gh} may in principle be explained by a non-renormalisation theorem, let us point out  that the uniqueness of the invariant (\ref{5DR4invariant}) implies in principle that the finiteness of the four-graviton scattering amplitude in five dimensions extends to all scattering amplitudes at the two-loop order, including higher-point amplitudes and ones with external vector-multiplet states. Indeed, the only possible alternative invariants can be written only as $(4,2)$ harmonic-superspacee integrals, and there are several reasons to believe that they cannot support logarithmic divergences. First of all, there exists an $\cN=3$ harmonic-superspacee formulation of Yang--Mills theory in four dimensions, and so extrapolating it to five-dimensional $\cN=2$ supergravity, one expects only invariants that can be written at least as $\int d^{12} \theta $ superspace integrals to contribute to logarithmic divergences. In components, one knows that genuine $(4,2)$ superspace integrals are associated to long cocycles, and therefore, applying the algebraic renormalisation arguments of \cite{Bossard:2009sy}, one would conclude as well that such invariants cannot be associated to logarithmic divergences. 

A similar argument to that given in \cite{Rivelles:1982gn} for the existence of auxiliary fields in four-dimensional theories implies that $\cN=2$ supergravity in five dimensions can only admit an off-shell realisation with finitely many auxiliary fields when coupled to five modulo eight vector multiplets. In particular, the linearised theory with five vector multiplets can be obtained by dimensional reduction of the off-shell formulation of ten-dimensional  supergravity \cite{Howe:1982mt}. Nonetheless, the argument of  \cite{Rivelles:1982gn} does not rule out the existence of an off-shell formulation of the theory in harmonic superspace for an arbitrary number of vector multiplets. In this subsection we shall assume that such an off-shell formulation exists. One must note that the situation is much simpler in five dimensions, in the sense that realising the shift symmetry of the dilaton does not require us to consider a Lorentz-harmonic formulation of the theory. Moreover, if we rely on the $SO(n)$ symmetry to fix uniquely the allowed candidate counterterm, we do not require this symmetry at the level of the integrand in order to prove the non-renormalisation theorem. Therefore, the existence of an off-shell formulation of the theory for $n=5 + 8k$ vector multiplets would be enough to prove the non-renormalisation theorem in these special cases. In such a conventional superspace formulation of the theory, the non-renormalisation theorem sketched in Section 4.4 would apply directly. Moreover, owing to the property that the coefficient for the logarithmic divergence is necessarily a polynomial in the number of vector multiplets, its vanishing for $n=5$ modulo $8$ then implies its vanishing for all $n$.

Assuming the existence of such a formulation of the theory in superspace with all supercharges realised linearly, the beta function associated to a potential two-loop divergence will be, using the same argument as in Section \ref{ImplicationUV} \cite{Bossard:2009sy}, equal to the anomalous dimension of the classical Lagrange density in superspace for mixing under renormalisation with the density $\,E \Phi$. But the variation of this integrand with respect to a dilaton shift gives rise to an integrand with vanishing integral, \ie the Berezinian of the supervielbeins. In an off-shell formulation, the latter will be a total derivative of a degree one co-form that will necessarily depend non-trivially on a prepotential. Assuming the existence of Feynman rules within the background field method that lead only to possible logarithmic divergences associated with functions of the potentials themselves (and not prepotentials), we conclude that the associated anomalous dimension must vanish. And in consequence,  so must the beta function.  

Note that although the shift symmetry of the dilaton is subject to potential anomalies, there is no supersymmetry invariant with the appropriate power counting to define a one-loop anomaly for the shift symmetry. Although one expects this symmetry to become anomalous at two loops, this would not affect the potential logarithmic divergences at this order.

\subsection{$D=5$ Heterotic string theory non-renormalisation theorem}
\label{Heterotic} 
Considering the full-superspace integral of a general function of both the dilaton $\Phi$ and the scalar fields $t^m$ parametrising the $SO(5,n)/(SO(5) \times SO(n))$ symmetric space, one computes in the same way that the only contribution to the $R^4$ coupling comes from the term 
\be \Scal{ \frac{ \partial \ }{\partial \phi} - 3 } \Scal{ \frac{ \partial \ }{\partial \phi} - \frac{3}{2} } \Scal{ \frac{ \partial \ }{\partial \phi} }  \Scal{ \frac{ \partial \ }{\partial \phi} + \frac{3}{2}  } G(\phi,t^m) \ . \ee 
In perturbative heterotic string theory, the $\ell$-loop contribution to the effective action $R^4$ coupling in five dimensions appears with a factor 
\be G_\ell(\phi,t^m) = e^{\frac{3}{2} ( \ell - 2 )  \phi }  K(t^m)  \ . \ee
It follows that the one-, two-, three- and four-loop contributions to the effective action $R^4$ coupling cannot be written as full-superspace integrals in five dimensions. In general, they cannot be written as $(4,1)$ harmonic-superspacee integrals either, except in the marginal case $\ell = 2$ and $K(t^m) = 1$. Therefore such couplings can only be defined as $(4,2)$ harmonic-superspacee integrals, or as closed superforms, and can be considered as being one-half BPS protected. 

It is striking that we obtain precisely the same conclusion in four dimensions, for which the $R^4$ coupling of the full-superspace integral of a function $G(\tau,t^m)$ of the complex scalar $\tau$ and the scalar fields $t^m$ parametrising the $SO(6,n)/(SO(6) \times SO(n))$ symmetric space is multiplied by 
\be \scal{ \Delta - 2 } \Delta G(\tau,t^m) \ . \ee
In perturbative heterotic string theory, the $\ell$-loop contribution to the effective action $R^4$ coupling in four dimensions appears with a factor 
\be G_\ell(\phi,t^m) = e^{2( \ell - 3 )  \phi }  K(t^m)  \ , \ee
with $\tau = a + i e^{-2\phi}$. It follows that the one-, two-, three- and four-loop contributions to the effective action $R^4$ coupling cannot be written as full-superspace integrals in four dimensions either. 

The fact that the string-theory interpretation is the same in both four and five dimensions suggests that the same property should hold in ten dimensions, \ie 
\begin{multline}  \int d^{10} x d^{16} \theta \, \,E \, G(\Phi)  \\ \sim \int d^{10} x \sqrt{-g} \Scal{ \frac{ \partial \ }{\partial \phi} - \frac{1}{2}  } \Scal{ \frac{ \partial \ }{\partial \phi} - \frac{5}{2} } \Scal{ \frac{ \partial \ }{\partial \phi} - \frac{9}{2} }  \Scal{ \frac{ \partial \ }{\partial \phi} - \frac{13}{2}  } G(\Phi) R^4 + \dots  \end{multline} 
Since the $\ell$-loop contribution gives rise to a factor 
\be G_\ell(\phi) = e^{( 2 \ell - \frac{3}{2} ) \phi }  \ , \ee
it would then follow that the $R^4$ coupling cannot appear as a full-superspace integral before five loops in ten dimensions. If it were possible to write $e^{( 2 \ell - \frac{3}{2} ) \phi } R^4$ couplings for $\ell = 1,2,3,4$ as full-superspace integrals in ten dimensions, then one would obtain by dimensional reduction that the corresponding dimensionally reduced invariants are also full-superspace integrals, which is in contradiction with our results. 

The only other invariants that include an $R^4$ coupling are the Chern--Simons like invariants that are obtained from the $d$-exact  11-superforms 
\be H \wedge R^a{}_b \wedge R^b{}_c \wedge R^c{}_d \wedge R^d{}_a \ , \quad   H \wedge R^{ab} \wedge R_{ab} \wedge R^{cd} \wedge R_{cd} \ . \ee
These can only appear at one loop in string theory, because their $R^4$ couplings come with a factor $e^{\frac{1}{2}\phi}$. This property is understood in string theory because these invariants are required as counterterms in order to cancel the gauge anomaly, which is itself subject to a non-renormalisation theorem  \cite{Yasuda:1988fi}. The explicit computation in string theory indeed confirms that there is no $R^4$ correction to the affective action at the 2-loop level \cite{D'Hoker:2005jc}, whereas only the Chern--Simons invariants appear at the 1-loop level. Nonetheless, it has been argued in \cite{Tseytlin:1995bi} that this state of affairs cannot extend to all orders in perturbation theory, because it would be in contradiction with heterotic / type I duality. 

The structure of the supersymmetry invariants in four and five dimensions shows that $R^4$ couplings can only be considered as being protected  until four loops in string theory, and the above results suggest that the non-renormalisation theorem for the $t_8 t_8 R^4 - \tfrac{1}{8} \varepsilon_{10} \varepsilon_{10} R^4$ term in the effective action will apply until four loops, but not beyond.

%%%%%%%%%%%%%%%%%%
\section{Conclusions}

In this paper we have discussed the possible ultra-violet divergences that can arise in half-maximal supergravity theories at three and two loops, in $D=4$ and $5$ respectively. We have shown, provided that  some assumptions regarding off-shell formalisms are made, that the pure half-maximal supergravity theories should be finite in these cases, in agreement with the amplitude results and string theory. In the presence of vector multiplets, this conclusion remains unchanged in $D=5$, but cannot be justified in $D=4$ owing to appearance of an $F^4$ term in the $SL(2,\bbR)$ anomaly. 

The key observation is that, although the candidate counterterms seem superficially to be F-terms, they can be rewritten as D-terms,
\ie integrals over the full sixteen-theta superspaces. The fact that the volume of superspace vanishes, for both $D=4$ and 5, implies that these full-superspace integrals are duality-invariant. There are, however, no candidate counterterms with manifestly duality-invariant full-superspace integrands. The relevant duality-invariant integrals can either be written as full-superspace integrals of integrands that are not themselves invariant, or they can be written as sub-superspace integrals of invariant integrands. We have called this situation the F/D borderline, since the status of these invariants is ambiguous. Given the existence of suitable off-shell versions of the theories preserving all of the supersymmetries linearly, as well as duality, we have argued that the F-term character wins out and that these invariants are therefore protected. This result is vitiated in the case of vector multiplets in $D=4$ because the three-loop counterterm no longer needs to be fully duality-invariant.

The recognition of F/D marginal structure further expands the class of special invariant structures that have a bearing on non-renormalisation properties of supersymmetric theories. By combining duality properties with supersymmetry structure, they expand the class of Chern--Simons-type invariants vulnerable to exclusion as candidate counterterms. Other examples of special structure that have been found in the now decade-long to- and fro- discussion of counterterm analysis versus unitarity-method loop calculations include special cohomology types in higher-dimensional super Yang--Mills theory \cite{Bossard:2010pk}. 

For the future, it would clearly be of interest to construct the off-shell formalisms whose existence we have relied upon in our arguments. This is not an easy problem. We know that there is an off-shell version of $\cN=4$ Yang--Mills theory in harmonic superspace, but that it only has linearly realised $\cN=3$ supersymmetry. Moreover, this construction is rather special in that it relies upon the fact that the harmonic coset can be regarded as three-dimensional, so that the Chern--Simons action in this sector can be used to set the corresponding field strength to zero,  thereby leading to the usual constraints in ordinary superspace, which are well-known to imply the equations of motion. So far, no other construction of this type has been made, except for the closely related $D=3,\ \cN=6$ Yang--Mills where such a construction leads to an off-shell version of non-abelian Chern--Simons theory \cite{Howe:1994ms}. An additional complication is the requirement that duality symmetry be preserved. In four dimensions this is incompatible with manifest Lorentz symmetry and is therefore likely to require the use of Lorentz harmonics as well as those associated with R-symmetry. It may therefore be easier to try to tackle the five-dimensional case first where this last problem does not arise.

Whether or not the above above programme can be implemented successfully, it seems difficult to imagine any purely field-theoretic argument that could protect yet higher-loop counterterms against ultra-violet divergences. This is because there are no further obstructions to the construction of counterterms that are manifestly invariant under all symmetries. This being the case, there is an obvious challenge on the computational side. If it turns out that, {\it e.g.}, $\cN=4,\ D=4$ supergravity is finite at four loops, then all bets would be off regarding the perturbative finiteness of $\cN=8$ supergravity.

It is also important to stress the implications of the counterterm structures that we find, without relying on a hypothetical full-superspace off-shell formulation of the theory. In pure supergravity in four dimensions, we have shown that there is a unique duality-invariant candidate counterterm at three loops. It  therefore follows that the vanishing of the four-graviton amplitude at this order  \cite{Bern:2012cd}, implies the finiteness of all amplitudes at that order. In the presence of vector multiplets, in addition to the one-loop $F^4$ divergence, we find a unique potential two-loop candidate of generic form $\partial^2 F^4$ (whose invariance under $SO(6,n)$ still remains to be established), and only two independent duality invariants at three loops,  involving either $R^4$ or $\partial^4 F^4$. Moreover, the one-loop $SL(2,\bbR)$  anomaly allows in principle for one additional non-duality-invariant $R^4$ type counterterm. According to the string-theory arguments of \cite{Tourkine:2012ip}, which are further strengthened by our analysis of Section  \ref{Heterotic}, the four-graviton amplitude is also finite in the presence of vector multiplets. The four-matter-photon amplitude would  in principle be allowed to admit genuine two- and three-loop divergences should there be no full off-shell formulation of the theory with sixteen supercharges realised linearly.  It should be possible to check computationally whether or not these amplitudes diverge, which would in turn shed light on the validity of the conjecture regarding the existence of an off-shell formulation of the theory.   

In five dimensions there is a unique duality-invariant two-loop candidate, which can be written as a harmonic-superspacee integral over twelve of the sixteen fermionic coordinates. It can also be written as the full-superspace integral of the dilaton superfield. Assuming that there exists an off-shell $D=5$ formulation of the theory with at least twelve supercharges realised linearly, one could already conclude that the finiteness of the four-graviton amplitude \cite{Bern:2012gh} implies the finiteness of all amplitudes at two loops. If the four-graviton amplitude were finite but if the four-matter-photon amplitude were to diverge in half-maximal $\cN=2$ supergravity in five dimensions, one would then be able to infer that there is no off-shell formulation of the theory with more than eight supercharges realised linearly. 

What lesson do we learn from this analysis concerning $\cN=8$ supergravity? First of all, note that the $\cN=4$ theory's property that the duality invariant $(4,1,1)$ harmonic-superspacee integral can be rewritten as the sixteen-theta full-superspace integral of a function of the complex scalar is only possible because the latter is chiral. It therefore appears that the duality-invariant seven-loop candidate counterterm expressed as a $(8,1,1)$ harmonic-superspacee integral \cite{Bossard:2011tq} cannot be rewritten as a thirty-two-theta full-superspace integral in $\cN=8$ supergravity. Although an off-shell formulation of the maximally supersymmetric theory with all thirty-two supercharges realised linearly would then permit one to conclude that the theory must be finite until eight loops, such a formalism is extremely unlikely to exist.

%%%%%%%%%%%%%%%%%%%%%%%%%%%%%%%%%%%%%%%%%%%%%%%%%%%%%%%%%%%%%%%%
\section*{Acknowledgements}
%%%%%%%%%%%%%%%%%%%%%%%%%%%%%%%%%%%%%%%%%%%%%%%%%%%%%%%%%%%%%%%%
We  would like  to thank Costas Bachas, Zvi Bern, Tristan Dennen, Emery Sokatchev,  Piotr Tourkine and Pierre Vanhove for useful discussions. G.B. and K.S.S. would like to thank INFN Frascati, and K.S.S. would like to thank The Mitchell Institute, Texas A\& M University, for hospitality during the course of the work. The work of G.B. was supported in part by the French ANR contract 05-BLAN-NT09-573739, the ERC Advanced Grant no. 226371 and the ITN programme PITN-GA-2009-237920. The work of K.S.S. was supported in part by the STFC under consolidated grant ST/J000353/1.

%%%%%%%%%%%%%%%%%%%%%%%%%%%%%%%%%%%%%%%%%%%%%%%%%%%%%%%%%%%%%%%%
%%%%%%%%%%%%%%%%%%%%%%%%%%%%%%%%%%%%%%%%%%%%%%%%%%%%%%%%%%%%%%%%


\begin{thebibliography}{10}

%\cite{Bern:2007hh}
\bibitem{Bern:2007hh}
  Z.~Bern, J.~J.~Carrasco, L.~J.~Dixon, H.~Johansson, D.~A.~Kosower and R.~Roiban,
  ``Three-loop superfiniteness of $\cN=8$ Supergravity,''
  Phys.\ Rev.\ Lett.\  {\bf 98} (2007) 161303 
  \eprint{hep-th/0702112}.
  %%CITATION = PRLTA,98,161303;%% 

%\cite{Bern:2009kd}
\bibitem{Bern:2009kd}
  Z.~Bern, J.~J.~Carrasco, L.~J.~Dixon, H.~Johansson and R.~Roiban,
  ``The ultraviolet behavior of $\cN=8$ supergravity at four loops,''
  Phys.\ Rev.\ Lett.\  {\bf 103} (2009) 081301,
  \eprintN{0905.2326}.
  %%CITATION = PRLTA,103,081301;%%

%\cite{Kallosh:1980fi}
\bibitem{Kallosh:1980fi}
  R.~E.~Kallosh,
  ``Counterterms in extended supergravities,''
  Phys.\ Lett.\  B {\bf 99} (1981) 122.
  %%CITATION = PHLTA,B99,122;%%

%\cite{Howe:1981xy}
\bibitem{Howe:1981xy}
  P.~S.~Howe, K.~S.~Stelle and P.~K.~Townsend,
  ``Superactions,''
  Nucl.\ Phys.\  B {\bf 191} (1981) 445.
  %%CITATION = NUPHA,B191,445;%%
  
%\cite{Bossard:2009sy}
\bibitem{Bossard:2009sy}
  G.~Bossard, P.~S.~Howe and K.~S.~Stelle,
  ``The ultra-violet question in maximally supersymmetric field theories,''
  Gen.\ Rel.\ Grav.\  {\bf 41} (2009) 919,
  \eprintN{0901.4661}.
  %%CITATION = GRGVA,41,919;%% 
  
%\cite{Bern:1998ug}
\bibitem{Bern:1998ug}
  Z.~Bern, L.~J.~Dixon, D.~C.~Dunbar, M.~Perelstein and J.~S.~Rozowsky,
  ``On the relationship between Yang--Mills theory and gravity and its
  implication for ultraviolet divergences,''
  Nucl.\ Phys.\  B {\bf 530}, 401 (1998)
  \eprint{hep-th/9802162}.
  %%CITATION = NUPHA,B530,401;%% 
  
  %\cite{Hillmann:2009zf}
\bibitem{Hillmann:2009zf} 
  C.~Hillmann,
  ``$E_{7(7)}$ invariant Lagrangian of $d=4$ $\cN=8$ supergravity,''
  JHEP {\bf 1004} (2010) 010,  
  \eprintN{0911.5225}.
  %%CITATION = ARXIV:0911.5225;%%
   
  
%\cite{Bossard:2010dq}
\bibitem{Bossard:2010dq}
  G.~Bossard, C.~Hillmann and H.~Nicolai,
  ``$E_{7(7)}$ symmetry in perturbatively quantised $\cN=8$ supergravity,''
  JHEP {\bf 1012} (2010) 052, 
  \eprintN{1007.5472}.
  %%CITATION = ARXIV:1007.5472;%% 
  

%\cite{Elvang:2010kc}
\bibitem{Elvang:2010kc}
H.~Elvang and M.~Kiermaier,
``Stringy KLT relations, global symmetries, and $E_{7(7)}$ violation,''
JHEP {\bf 1010} (2010) 108, 
\eprintN{1007.4813}.
%%CITATION = JHEPA,1101,020;%%

%\cite{Bossard:2010bd}
\bibitem{Bossard:2010bd}
G.~Bossard, P.~S.~Howe and K.~S.~Stelle,
``On duality symmetries of supergravity invariants,''
JHEP {\bf 1101} (2011) 020, 
\eprintN{1009.0743}.
%%CITATION = JHEPA,1010,108;%% %\cite{Bossard:2010bd} 

%\cite{Drummond:2003ex}
\bibitem{Drummond:2003ex}
J.~M.~Drummond, P.~J.~Heslop, P.~S.~Howe and S.~F.~Kerstan,
``Integral invariants in ${\mathcal{N}}\!=4$ SYM and the effective action for coincident D-branes,''
JHEP {\bf 0308} 2003) 016 
\eprint{hep-th/0305202}.
%%CITATION = JHEPA,0308,016;%%     



%\cite{Bossard:2011tq}
\bibitem{Bossard:2011tq} 
  G.~Bossard, P.~S.~Howe, K.~S.~Stelle and P.~Vanhove,
  ``The vanishing volume of $D=4$ superspace,''
  Class.\ Quant.\ Grav.\  {\bf 28} (2011) 215005,  
  \eprintN{1105.6087}.
  %%CITATION = ARXIV:1105.6087;%%



%\cite{Beisert:2010jx}
\bibitem{Beisert:2010jx}
N.~Beisert, H.~Elvang, D.~Z.~Freedman, M.~Kiermaier, A.~Morales and S.~Stieberger,
``$E_{7(7)}$ constraints on counterterms in ${\mathcal{N}}\!=8$ supergravity,''
Phys.\ Lett.\ B {\bf 694} (2010) 265, 
\eprintN{1009.1643}.
%%CITATION = PHLTA,B694,265;%%


%\cite{Green:2010sp}
\bibitem{Green:2010sp}
  M.~B.~Green, J.~G.~Russo and P.~Vanhove,
  ``String theory dualities and supergravity divergences,''
  JHEP {\bf 1006} (2010) 075, 
  \eprintN{1002.3805}.
  %%CITATION = JHEPA,1006,075;%%

%\cite{Green:2010kv}
\bibitem{Green:2010kv} 
  M.~B.~Green, S.~D.~Miller, J.~G.~Russo and P.~Vanhove,
  ``Eisenstein series for higher-rank groups and string theory amplitudes,''
  Commun.\ Num.\ Theor.\ Phys.\  {\bf 4} (2010) 551,
   \eprintN{1004.0163}.
  %%CITATION = ARXIV:1004.0163;%%
  
 
%\cite{Elvang:2010jv}
\bibitem{Elvang:2010jv}
H.~Elvang, D.~Z.~Freedman and M.~Kiermaier,
``A simple approach to counterterms in ${\mathcal{N}}\!=8$ supergravity,''
JHEP {\bf 1011} (2010) 016, 
\eprintN{1003.5018}.
%%CITATION = JHEPA,1011,016;%%


%\cite{Drummond:2010fp}
\bibitem{Drummond:2010fp}
J.~M.~Drummond, P.~J.~Heslop and P.~S.~Howe,
``A note on ${\mathcal{N}}\!=8$ counterterms,''
\eprintN{1008.4939}.
%%CITATION = ARXIV:1008.4939;%%


%\cite{Bern:2012cd}
\bibitem{Bern:2012cd}
  Z.~Bern, S.~Davies, T.~Dennen and Y.~-t.~Huang,
  ``Absence of three-loop four-point divergences in $\cN=4$ supergravity,''
  Phys.\ Rev.\ Lett.\  {\bf 108} (2012) 201301, 
  \eprintN{1202.3423}.
  %%CITATION = ARXIV:1202.3423;%%
 
  
%\cite{Tourkine:2012ip}
\bibitem{Tourkine:2012ip}
  P.~Tourkine and P.~Vanhove,
  ``An $R^4$ non-renormalisation theorem in $\cN=4$ supergravity,''
  Class.\ Quant.\ Grav.\  {\bf 29} (2012) 115006, 
  \eprintN{1202.3692}.
  %%CITATION = ARXIV:1202.3692;%%
  
%\cite{Tourkine:2012vx}
\bibitem{Tourkine:2012vx}
  P.~Tourkine and P.~Vanhove,
  ``One-loop four-graviton amplitudes in $\cN=4$ supergravity models,''
  Phys.\ Rev.\ D {\bf 87} (2013) 045001, 
  \eprintN{1208.1255}.
  %%CITATION = ARXIV:1208.1255;%%
  
  
  %\cite{Bern:2012gh}
\bibitem{Bern:2012gh}
  Z.~Bern, S.~Davies, T.~Dennen and Y.~-t.~Huang,
  ``Ultraviolet cancellations in half-maximal supergravity as a consequence of the double-copy structure,''
  Phys.\ Rev.\ D {\bf 86} (2012) 105014
  \eprintN{1209.2472}.
  %%CITATION = ARXIV:1209.2472;%%
  
 %\cite{Kallosh:2011dp}
\bibitem{Kallosh:2011dp}
  R.~Kallosh,
  ``$E_{7(7)}$ symmetry and finiteness of $\cN=8$ supergravity,''
  JHEP {\bf 1203} (2012) 083, 
 \eprintN{1103.4115}.
  %%CITATION = ARXIV:1103.4115;%%
  
%\cite{Kallosh:2011qt}
\bibitem{Kallosh:2011qt}
  R.~Kallosh,
  ``$\cN=8$ counterterms and $E_{7(7)}$ current conservation,''
  JHEP {\bf 1106} (2011) 073, 
  \eprintN{1104.5480}.
  %%CITATION = ARXIV:1104.5480;%%
  
\bibitem{Howe:1980th}
  P.~S.~Howe and U.~Lindstrom,
  ``Higher order invariants in extended supergravity,''
  Nucl.\ Phys.\ B {\bf 181} (1981) 487.
  %%CITATION = NUPHA,B181,487;%%   
   
  
 %\cite{Kallosh:2012ei}
\bibitem{Kallosh:2012ei}
  R.~Kallosh,
  ``On absence of 3-loop divergence in $\cN=4$ supergravity,''
  Phys.\ Rev.\ D {\bf 85} (2012) 081702, 
  \eprintN{1202.4690}.
  %%CITATION = ARXIV:1202.4690;%% 

%\cite{Ferrara:2012ui}
\bibitem{Ferrara:2012ui}
  S.~Ferrara, R.~Kallosh and A.~Van Proeyen,
  ``Conjecture on hidden superconformal symmetry of $\cN=4$ supergravity,''
  Phys.\ Rev.\ D {\bf 87} (2013) 025004, 
  \eprintN{1209.0418}.
  %%CITATION = ARXIV:1209.0418;%%
  
 %\cite{Marcus:1985yy}
\bibitem{Marcus:1985yy}
  N.~Marcus,
  ``Composite anomalies in supergravity,''
  Phys.\ Lett.\  B {\bf 157} (1985) 383.
  %%CITATION = PHLTA,B157,383;%%
  
  %\cite{Howe:1986ys}
\bibitem{Howe:1986ys} 
  P.~S.~Howe, G.~Papadopoulos and K.~S.~Stelle,
  ``Quantizing the $\cN=2$ super sigma model in two dimensions,''
  Phys.\ Lett.\ B {\bf 174}, 405 (1986).
  %%CITATION = PHLTA,B174,405;%%
  
 %\cite{Fischler:1979yk}
\bibitem{Fischler:1979yk}
  M.~Fischler,
  ``Finiteness calculations for $O(4)$ through $O(8)$ extended supergravity and $O(4)$ supergravity coupled to selfdual $O(4)$ matter,''
  Phys.\ Rev.\ D {\bf 20} (1979) 396.
  %%CITATION = PHRVA,D20,396;%%
  
 %\cite{Siegel:1981dx}
\bibitem{Siegel:1981dx}
  W.~Siegel and M.~Rocek,
  ``On off-shell supermultiplets,''
ÊÊPhys.\ Lett.\ B {\bf 105} (1981) 275.
ÊÊ%%CITATION = PHLTA,B105,275;%%

%\cite{Rivelles:1982gn}
\bibitem{Rivelles:1982gn}
  V.~O.~Rivelles and J.~G.~Taylor,
  ``Off-shell no go theorems for higher dimensional supersymmetries and supergravities,''
ÊÊPhys.\ Lett.\ B {\bf 121} (1983) 37.
ÊÊ%%CITATION = PHLTA,B121,37;%%

%\cite{Howe:1982mt}
\bibitem{Howe:1982mt}
  P.~S.~Howe, H.~Nicolai and A.~Van Proeyen,
  ``Auxiliary fields and a superspace Lagrangian for linearised ten-dimensional supergravity,''
ÊÊPhys.\ Lett.\ B {\bf 112} (1982) 446.
ÊÊ%%CITATION = PHLTA,B112,446;%%


%\cite{Galperin:1985uw}
\bibitem{Galperin:1985uw}
  A.~Galperin, E.~Ivanov, S.~Kalitsyn, V.~Ogievetsky and E.~Sokatchev,
  ``$\cN = 3$ supersymmetric gauge theory,''
  Phys.\ Lett.\  B {\bf 151}  (1985)  215.
  %%CITATION = PHLTA,B151,215;%%


%\cite{Howe:1985ar}
\bibitem{Howe:1985ar}
  P.~S.~Howe, K.~S.~Stelle and P.~C.~West,
  ``$\cN=1$ $d = 6$ harmonic superspace,''
ÊÊClass.\ Quant.\ Grav.\  {\bf 2} (1985) 815.
ÊÊ%%CITATION = CQGRD,2,815;%%

%\cite{Galperin:1984av}
\bibitem{Galperin:1984av}
  A.~Galperin, E.~Ivanov, S.~Kalitsyn, V.~Ogievetsky and E.~Sokatchev,
  ``Unconstrained $\cN=2$ matter, Yang--Mills and supergravity theories in harmonic
  superspace,''
  Class.\ Quant.\ Grav.\  {\bf 1}  (1984) 469.
  %%CITATION = CQGRD,1,469;%%

%\cite{Karlhede:1984vr}
\bibitem{Karlhede:1984vr}
  A.~Karlhede, U.~Lindstrom and M.~Rocek,
  ``Selfinteracting tensor multiplets In $\cN=2$ superspace,''
  Phys.\ Lett.\  B {\bf 147}  (1984)  297.
  %%CITATION = PHLTA,B147,297;%%
  
%\cite{Henneaux:1988gg}
\bibitem{Henneaux:1988gg}
  M.~Henneaux and C.~Teitelboim,
  ``Dynamics of chiral (selfdual) $p$-forms,''
  Phys.\ Lett.\ B {\bf 206} (1988) 650.
  %%CITATION = PHLTA,B206,650;%% 
  
%\cite{Sokatchev:1985tc}
\bibitem{Sokatchev:1985tc}
  E.~Sokatchev,
  ``Light cone harmonic superspace and its applications,''
  Phys.\ Lett.\ B {\bf 169} (1986) 209.
  %%CITATION = PHLTA,B169,209;%% 
  
%\cite{Delduc:1989ah}
\bibitem{Delduc:1989ah}
  F.~Delduc, S.~Kalitsyn and E.~Sokatchev,
  ``Learning the ABC of light cone harmonic space,''
  Class.\ Quant.\ Grav.\  {\bf 6} (1989) 1561.
  %%CITATION = CQGRD,6,1561;%% 
  
%\cite{Galperin:1991gk}
\bibitem{Galperin:1991gk}
  A.~S.~Galperin, P.~S.~Howe and K.~S.~Stelle,
  ``The superparticle and the Lorentz group,''
  Nucl.\ Phys.\ B {\bf 368} (1992) 248
  \eprint{hep-th/9201020}.
  %%CITATION = HEP-TH/9201020;%%  
  
%\cite{Bossard:2012xs}
\bibitem{Bossard:2012xs}
  G.~Bossard, P.~S.~Howe and K.~S.~Stelle,
  ``Anomalies and divergences in $\cN=4$ supergravity,''
  Phys.\ Lett.\ B {\bf 719} (2013) 424, 
  \eprintN{1212.0841}.
  %%CITATION = ARXIV:1212.0841;%% 
  
 %\cite{Howe:1981gz} 
\bibitem{Howe:1981gz} 
P.~S.~Howe, 
``Supergravity in superspace,'' 
Nucl.\ Phys.\  B {\bf 199} (1982) 309. 
%%CITATION = NUPHA,B199,309;%%  

%\cite{Howe:1980sy}
\bibitem{Howe:1980sy}
  P.~S.~Howe,
  ``A superspace approach to extended conformal supergravity,''
  Phys.\ Lett.\ B {\bf 100} (1981) 389.
  %%CITATION = PHLTA,B100,389;%%
  
%\cite{Gates:1982ae}
\bibitem{Gates:1982ae}
  S.~J.~Gates, Jr. and R.~Grimm,
  ``Consequences of conformally covariant constraints for $\cN>4$ superspace,''
  Phys.\ Lett.\ B {\bf 133} (1983) 192.
  %%CITATION = PHLTA,B133,192;%% 
  
%\cite{Brink:1979nt}
\bibitem{Brink:1979nt}
  L.~Brink and P.~S.~Howe,
  ``The $\cN=8$ supergravity in superspace,''
  Phys.\ Lett.\  B {\bf 88} (1979) 268.
  %%CITATION = PHLTA,B88,268;%%   
  
\bibitem{Rosly:1982}
 A.~A.~Rosly,
 ``Super Yang--Mills  constraints
 as integrability conditions,'' in {\it Proceedings of the International
 Seminar on Group Theoretical
 Methods in Physics},'' (Zvenigorod, USSR, 1982),
 M. A. Markov  (Ed.),
 Nauka, Moscow, 1983, Vol. 1, p. 263 (in Russian);
 English translation: in {\it Group Theoretical
 Methods in Physics},'' M. A. Markov, V. I. Man'ko
 and A. E. Shabad  (Eds.), Harwood Academic Publishers,
 London, Vol. 3, 1987, p. 587  
  
%\cite{Howe:1995md}
\bibitem{Howe:1995md}
  P.~S.~Howe and G.~G.~Hartwell,
  ``A superspace survey,''
  Class.\ Quant.\ Grav.\  {\bf 12} (1995) 1823.
  %%CITATION = CQGRD,12,1823;%% 
  
%\cite{Galperin:1987ek}
\bibitem{Galperin:1987ek}
  A.~S.~Galperin, E.~A.~Ivanov, V.~I.~Ogievetsky and E.~Sokatchev,
  ``$\cN=2$ supergravity in superspace: different versions and matter couplings,''
  Class.\ Quant.\ Grav.\  {\bf 4} (1987) 1255.
  %%CITATION = CQGRD,4,1255;%%
  
  
%\cite{Kuzenko:2008ep}
\bibitem{Kuzenko:2008ep}
  S.~M.~Kuzenko, U.~Lindstrom, M.~Rocek and G.~Tartaglino-Mazzucchelli,
  ``4D $\cN = 2$ supergravity and projective superspace,''
  JHEP {\bf 0809} (2008) 051, 
  \eprintN{0805.4683}.
  %%CITATION = ARXIV:0805.4683;%%  
  
%\cite{Hartwell:1994rp}
\bibitem{Hartwell:1994rp}
  G.~G.~Hartwell and P.~S.~Howe,
  ``$(N, p, q)$ harmonic superspace,''
  Int.\ J.\ Mod.\ Phys.\ A {\bf 10} (1995) 3901
  \eprint{hep-th/9412147}.
  %%CITATION = HEP-TH/9412147;%% 
  
%\cite{Intriligator:1998ig}
\bibitem{Intriligator:1998ig}
  K.~A.~Intriligator,
  ``Bonus symmetries of $\cN=4$ superYang--Mills correlation functions via AdS duality,''
  Nucl.\ Phys.\ B {\bf 551} (1999) 575
  \eprint{hep-th/9811047}.
  %%CITATION = HEP-TH/9811047;%%  
  
%\cite{Heslop:2003xu}
\bibitem{Heslop:2003xu}
  P.~J.~Heslop and P.~S.~Howe,
  ``Aspects of $\cN=4$ SYM,''
  JHEP {\bf 0401} (2004) 058
  \eprint{hep-th/0307210}.
  %%CITATION = HEP-TH/0307210;%%        
  
 %\cite{Dobrev:1985qv}
\bibitem{DobrevSC}
  V.~K.~Dobrev and V.~B.~Petkova,
 ``All positive energy unitary irreducible representations of extended
  conformal supersymmetry,''
  Phys.\ Lett.\  B {\bf 162} (1985) 127.
%%CITATION = PHLTA,B162,127;%%
  

%\cite{Antoniadis:2007cw}
\bibitem{Antoniadis:2007cw} 
  I.~Antoniadis, S.~Hohenegger, K.~S.~Narain and E.~Sokatchev,
  ``Harmonicity in $\cN=4$ supersymmetry and its quantum anomaly,''
  Nucl.\ Phys.\ B {\bf 794} (2008) 348,  
  \eprintN{0708.0482}.
  %%CITATION = ARXIV:0708.0482;%%
  
   
 %\cite{deHaro:2002vk}
\bibitem{Skenderis} 
  S.~de Haro, A.~Sinkovics and K.~Skenderis,
  ``On a supersymmetric completion of the $R^4$ term in 2B supergravity,''
  Phys.\ Rev.\ D {\bf 67} (2003) 084010 
  \eprint{hep-th/0210080}.
  %%CITATION = HEP-TH/0210080;%%

%\cite{Howe:1983sra}
\bibitem{Howe:1983sra}
  P.~S.~Howe and P.~C.~West,
  ``The complete $\cN=2$, $D=10$ supergravity,''
  Nucl.\ Phys.\ B {\bf 238} 181 (1984).
  %%CITATION = NUPHA,B238,181;%%
  
%\cite{Berkovits:2001ue}
\bibitem{Berkovits:2001ue}
  N.~Berkovits and P.~S.~Howe,
  ``Ten-dimensional supergravity constraints from the pure spinor formalism for the superstring,''
  Nucl.\ Phys.\ B {\bf 635} (2002) 75
  \eprint{hep-th/0112160}.
  %%CITATION = HEP-TH/0112160;%%  
  

%\cite{Voronov}
\bibitem{Voronov}
 T. Voronov, 
 ``Geometric integration theory on supermanifolds'',
 Sov. Sci. Rev. C: Maths. Phys {\bf 9} (1992) 1.

%\cite{Gates:1997kr}
\bibitem{Gates:1997kr}
  S.~J.~Gates,
  ``Ectoplasm has no topology: The prelude,''
  \eprint{hep-th/9709104}.
  %%CITATION = HEP-TH/9709104;%%

%\cite{Gates:1997ag}
\bibitem{Gates:1997ag}
  S.~J.~Gates, M.~T.~Grisaru, M.~E.~Knutt-Wehlau and W.~Siegel,
  ``Component actions from curved superspace: Normal coordinates and
  ectoplasm,''
  Phys.\ Lett.\  B {\bf 421} (1998) 203,
  \eprint{hep-th/9711151}.
  %%CITATION = PHLTA,B421,203;%%
  
%\cite{Bonora:1986ix}
\bibitem{Bonora:1986ix}
  L.~Bonora, P.~Pasti and M.~Tonin,
  ``Superspace formulation of 10D sugra+SYM theory \`a la Green--Schwarz,''
  Phys.\ Lett.\  B {\bf 188} (1987) 335.
  %%CITATION = PHLTA,B188,335;%% 
  
 %\cite{Cederwall:2001bt}
\bibitem{Cederwall:2001bt}
  M.~Cederwall, B.~E.~W.~Nilsson and D.~Tsimpis,
  ``The structure of maximally supersymmetric Yang--Mills theory: constraining
  higher-order corrections,''
  JHEP {\bf 0106} (2001) 034
  \eprint{hep-th/0102009}.
  %%CITATION = JHEPA,0106,034;%%   
  
%\cite{Cederwall:2001dx}
\bibitem{Cederwall:2001dx}
  M.~Cederwall, B.~E.~W.~Nilsson and D.~Tsimpis,
  ``Spinorial cohomology and maximally supersymmetric theories,''
  JHEP {\bf 0202} (2002) 009
  \eprint{hep-th/0110069}.
  %%CITATION = JHEPA,0202,009;%%
  
%\cite{Howe:2003cy}
\bibitem{Howe:2003cy}
  P.~S.~Howe and D.~Tsimpis,
  ``On higher order corrections in M theory,''
  JHEP {\bf 0309} (2003) 038
  \eprint{hep-th/0305129}.
  %%CITATION = HEP-TH/0305129;%%    
  
%\cite{Berkovits:2008qw}
\bibitem{Berkovits:2008qw}
  N.~Berkovits and P.~S.~Howe,
  ``The cohomology of superspace, pure spinors and invariant integrals,''
ÊÊJHEP {\bf 0806} (2008) 046, 
ÊÊ\eprintN{0803.3024}.
ÊÊ%%CITATION = ARXIV:0803.3024;%% 
  
%\cite{Howe:1991mf}
\bibitem{Howe:1991mf}
  P.~S.~Howe,
  ``Pure spinors lines in superspace and ten-dimensional supersymmetric
  theories,''
  Phys.\ Lett.\  B {\bf 258} (1991) 141
  [Addendum-ibid.\  B {\bf 259} (1991) 511].
  %%CITATION = PHLTA,B258,141;%%

%\cite{Howe:1991bx}
\bibitem{Howe:1991bx}
  P.~S.~Howe,
  ``Pure spinors, function superspaces and supergravity theories in
  ten-dimensions and eleven-dimensions,''
  Phys.\ Lett.\  B {\bf 273} (1991) 90.
  %%CITATION = PHLTA,B273,90;%%  
  
%\cite{Berkovits:2002zk}
\bibitem{Berkovits:2002zk}
  N.~Berkovits,
  ``ICTP lectures on covariant quantization of the superstring,''
  \eprint{hep-th/0209059}.
  %%CITATION = HEP-TH/0209059;%%  
  
%\cite{Brandt:2009xv}
\bibitem{Brandt:2009xv}
  F.~Brandt,
  ``Supersymmetry algebra cohomology I: Definition and general structure,''
  J.\ Math.\ Phys.\  {\bf 51} (2010) 122302, 
  \eprintN{0911.2118}.
  %%CITATION = ARXIV:0911.2118;%%  
  
%\cite{Brandt:2010tz}
\bibitem{Brandt:2010tz}
  F.~Brandt,
  ``Supersymmetry algebra cohomology III: Primitive elements in four and five dimensions,''
  J.\ Math.\ Phys.\  {\bf 52} (2011) 052301, 
 \eprintN{1005.2102}.
  %%CITATION = ARXIV:1005.2102;%%   

%\cite{Movshev:2010mf}
\bibitem{Movshev:2010mf}
  M.~V.~Movshev, A.~Schwarz and R.~Xu,
  ``Homology of Lie algebra of supersymmetries,''
  \eprintN{1011.4731}.
  %%CITATION = ARXIV:1011.4731;%%
    
    

%\cite{Green:1998by}
\bibitem{GreenSethi} 
  M.~B.~Green and S.~Sethi,
  ``Supersymmetry constraints on type IIB supergravity,''
  Phys.\ Rev.\ D {\bf 59}  (1999) 046006
  \eprint{hep-th/9808061}.
  %%CITATION = HEP-TH/9808061;%%
  

  
%\cite{Grisaru:1976nn}
\bibitem{Grisaru:1976nn}
  M.~T.~Grisaru,
  ``Two loop renormalizability of supergravity,''
  Phys.\ Lett.\ B {\bf 66} (1977) 75.
  %%CITATION = PHLTA,B66,75;%%  

%\cite{Deser:1977nt}
\bibitem{Deser:1977nt}
  S.~Deser, J.~H.~Kay and K.~S.~Stelle,
  ``Renormalizability properties of supergravity,''
  Phys.\ Rev.\ Lett.\  {\bf 38} (1977) 527.
  %%CITATION = PRLTA,38,527;%%

%\cite{Deser:1978br}
\bibitem{Deser:1978br}
  S.~Deser and J.~H.~Kay,
  ``Three loop counterterms for extended supergravity,''
  Phys.\ Lett.\  B {\bf 76} (1978) 400.
  %%CITATION = PHLTA,B76,400;%%
  
    %\KuzenkoRY 
\bibitem{KuzenkoRY} 
  S.~M.~Kuzenko and G.~Tartaglino-Mazzucchelli, 
  ``Different representations for the action principle in $4D$ $\cN = 2$ 
  supergravity,'' 
  JHEP {\bf 0904} (2009) 007, 
  \eprintN{0812.3464}. 
  %%CITATION = JHEPA,0904,007;%% 
  
  %\cite{Antoniadis:2006mr}
\bibitem{Antoniadis:2006mr} 
  I.~Antoniadis, S.~Hohenegger and K.~S.~Narain,
  ``$\cN=4$ topological amplitudes and string effective action,''
  Nucl.\ Phys.\ B {\bf 771} (2007) 40 
  \eprint{hep-th/0610258}.
  %%CITATION = HEP-TH/0610258;%%
  
%\cite{Alvarez:1984yi}
\bibitem{Singer}
  O.~Alvarez, I.~M.~Singer and B.~Zumino,
  ``Gravitational anomalies and the family's index theorem,''
  Commun.\ Math.\ Phys.\  {\bf 96}, 409 (1984).
  %%CITATION = CMPHA,96,409;%%
  
%\cite{Bern:2010ue}
\bibitem{Bern:2010ue} 
  Z.~Bern, J.~J.~M.~Carrasco and H.~Johansson,
  ``Perturbative quantum gravity as a double copy of gauge theory,''
  Phys.\ Rev.\ Lett.\  {\bf 105} (2010) 061602 
  \eprintN{1004.0476}.
  %%CITATION = ARXIV:1004.0476;%%  

   \bibitem{LanceZvi}
Z.~Bern and L.~J.~Dixon, Private communication. 

%\cite{Carrasco:2013ypa}
\bibitem{Carrasco:2013ypa} 
  J.~J.~M.~Carrasco, R.~Kallosh, R.~Roiban and A.~A.~Tseytlin,
  ``On the $U(1)$ duality anomaly and the S-matrix of $\cN=4$ supergravity,''
  \eprintN{1303.6219}.
  %%CITATION = ARXIV:1303.6219;%%
  

%\cite{diVecchia:1984jh}
\bibitem{diVecchia:1984jh}
  P.~di Vecchia, S.~Ferrara and L.~Girardello,
  ``Anomalies of hidden local chiral symmetries in sigma models and extended supergravities,''
  Phys.\ Lett.\ B {\bf 151} (1985) 199.
  %%CITATION = PHLTA,B151,199;%%


  %\cite{deWit:1985bn}
\bibitem{deWit}
  B.~de Wit and M.~T.~Grisaru, 
  ``Compensating fields and anomalies,'' In {\it Essays in
  Honor of 60th birthday of E.S.~Fradkin},
  Quantum field theory and quantum statistics, Vol. 2, 411 
    %%CITATION = PRINT-85-0499-UTRECHT-;%%

 %\cite{Pugh:2010ii}
\bibitem{Pugh:2010ii}
  T.~G.~Pugh, E.~Sezgin and K.~S.~Stelle,
  ``D=7 / D=6 Heterotic Supergravity with Gauged R-Symmetry,''
  JHEP {\bf 1102} (2011) 115
  \eprintN{1008.0726}.
  %%CITATION = ARXIV:1008.0726;%%


\bibitem{PS}
  O.~Piguet and S.~P.~Sorella,
  ``Algebraic renormalization: Perturbative renormalization, symmetries and
  anomalies,''
  Lect.\ Notes Phys.\  {\bf M28}, 1 (1995).
  %%CITATION = LNPHA,M28,1;%%
  
\bibitem{Harvey:1996ir} 
   J.~A.~Harvey and G.~W.~Moore,
  ``Five-brane instantons and $R^2 $ couplings in $\cN=4$ string theory,''
  Phys.\ Rev.\ D {\bf 57}  (1998) 2323
  \eprint{hep-th/9610237}.
  %%CITATION = HEP-TH/9610237;%%  
  
  
\bibitem{Gregori:1997hi} 
  A.~Gregori, E.~Kiritsis, C.~Kounnas, N.~A.~Obers, P.~M.~Petropoulos and B.~Pioline,
  ``$R^2$ corrections and nonperturbative dualities of $\cN=4$ string ground states,''
  Nucl.\ Phys.\ B {\bf 510} (1998) 423 
  \eprint{hep-th/9708062}.
  %%CITATION = HEP-TH/9708062;%%
  
\bibitem{Yasuda:1988fi}
  O.~Yasuda,
  ``Nonrenormalization theorem for the Green--Schwarz counterterm and the low-energy effective action,''
  Phys.\ Lett.\ B {\bf 218} (1989) 455.
  %%CITATION = PHLTA,B218,455;%%
  
  
%\cite{Kiritsis:2000zi}
\bibitem{Kiritsis:2000zi} 
  E.~Kiritsis, N.~A.~Obers and B.~Pioline,
  ``Heterotic / type II triality and instantons on K3,''
  JHEP {\bf 0001}  (2000) 029
 \eprint{hep-th/0001083}.
  %%CITATION = HEP-TH/0001083;%%

 %\cite{Fradkin:1983xs}
\bibitem{Fradkin}
  E.~S.~Fradkin and A.~A.~Tseytlin,
  ``One loop infinities in dimensionally reduced supergravities,''
  Phys.\ Lett.\ B {\bf 137} (1984) 357.
  %%CITATION = PHLTA,B137,357;%%


  %\cite{Howe:2010nu}
\bibitem{Howe:2010nu} 
  P.~S.~Howe, U.~Lindstrom and L.~Wulff,
  ``D=10 supersymmetric Yang--Mills theory at $\alpha^\prime{}^4$,''
  JHEP {\bf 1007} (2010) 028,  
  \eprintN{1004.3466}.
  %%CITATION = ARXIV:1004.3466;%%
  
%\cite{Bergshoeff:1986jm}
\bibitem{Bergshoeff:1986jm}
  E.~Bergshoeff, M.~Rakowski and E.~Sezgin,
  ``Higher Derivative Superyang-mills Theories,''
  Phys.\ Lett.\ B {\bf 185} (1987) 371.
  %%CITATION = PHLTA,B185,371;%%  
  
%\cite{Koerber:2002zb}
\bibitem{Koerber:2002zb}
  P.~Koerber and A.~Sevrin,
  ``The non-abelian D-brane effective action through order $\alpha^{\prime}{}^4$,''
  JHEP {\bf 0210} (2002) 046
  \eprint{hep-th/0208044}.
  %%CITATION = JHEPA,0210,046;%% 
  
  
  %\cite{Green:2005ba}
\bibitem{Green:2005ba}
  M.~B.~Green and P.~Vanhove,
  ``Duality and higher derivative terms in M theory,''
  JHEP {\bf 0601} (2006) 093
  \eprint{hep-th/0510027}.
  %%CITATION = HEP-TH/0510027;%%

  
%\cite{SokZup}
\bibitem{SokZup}
E. Sokatchev and B. Zupnik, unpublished.   

%\cite{Howe:1988qz}
\bibitem{Howe:1988qz}
  P.~S.~Howe and K.~S.~Stelle,
  ``The ultraviolet properties of supersymmetric field theories,''
ÊÊInt.\ J.\ Mod.\ Phys.\ A {\bf 4} (1989) 1871.
ÊÊ%%CITATION = IMPAE,A4,1871;%%%\cite{Gates:1979wg}


\bibitem{Gates:1979wg}
  S.~J.~Gates, Jr., K.~S.~Stelle and P.~C.~West,
  ``Algebraic origins of superspace constraints in supergravity,''
ÊÊNucl.\ Phys.\ B {\bf 169} (1980) 347.
ÊÊ%%CITATION = NUPHA,B169,347;%%

%\cite{Stelle:1980uw}
\bibitem{Stelle:1980uw}
  K.~S.~Stelle and P.~C.~West,
  ``Algebraic derivation of $\cN=2$ supergravity constraints,''
ÊÊPhys.\ Lett.\ B {\bf 90} (1980) 393.
ÊÊ%%CITATION = PHLTA,B90,393;%%

%\cite{Siegel:1978mj}
\bibitem{Siegel:1978mj}
  W.~Siegel and S.~J.~Gates, Jr.,
  ``Superfield supergravity,''
ÊÊNucl.\ Phys.\ B {\bf 147} (1979) 77.
ÊÊ%%CITATION = NUPHA,B147,77;%%

%\cite{Grisaru:1981xm}
\bibitem{Grisaru:1981xm}
 M.~T.~Grisaru and W.~Siegel,
  ``Supergraphity: (I). Background field formalism,''
  Nucl.\ Phys.\ B {\bf 187} (1981) 149.
  %%CITATION = NUPHA,B187,149;%%


%\cite{Grisaru:1982zh}
\bibitem{Grisaru:1982zh}
  M.~T.~Grisaru and W.~Siegel,
  ``Supergraphity: (2). Manifestly covariant rules and higher loop finiteness,''
  Nucl.\ Phys.\ B {\bf 201} (1982) 292
   [Erratum-ibid.\ B {\bf 206} (1982) 496].
  %%CITATION = NUPHA,B201,292;%%

%\cite{Howe:1983sr}
\bibitem{Howe:1983sr}
  P.~S.~Howe, K.~S.~Stelle and P.~K.~Townsend,
  ``Miraculous ultraviolet cancellations in supersymmetry made manifest,''
ÊÊNucl.\ Phys.\ B {\bf 236} (1984) 125.
ÊÊ%%CITATION = NUPHA,B236,125;%%

%\cite{Baulieu:1985gy}
\bibitem{Baulieu:1985gy} 
  L.~Baulieu and M.~P.~Bellon,
  ``A simple algebraic construction of the symmetries of supergravity,''
  Phys.\ Lett.\ B {\bf 161} (1985) 96 .
  %%CITATION = PHLTA,B161,96;%%

%\cite{Blasi:2000qw}
\bibitem{Sorella1}
  A.~Blasi, V.~E.~R.~Lemes, N.~Maggiore, S.~P.~Sorella, A.~Tanzini, O.~S.~Ventura and L.~C.~Q.~Vilar,
  ``Perturbative beta function of $\cN=2$ superYang--Mills theories,''
  JHEP {\bf 0005} (2000) 039
  \eprint{hep-th/0004048}.
  %%CITATION = HEP-TH/0004048;%%  
  
  %\cite{Lemes:2001vf}
\bibitem{Sorella2}
  V.~E.~R.~Lemes, M.~S.~Sarandy, S.~P.~Sorella, O.~S.~Ventura and L.~C.~Q.~Vilar,
  ``An algebraic criterion for the ultraviolet finiteness of quantum field theories,''
  J.\ Phys.\ A A {\bf 34} (2001) 9485
  \eprint{hep-th/0103110}.
  %%CITATION = HEP-TH/0103110;%% 
  
%\cite{Schwarz:1993vs}
\bibitem{Schwarz:1993vs}
  J.~H.~Schwarz and A.~Sen,
  ``Duality symmetric actions,''
  Nucl.\ Phys.\ B {\bf 411} (1994) 35
  \eprint{hep-th/9304154}.
  %%CITATION = HEP-TH/9304154;%%
  
%\cite{Sokatchev:1988qr}
\bibitem{Sokatchev:1988qr}
  E.~Sokatchev,
  ``An off-shell formulation of $\cN=4$ supersymmetric Yang--mills theory in twistor harmonic superspace,''
  Phys.\ Lett.\ B {\bf 217} (1989) 489.
  %%CITATION = PHLTA,B217,489;%%
  
%\cite{Howe:1981nz}
\bibitem{Howe:1981nz}
  P.~S.~Howe and U.~Lindstrom,
  ``The supercurrent in five-dimensions,''
  Phys.\ Lett.\ B {\bf 103} (1981) 422.
  %%CITATION = PHLTA,B103,422;%%   
  
   \bibitem{D'Hoker:2005jc}
  E.~D'Hoker and D.~H.~Phong,
  ``Two-loop superstrings VI: Non-renormalization theorems and the 4-point function,''
  Nucl.\ Phys.\ B {\bf 715} (2005) 3
  \eprint{hep-th/0501197}.
  %%CITATION = HEP-TH/0501197;%%
  
  %\cite{Tseytlin:1995bi}
\bibitem{Tseytlin:1995bi} 
  A.~A.~Tseytlin,
  ``Heterotic type I superstring duality and low-energy effective actions,''
  Nucl.\ Phys.\ B {\bf 467}  (1996) 383
  \eprint{hep-th/9512081}.
  %%CITATION = HEP-TH/9512081;%%
  
 %\cite{Bossard:2010pk}
\bibitem{Bossard:2010pk}
  G.~Bossard, P.~S.~Howe, U.~Lindstrom, K.~S.~Stelle and L.~Wulff,
  ``Integral invariants in maximally supersymmetric Yang--Mills theories,''
  JHEP {\bf 1105} (2011) 021, 
  \eprintN{1012.3142}.
  %%CITATION = ARXIV:1012.3142;%%  
  
%\cite{Howe:1994ms}
\bibitem{Howe:1994ms}
  P.~S.~Howe and M.~I.~Leeming,
  ``Harmonic superspaces in low dimensions,''
  Class.\ Quant.\ Grav.\  {\bf 11} (1994) 2843
  \eprint{hep-th/9408062}.
  %%CITATION = HEP-TH/9408062;%% 

  
  
    
\end{thebibliography}
 \end{document}